\documentclass[12pt,a4paper,german,english]{scrartcl}

\usepackage{hyperref}

\usepackage{babel}
\usepackage{amsmath}
\usepackage{amsthm}
\usepackage{amssymb}
\usepackage{amsxtra}
\usepackage{yhmath}
\usepackage{eulervm}
\usepackage{wasysym}
\usepackage{makeidx}
\usepackage{wrapfig}
\usepackage[latin1]{inputenc}
\usepackage[T1]{fontenc}
\usepackage{url}
\usepackage{mathrsfs}
\usepackage{fixltx2e}
\usepackage{mparhack}
\newcommand{\osumD}{\makebox [0pt][l]{\,$\bigcirc$}\sum}
\newcommand{\osumT}{\parbox {0pt} {\makebox[0pt][l]{$\,\scriptstyle\bigcirc$}}\sum}
\newcommand{\osumS}{\parbox {0pt} {\kern0.1ex\makebox[0pt][l]{$\,\textstyle \circ $}}\sum}
\newcommand{\osumSS}{\parbox {0pt} {\kern0.15ex\makebox[0pt][l]{$\,\textstyle \circ$}}\sum}
\newcommand{\osum}{\mathop{\mathchoice {\osumD} {\osumT} {\osumS} {\osumSS}}}
\newcommand{\ir}{\mathrm{i}}
\newcommand{\e}{\mathrm{e}}
\newcommand{\dr}{\mathrm{d}}

\DeclareMathOperator{\ghost}{gh}
\DeclareMathOperator{\dv}{\mathrm d}
\DeclareMathOperator{\s}{\mathrm s}

\DeclareMathOperator{\ro}{\mathrm r}
\DeclareMathOperator{\st}{\tilde{\mathrm s}}
\DeclareMathOperator{\bv}{\mathrm b}
\DeclareMathOperator{\tr}{tr}

\DeclareMathOperator{\modulo}{mod\,}

\ifx\KOMAScript\undefined%
  \DeclareRobustCommand{\KOMAScript}{\textsf{K\kern.05em O\kern.05em%
      M\kern.05em A\kern.1em-\kern.1em Script}}
\fi

\newcommand{\dvp}{\,\mathrm d_\phi}
\DeclareMathOperator{\dvt}{\mathrm d_{\tilde T}}
\DeclareMathOperator{\pgh}{pgh}
\DeclareMathOperator{\af}{af}

\numberwithin{equation}{section}
\makeindex
\newtheoremstyle{note}
  {3pt}
  {3pt}
  {\itshape}
  {}
  {\bfseries}
  {:}
  {.5em}
  {}

\theoremstyle{note}

\def\Real{\mbox{R\hspace{-.9em}I}\,\ }
\newtheorem{theorem}{Theorem}
\def\A#1#2#3{\mbox{${#1}_{#2}{}^{#3}$}}
\def\F#1#2{\A{\cal F}{#1}{#2}}

\def\T#1#2{\A{T}{#1}{#2}}
\begin{document}
\addcontentsline{toc}{section}{Overview}
\begin{flushright}
ITP--UH--11/12
\end{flushright}

\begin{center}
 {\large\bfseries BRST Symmetry and Cohomology}
 \\[5mm]
 Norbert Dragon and Friedemann Brandt \\[2mm]
 \textit{Institut f\"ur Theoretische Physik, Leibniz Universit\"at Hannover, Appelstra\ss e 2, D-30167 Hannover, Germany}
\end{center}

\begin{abstract}
We present the mathematical considerations
which determine all gauge invariant actions and anomaly candidates in 
gauge theories of standard type  such as ordinary or gravitational Yang Mills theories. 
Starting from elementary concepts of field theory the discussion tries to be explicit 
and complete, only the cohomology of simple Lie algebras it quoted from the literature.
\end{abstract}
\pagenumbering{roman}

\selectlanguage{english}

\section*{Overview}

After a short introduction to jet spaces section \ref{sec1} deals with the ``raison d'\^etre'' 
of gauge symmetries: the problem to define the subspace of physical states in a Lorentz invariant
theory with higher spin. The operator $Q_s$ which characterizes the physical
states was found by Carlo Becchi, Alain Rouet and Raymond Stora as a symmetry generator of a
fermionic symmetry, the \textsc{brst} symmetry, in gauge theories with covariant gauge
fixing \cite{brs}. Independently Igor Tjutin described the symmetry in a Lebedev Institute report 
which however remained unpublished for political reasons.
For a derivation of the \textsc{brst} symmetry from the gauge
fixing in path integrals the reader may consult the
literature \cite{beau,henteit,Gomis:1995he}. Section \ref{sec1} is supplemented by a discussion of
free vectorfields for gauge parameter $\lambda \ne 1$. This is not a
completely trivial exercise \cite{itzub} and rarely  discussed in detail \cite{henteit}.

Section \ref{sec2} deals with the requirement that the physical subspace
remains physical if interactions are switched on. This restricts the action to
be \textsc{brst} invariant. Consequently the Lagrange density has to satisfy a
cohomological equation similar to the physical states. Quantum
corrections may violate the requirement of \textsc{brst} symmetry because
the naive evaluation of Feynman diagrams leads to divergent loop
integrals which have to be regularized. This regularization can lead to
an anomalous symmetry breaking. It has to satisfy a cohomological equation, the 
Wess Zumino consistency condition \cite{wesszumino}.

In section \ref{sec3} we study some elementary cohomological problems of a
nilpotent fermionic derivative $\dv$,
\begin{equation}
\nonumber
\dv^2 = 0\quad ,\quad \dv \omega = 0\quad,\quad \omega \modulo \dr \eta\ .
\end{equation}
We derive the Poincar\'e lemma as the basic lemma of all the investigations to come.
In particular one has to consider Lagrange densities as jet functions, i.e. functions
of the fields and  their derivatives and not only of the coordinates. We
investigate differential forms depending on these jet variables and derive
the algebraic Poincar\'e lemma which is where Lagrangians of local actions enter the stage.
The relative cohomology, which characterizes Lagrange densities and candidate anomalies, is shown to
lead to the descent equations which can again be written compactly as a
cohomological problem. The section concludes with K\"unneth's formula
which allows to tackle cohomological problems in smaller bits if the
complete problem factorizes.

Section \ref{sec4} presents a formulation \cite{Brandt:1993xq} of the gravitational \textsc{brst} transformations in
which the cohomology factorizes. Consequently one has to deal
only with the subalgebra of tensors and undifferentiated ghosts. 
It is shown that the ghosts which correspond to translations can be removed
from anomalies (if the space-time dimension exceeds two)\footnote{The two-dimensional case can be special \cite{Brandt:1995gu}.}, i.e. coordinate transformations are not anomalous.

In section \ref{sec5} we solve the cohomology of the \textsc{brst} transformations acting
on ghosts and tensors. The tensors have to couple together with the
translation ghosts to invariants and also the ghosts for spin and
isospin transformations have to couple to invariants. The invariant
ghost polynomials generate the Lie algebra cohomology which we quote
from the mathematical literature \cite{greub3}.  Moreover the tensors
are restricted by the covariant Poincar\'e lemma \cite{Brandt:1989et}, for which we give a
simplified proof. This lemma introduces the Chern forms. They are the integrands of
all local actions which do not change under a smooth change of the fields 
and therefore give topological informations about classes of fields
which are related by smooth deformations.

In section \ref{sec6} we exhibit the Chern forms as the \textsc{brst} transformation of the 
Chern Simons forms. Chern Simons forms can contribute to local gauge invariant actions though they
are not gauge invariant. They are independent of the metric and do
not contribute to the energy momentum tensor but nevertheless influence the field equations.
We conclude by giving examples of Lagrange densities and anomaly candidates.

In section \ref{sec7} we sketch how the cohomological analysis presented in sections \ref{sec3} to \ref{sec6} can be 
extended to include antifields and how the cohomology is affected by the inclusion of antifields.

\newpage

\tableofcontents

\newpage


\pagenumbering{arabic}

\section{The Space of Physical States}\label{sec1}
\subsection{Indefinite Fock Space}
\textsc{brst} symmetry is indispensable  in Lorentz covariant
theories with fields with higher spin because it allows to
construct an acceptable space of physical states out of the Fock space
which contains states with negative norm.

Before we demonstrate the problem, we recollect some elementary definitions and concepts. 
A (bosonic) field $\phi$ is a map of a base space, which locally is some domain of $\mathbb{R}^D$ 
with points $x=(x^0,x^1,x^2,\dots x^{D-1})$
to a target space $\mathbb{R}^d$, 
\begin{equation}
\phi:\left \{
\begin{array}{c c l}
\mathbb{R}^D &\rightarrow & \mathbb{R}^d\\
x  &\mapsto & \phi(x) =(\phi^1(x), \phi^2(x) \dots  \phi^{d}(x))\ .
\end{array}
\right .
\end{equation}
By assumption we consider fields, which are sufficiently differentiable. Each field defines a field $\hat{\phi}$,
the prolongation of $\phi$ to the jet space $\mathcal{J}_1$\,,  
\begin{equation}
\hat{\phi}:\left \{
\begin{array}{c c l}
\mathbb{R}^D &\rightarrow & \mathbb{R}^{D + d + D\,d}\\
x &\mapsto & (x, \phi(x), \partial_0 \phi(x), \partial_1 \phi(x), \partial_2 \phi(x) \dots  \partial_{D-1} \phi(x))
\end{array}
\right .\ .
\end{equation}
Locally the jet space $\mathcal J_1$ is the cartesian product of some domain of the base space, 
the target space and the tangent space of a point. 

Analogously, the prolongation of $\phi$ to the jet space $\mathcal{J}_k$ maps $x$ to $x$, the field $\phi(x)$ 
and its partial derivatives $\partial\dots \partial \phi(x)$ up to $k$\textsuperscript{th} order. The 
prolongation $\hat \phi$ of an infinitely differentiable field maps the base space to 
$\mathcal J = \mathcal J_\infty$ and each point $x$ to $x$, $\phi(x)$ and all its derivatives at $x$\,. 

Jet functions $\mathcal L$ are maps from some $\mathcal J_k$, 
where $k$ is \emph{finite}, to $\mathbb R$\,. By composition with the projection 
\begin{equation}
\pi_k:\left \{
\begin{array}{c c l}
\mathcal J & \rightarrow & \mathcal J_k\\
(x, \phi, \partial \phi, \dots,\partial^k \phi,\dots)  &\mapsto & (x, \phi,\partial \phi,\dots,  \partial^k \phi) 
\end{array}
\right .
\end{equation}
each jet function can be constantly continued to the function $\mathcal L \circ \pi_k$ of $\mathcal J$. 

In notation we do not distinguish between $\mathcal L$ and its constant continuation but consider jet functions
as functions of some $\mathcal J_k$ or of $\mathcal J$ as needed.

The action $W$ is a local functional of fields $\phi$, which is to say it maps fields to the integral 
over a jet function, 
the Lagrange density $\mathcal L$, evaluated on the prologation of the fields,
\begin{equation}
W: \phi \mapsto W[\phi] = \int\! \dr^D x \,({\mathcal L}\circ \hat{\phi})(x)\ .
\end{equation}
The equations of motion are derived from the variational principle 
that for physical fields the action $W$ be stationary up to boundary terms
under all variations of the fields. This holds if and only if the Euler derivative of the
Lagrangian \footnote{The dots denote terms which occur if $\mathcal L$ 
depends on second or higher derivatives of $\phi$\,.}
\begin{equation}
\label{eulerableit}
\frac{\hat\partial \mathcal L}{\hat \partial \phi^i}= 
\frac{\partial \mathcal L}{\partial \phi^i} - \partial_n
\frac{\partial \mathcal L}{\partial (\partial_n\phi^i)} + \dots\ ,
\end{equation}
vanishes on the prolongation of the physical field,
\begin{equation}
\frac{\hat\partial \mathcal L}{\hat \partial \phi}\circ \hat \phi_{\text{physical}}= 0\ .
\end{equation}

In case of the massless vectorfield $A$, $D=d=4$, and the Lagrangian is 
\begin{equation}
\label{wmax}
{\cal L}:\left \{
\begin{array}{c c l}
\mathcal{J}_1 &\rightarrow & \mathbb{R}\\
(x, A, \partial A) &\mapsto & 
= - \frac{1}{4e^2} (\partial_m A_n -
\partial_n A_m)(\partial^m A^n - \partial^n A^m)
-\frac{\lambda}{2e^2}(\partial_m A^m)^2
\end{array}
\right .
\end{equation}
Here we use the shorthand $A^m = \eta^{mk}A_k$ and $\partial^n = \eta^{nl}\partial_l$ where 
$\eta$ is the diagonal matrix $\eta=\text{diag}(1,-1,-1,-1)$\,.
To avoid technical complications at this stage we consider
the case $\lambda = 1$, $\lambda\ne 1$ is discussed at the end of this section.
We choose to introduce the gauge coupling~$e$ as normalization of the
kinetic energies to avoid its appearance in Lie algebras, which we have to consider later.

The physical vectorfield has to satisfy the wave equation,
\begin{equation} 
\frac{1}{e^2}\Box A_n(x)
= 0 \quad,\quad \Box=\eta^{mn}\partial_m\partial_n=\partial_0{}^2
- \partial_1{}^2 - \partial_2{}^2 - \partial_3{}^2\ ,
\end{equation}
with the solution
\begin{equation}
\label{free}
A_n(x) = e \int\! \tilde{\dr}k\,
(\e^{\ir kx}
a_n^\dagger(\vec{k})
 +
\e^{-\ir kx}
a_n(\vec{k})
)_{\displaystyle |_{k^0 = \sqrt{\vec{k}^2}}}\ .
\end{equation}
Here we use the notation
\begin{equation}
\tilde{\dr}k=\frac{\dr^3k}{(2\pi)^3 2|\vec{k}|}\ ,\ 
k x = k^0 x^0 - k^1 x^1 - k^2 x^2 - k^3 x^3 = k^m x^n \eta_{mn}\ .
\end{equation}

The vectorfield is quantized by the requirement that the propagator, the
vacuumexpectation value of the time ordered product of two fields,
\begin{equation}
\label{prop}
\langle \Omega | T A^m(x)A_n(0) \, \Omega\rangle\ ,
\end{equation}
be  a Green function corresponding to the Euler derivative,
\begin{equation}
\label{green}
\frac{1}{e^2} \Box
\langle \Omega | T A^m(x)A_n(0) \, \Omega \rangle =
\ir\, \delta^4(x)\,\delta^m{}_n\ .
\end{equation}
The creation and annihilation operators
$a^\dagger(\vec{k})$  and $a(\vec{k})$ are identified by their commutation
relations with the momentum operators $P^m$\,,
\begin{equation}
\left[ P_m , a^\dagger_n(\vec{k})\right] = k_m a^\dagger_n(\vec{k})\quad,\quad
\left[ P_m , a_n(\vec{k})\right] = - k_m a_n(\vec{k})\ ,
\end{equation}
which follow because by definition the momentum operators $P_m$ generate translations,
\begin{equation}
\label{heisen}
\left[\ir P_m , A_n(x)\right] = \partial_m A_n(x)\ .
\end{equation}
$a^\dagger_n(\vec{k})$ adds and $a_n(\vec{k})$ subtracts energy
$k_0=\sqrt{\vec{k}^2}\ge 0$.
Consequently the annihilation operators annihilate the lowest energy state, the vacuum
$|\Omega\rangle$\,, and justify their denomination,
\begin{equation}
P_m |\Omega\rangle = 0 \ ,\   a(\vec{k}) |\Omega\rangle = 0\ .
\end{equation}
For $x^0 > 0$ the propagator (\ref{prop}) contains only positive frequencies from
$\e^{-\ir kx}a_m(\vec{k})$, for $x^0 < 0$ only negative frequencies from
$\e^{\ir kx}a^\dagger _m(\vec{k})$. These boundary conditions fix the solution to
(\ref{green}) to be
\begin{equation}
\langle \Omega | T A_m(x)A_n(0) \, \Omega\rangle =
-\ir \,e^2\ \eta_{mn} \lim_{\epsilon\rightarrow 0+}  
\int\! \frac{\dr^4p}{(2\pi)^4}\,\frac{\e^{\ir px}}{p^2 + \ir \epsilon}
\end{equation}
with $\eta = \text{diag}(1,-1,-1,-1)$.
Evaluating the $p^0$ integral for positive and for negative $x^0$ and
comparing with the explicit expression for the propagator (\ref{prop}) which
results if one inputs the free fields (\ref{free}) one can read off
$\langle \Omega | a_m(\vec{k}) a^\dagger_n(\vec{k^\prime})\,\Omega \rangle$ and the value
of the commutator
\begin{equation}
\label{comm}
\left[a_m(\vec{k}),a^\dagger_n(\vec{k^\prime})\right]
= -\ \eta_{mn}(2\pi)^3 2 k^0 \delta^3 (\vec{k}-\vec{k^\prime})\ .
\end{equation}

It is inevitable that the Lorentz metric $\eta$
appears in such commutation relations in Lorentz covariant theories
with fields with higher spin.  The Fock space which results from
such commutation relations necessarily contains negative
norm states because the Lorentz metric is indefinite and contains both signs.
In particular the state
\begin{equation}
|f_0\rangle =
\int\! \tilde{\dr}k\,f(\vec{k})\,a^\dagger_0(\vec{k})\,|\Omega\rangle
\end{equation}
has negative norm
\begin{equation}
\langle f_0 | f_0\rangle
= - \eta_{00} \int\!\tilde{\dr}k\,|f(\vec{k})|^2 < 0 \ .
\end{equation}

\subsection{Definiteness of the Scalar Product of Physical States }

Such a space with an indefinite scalar product cannot be the space of physical states because 
in quantum mechanics
\begin{equation}
\label{wahr}
w(i,A,\Psi)=|\langle\Lambda_i |\Psi \rangle|^2
\end{equation}
is the probability for the measurement to yield the result number $i$ (which for simplicity we take to be nondegenerate 
and discrete), if the state $\Psi$ is measured with the apparatus~$A$. Here the states $\Lambda_j$ are the
eigenstates of $A$, which yield the corresponding result number~$j$ with certainty
\begin{equation}
|\langle \Lambda_i | \Lambda_j\rangle|^2 = \delta^i{}_j\ .
\end{equation}
Therefore different $\Lambda_i$ are orthogonal to each other (and therefore linearly independent)
\begin{equation}
\langle \Lambda_i | \Lambda_j \rangle =
0 \ ,\quad  \text{ if } i\ne j\ .
\end{equation}
For $i=j$ the scalar product of the eigenstates is  real, 
$\langle \Phi |\Psi\rangle^\ast = \langle \Psi |\Phi\rangle$\,,
and has modulus~$1$,
\begin{equation}
\langle \Lambda_i | \Lambda_j \rangle = \eta_{ij}\ ,\quad \eta = \text{diag}(1,1,\dots, -1,-1,\dots)\ .
\end{equation}
In the space, which is spanned by the eigenstates, the scalar product therefore is of the form
\begin{equation} 
\langle \Lambda | \Psi \rangle = ( \Lambda |\eta \Psi)\ ,
\end{equation}
where the scalar product $(\Lambda | \Psi)$ is positive definite  and $\eta$ is the linear map which maps
$\Lambda_i$ to $\sum_j \Lambda_j \eta_{ji}$\,.
In particular the eigenstates $\Lambda_i$ of the measuring apparatus $A$ and each other apparatus are 
eigenvectors of $\eta$. 

But a superposition $\Gamma=a\Lambda_1+b\Lambda_2$ with $a\,b\ne 0$ is an eigenvector of~$\eta$
only if the eigenvalues $\eta_{11}$ and $\eta_{22}$ coincide. Therefore, in the space of physical
states, which contains the eigenvectors of all measuring devices and their superpositions, the scalar
product has to be definite.

\subsection{Classical Electrodynamics}
In classical electrodynamics (in the vacuum) one does not have the
troublesome amplitude $a^\dagger_0(\vec{k})$. There the wave equation
$\Box A_n =0$ results from Maxwell's equation
$\partial_m(\partial^m A^n - \partial^n A^m) = 0$ and the Lorenz condition
$\partial_m A^m = 0$. This gauge condition fixes the vectorfield up to
the gauge transformation $A_m\mapsto A^\prime_m = A_m + \partial_m C$
where $C$ satisfies the wave equation $\Box C = 0$.
In terms of the free fields $A$ and $C$
\begin{equation}
\label{freec}
C(x)= e \int\!\tilde{\dr}k\,
\Bigl(\e^{\ir kx}c^\dagger(\vec{k}) +
\e^{-\ir kx}c(\vec{k})\Bigr)_{\displaystyle |_{k^0 = \sqrt{\vec{k}^2}}}
\end{equation}
the Lorenz condition concerns the linear combination $k^m a_m $ of the amplitudes 
\begin{equation}
\partial_m A^m =
\ir\,e \int\!\tilde{\dr}k\,
\Bigl(\e^{\ir kx}k^m a_m^\dagger(\vec{k}) -
\e^{-\ir kx}k^m a_m(\vec{k})\Bigr)_{\displaystyle |_{k^0 = \sqrt{\vec{k}^2}}}
\end{equation}
and the gauge transformation changes the amplitudes by a contribution in direction~$k$
\begin{equation}
\label{gauge}
A^\prime_m - A_m = \partial_m C =
\ir\,e \int\!\tilde{\dr}k\,
\Bigl(\e^{\ir kx}k_m c^\dagger(\vec{k}) -
\e^{-\ir kx}k_m c(\vec{k})\Bigr)_{\displaystyle |_{k^0 = \sqrt{\vec{k}^2}}}\ .
\end{equation}
To make this even more explicit, we decompose the creation operator $a_m^\dagger(\vec{k})$ into parts in
the direction of the lightlike momentum $k$, in the direction $\bar{k}$
(which is $k$ with reflected 3-momentum)
\begin{equation}
(\bar{k}^0,\bar{k}^1,\bar{k}^2,\bar{k}^3)=(k^0,-k^1,-k^2,-k^3)\,
\end{equation}
and in two directions $n^1$ and $n^2$ which are orthogonal to $k$ and
$\bar{k}$ \footnote{For example $\vec{n}^1(\vec{k})\propto \vec{w}\times \vec{k}\ ,\ 
\vec{n}^2(\vec{k})\propto \vec{n}^{1\,*}\!\times \vec{k} $, where $\vec{w}$ is a constant complex
vector with linearly independent real and imaginary part.}
\begin{equation}
\label{ahel}
a_m^\dagger(\vec{k})=
\sum_{\tau= k,\bar{k},1,2}
\epsilon_m^*{}^{\tau} a_\tau^\dagger(\vec{k})\ .
\end{equation}
These polarization vectors $\epsilon^\tau(\vec{k})$ are functions of the lightcone $\mathbb R^3 - \{0\}$
\begin{equation}
\label{pola}
\epsilon_m^{*}{}^\tau(\vec{k}) = \Bigl(
\frac{1}{\sqrt2}\frac{k_m}{|\vec{k}|},
\frac{1}{\sqrt2}\frac{\bar{k}_m}{|\vec{k}|},
n_m^1,n_m^2\Bigr),\ \tau = k,\bar{k},1,2
\end{equation}
and have the scalar products
\begin{equation}
\epsilon^{*\,\tau} \cdot \epsilon^{\tau^\prime} =
\begin{pmatrix}
0&1&&\\
1&0&&\\
&&-1&\\
&&&-1
\end{pmatrix}\ .
\end{equation}
The field $\partial_m A^m$ contains the amplitudes
$a^\dagger_{\bar{k}},\ a_{\bar{k}}$.
The Lorenz gauge condition $\partial_m A^m = 0$ eliminates these
amplitudes in classical electrodynamics.

The fields $A^\prime_m$ and $A_m$ differ in the amplitudes
$a^\dagger_{k},\ a_{k}$ in the direction of the momentum~$k$.
An appropriate choice of the remaining gauge transformation (\ref{gauge})
cancels these amplitudes.

So in classical electrodynamics $a^\dagger_m$ can be restricted to 2 degrees
of freedom, the transverse oscillations
\begin{equation}
a_m^\dagger(\vec{k})=
\sum_{\tau= 1,2}
\epsilon_m^{\ast \,\tau} a_\tau^\dagger(\vec{k})\ .
\end{equation}

The corresponding quantized modes generate a positive definite Fock space.

We cannot, however, just require $a^\dagger_k=0$ and
$a^\dagger_{\bar{k}}=0$ in the quantized theory, this would contradict
the commutation relation
\begin{equation}
\left[a_k(\vec{k}),a^\dagger_{\bar{k}}(\vec{k^\prime})\right]
= -\ (2\pi)^3 2 k^0 \delta^3 (\vec{k}-\vec{k^\prime})\ne 0\ .
\end{equation}
 To get rid of the troublesome modes we
require, rather, that physical states do not
contain $a^\dagger_k$ and $a^\dagger_{\bar{k}}$ modes.
This requires the interactions to leave the
subspace of physical states invariant, a requirement, which is not at all
obviously satisfied, because the unphysical modes contribute to the
propagator. As we shall see, both the selection rule of physical states and
the restrictions on the interactions to respect the selection rule emerge from
the \textsc{brst} symmetry.

\subsection{The Physical States}

To single out a physical subspace of the Fock space $\cal F$ we require that
there exists a hermitean operator, the \textsc{brst} operator,
\begin{equation}
Q_s=Q^\dagger_s\ ,
\end{equation}
which defines a subspace $\cal N\subset F$, the gauge invariant states, by
\begin{equation}
{\cal N} = \left\{ |\Psi\rangle : |Q_s\Psi\rangle = 0\right\}\ .
\end{equation}
This requirement is no restriction at all, each subspace can be
characterized as kernel of some hermitean operator.

Inspired by gauge transformations (\ref{gauge}) we take the operator
$Q_s$ to act on one particle states according to
\begin{equation}
\label{sa}
Q_s\,a_m^\dagger(\vec{k})|\Omega\rangle = k_m c^\dagger(\vec{k})|\Omega\rangle\ .
\end{equation}
As a consequence the one particle states generated by
$a_\tau^\dagger(\vec{k})\ ,\  \tau = \bar{k},1,2\,,$ belong to~$\cal N$\,,
\begin{equation}
Q_s\,a_\tau^\dagger(\vec{k})|\Omega\rangle = 0 \ ,\  \tau = \bar{k},1,2\ .
\end{equation}
The states created by the creation operator
$a^\dagger_k$ in the direction of the momentum $k$ are not invariant
\begin{equation}
Q_s\, a_k^\dagger(\vec{k})|\Omega\rangle =
\sqrt{2} |\vec{k}| c^\dagger(\vec{k})|\Omega\rangle\ \ne 0
\end{equation}
and do not belong to $\cal N$.

The space $\cal N$ is not yet acceptable because it contains
nonvanishing zero-norm states
\begin{equation}
\label{zero}
|f\rangle =
\int\! \tilde{\dr}k\, f(\vec{k})\, a_{\bar{k}}^\dagger(\vec{k})|\Omega\rangle\ ,\ 
\langle f | f \rangle = 0\ ,\text{ because }\
\left[a_{\bar{k}}(\vec{k}),a^\dagger_{\bar{k}}(\vec{k^\prime})\right]
= 0\ .
\end{equation}
To get rid of these states the following observation is crucial:
\begin{theorem}{\quad}\\
\label{t1}
Scalar products of gauge invariant states
$|\psi\rangle \in \cal N$ and
$|\chi\rangle \in \cal N$
remain unchanged if the state $|\psi\rangle$ is replaced by
$|\psi + Q_s \Lambda\rangle $ .
\end{theorem}
Proof:
\begin{equation}
\label{scal}
\langle\chi |\psi +  Q_s \Lambda\rangle  =
\langle\chi|\psi\rangle + \langle\chi|Q_s \Lambda\rangle =
\langle\chi|\psi\rangle
\end{equation}
The term $\langle\chi|Q_s \Lambda\rangle$ vanishes, because $Q_s$ is
hermitean and $Q_s\chi = 0$ .

We obtain the \textsc{brst} algebra from the seemingly innocent requirement that
$|\psi +  Q_s\Lambda\rangle$ belongs to $\cal N$ whenever
$|\psi\rangle $ does. The requirement seems natural because
$|\psi +  Q_s\Lambda\rangle$ and $|\psi\rangle $ have the same
scalar products with gauge invariant states and therefore cannot be
distinguished experimentally. It is, nevertheless, a very restrictive condition,
because it requires $Q_s^2$ to vanish on each state $|\Lambda\rangle$, i.e. 
$Q_s$ is required to be nilpotent,
\begin{equation}
Q_s^2 = 0\ .
\end{equation}
Then the space $\cal N$ of gauge invariant states decomposes into equivalence classes
\begin{equation}
\label{equiv}
|\psi\rangle \sim |\psi +  Q_s\Lambda\rangle\ .
\end{equation}
These equivalence classes are the physical states,
\begin{equation}
\label{phys}
{\cal H}_{\text{phys}}=\frac{\cal N}{Q_s \cal F} =
\left\{ |\psi\rangle :  |Q_s\psi\rangle = 0 \ , |\psi\rangle\  {\modulo}\
|Q_s\Lambda\rangle \right\}\ .
\end{equation}
${\cal H}_{\text{phys}}$ inherits a scalar product from $\cal F$ because by theorem \ref{t1} 
the scalar product in $\cal N$ does not depend on the representative of the equivalence
class.

The construction of ${\cal H}_{\text{phys}}$ by itself does not guarantee that
${\cal H}_{\text{phys}}$ has a positive definite scalar product. This will hold
only if $Q_s$ acts on the space $\cal F$ in a suitable manner. One has to check this
positive definiteness in each class of models.

In the case at hand, the zero-norm states $|f\rangle$ (\ref{zero})
are equivalent to $0$
in ${\cal H}_{\text{phys}}$ if there exists a massless, real field $\bar{C}(x)$
\begin{equation}
\label{cq}
\bar{C}(x)= e \int\! \tilde{\dr}k\,
\Bigl(\e^{\ir kx}\bar{c}^\dagger(\vec{k}) +
\e^{-\ir kx}\bar{c}(\vec{k})\Bigr)_{\displaystyle |_{k^0 = \sqrt{\vec{k}^2}}}
\end{equation}
and if $Q_s$ transforms the one-particle states according to
\begin{equation}
\label{scq}
Q_s\, \bar{c}^\dagger(\vec{k})\, |\Omega\rangle =
\sqrt{2}\,\ir \, |\vec{k}|\, a^\dagger_{\bar{k}}(\vec{k})\,|\Omega\rangle\ .
\end{equation}
For the six one-particle states we conclude that
$\bar{c}^\dagger(\vec{k}) |\Omega\rangle$ and
$a^\dagger_{k}(\vec{k}) |\Omega\rangle$ are not invariant (not in $\cal N$),
$a^\dagger_{\bar{k}}(\vec{k}) |\Omega\rangle$  and
$c^\dagger(\vec{k}) |\Omega\rangle$ are of the form $Q_s |\Lambda\rangle$ and
equivalent to $0$, the remaining two transverse creation operators
generate the physical one particle space with positive norm.

Notice the following pattern: states from the Fock space $\cal F$ are
excluded in pairs from the physical Hilbert space ${\cal H}_{\text{phys}}$,
one state, $|n\rangle$, is not invariant
\begin{equation}
\label{nt}
Q_s|n\rangle = |t\rangle \ne 0
\end{equation}
and therefore not contained in $\cal N$, the other state, $|t\rangle$, is trivial
and equivalent to $0$ in ${\cal H}_{\text{phys}}$ because it is the \textsc{brst}
transformation of $|n\rangle$.

The algebra $Q_s^2=0$ enforces
\begin{equation}
\label{t0}
Q_s |t\rangle = 0\ .
\end{equation}
If one uses $|t\rangle$
and $|n\rangle$ as basis then $Q_s$ is represented by the matrix
\begin{equation}
\label{mat2}
Q_s = 
\begin{pmatrix}
0 & 1 \\
0 & 0
\end{pmatrix}\ .
\end{equation}
This is one of the two possible Jordan block matrices which can represent
a nilpotent operator $Q_s^2=0$. The only eigenvalue is 0, so a Jordan block
consists of a matrix with zeros and with 1 only in the upper diagonal
\begin{equation}
Q_{s\,ij}=\delta_{i+1,j}\ .
\end{equation}
Because of $Q_s^2 = 0$ the blocks can only have the size
$1\times 1$ or $2\times 2$. In the first case the corresponding vector on
which $Q_s$ acts is invariant and not trivial and contributes to
${\cal H}_{\text{phys}}$. The second case is given by (\ref{mat2}), the corresponding
vectors are not physical.

It is instructive to consider the scalar product of the states on which
$Q_s$ acts. If it is  positive definite then $Q_s$ has to vanish because $Q_s$
is hermitean and can be diagonalized in a space with positive definite scalar
product. Thereby the nondiagonalizable $2\times2$ block (\ref{mat2})
would be excluded. It is,
however, in Fock spaces with indefinite scalar product that we need the
\textsc{brst} operator and there it can act nontrivially. In the physical Hilbert space,
which has a positive definite scalar product, $Q_s$ vanishes.
Nevertheless the existence of the \textsc{brst} operator $Q_s$ in Fock space severely
restricts the possible actions of the models we are going to consider.

Reconsider the doublet (\ref{nt}, \ref{t0}): if the scalar product is nondegenerate 
then by a suitable choice of~$|n\rangle$ and~$|t\rangle$
it can be brought to  the standard form 
\begin{equation}
\langle n|n\rangle = 0 = \langle t|t\rangle\quad
\langle t|n\rangle =  \langle n|t\rangle = 1 \ .
\end{equation}
This is an indefinite scalar product of Lorentzian type
\begin{equation}
|e_\pm\rangle= \frac{1}{\sqrt{2}}(| n\rangle \pm | t\rangle) \quad
\langle e_+|e_-\rangle = 0 \quad
\langle e_+|e_+\rangle = - \langle e_-|e_-\rangle = 1\ .
\end{equation}
By the definition (\ref{phys}) pairs of states with
wrong sign norm and with acceptable norm are excluded from the 
space ${\cal H}_{\text{phys}}$ of physical states.

\subsection{Gauge Parameter $\lambda\ne 1$}

If the gauge parameter $\lambda$ is different from $1$, then the vectorfield
has to satisfy the coupled equations of motion
\begin{equation}
\label{eqnln1}
\frac{1}{e^2}(\Box A_n + (\lambda - 1)\partial_n \partial_m A^m )= 0\ ,
\end{equation}
which imply 
\begin{equation}
\label{box2}
\Box\Box A_m = 0
\end{equation}
and its Fourier transformed version $(p^2)^2\tilde{A}_m=0$\,.
Consequently the Fourier tansformed field $\tilde{A}$ vanishes outside the
light cone and the general solution $\tilde{A}$ contains a $\delta$-function 
and its derivative.
\begin{equation}
\tilde{A}_m = a_m(p)\delta(p^2) + b_m(p)\delta^\prime (p^2)
\end{equation}
However, the derivative of the $\delta$ function is ill defined because
spherical coordinates $p^2,v,\vartheta,\varphi$ are discontinuous
at $p=0$.

To solve $\Box\Box \phi = 0$ one can restrict $\phi(t,\vec{x})$ to
$\phi(t)\e^{\ir \vec{k}\vec{x}}$, the general solution can then be
obtained as a wavepacket which is superposed out of solutions of this
form. $\phi(t)$ has to satisfy the
ordinary differential equation
\begin{equation}
(\frac{\dr^2}{\dr t^2}+k^2)^2\phi=0
\end{equation}
which  has the general solution 
\begin{equation}
\phi(t)=(a +b\, t)\e^{ \ir k t}+ (c+d\, t)\e^{- \ir k t}\ .
\end{equation}
Therefore the equations (\ref{box2}) are solved by
\begin{equation}
A_n(x) = e \int\! \tilde{\dr}k\,
\e^{\ir kx} \Bigl(a_n^\dagger(\vec{k}) +
x^0  b^\dagger_ n(\vec{k})\Bigr)+
\e^{-\ir kx} \Bigl(a_n(\vec{k})
+x^0 b_n(\vec{k})
\Bigr)_{\displaystyle |_{k^0 = \sqrt{\vec{k}^2}}}\ .
\end{equation}

This equation makes the vague notion $\delta^\prime(p^2)$ explicit.
The amplitudes $b_n$, $b^\dagger_n$ are determined from the coupled
equations (\ref{eqnln1}),
\begin{equation}
\label{freeln1}
b^\dagger_ n(\vec{k})=
-\ir\, \frac{\lambda -1}{\lambda + 1}\, \frac{k_nk^m}{k_0}\,  a^\dagger_ m(\vec{k})\ ,\ 
b_ n(\vec{k})=
\ir\, \frac{\lambda -1}{\lambda + 1}\, \frac{k_nk^m}{k_0}\,  a_ m(\vec{k})\ .
\end{equation}
{}From (\ref{heisen}) one can deduce that the commutation relations
\begin{equation}
[P^i,a^\dagger_m(\vec{k})]=k^i\,a^\dagger_m(\vec{k})\ , \ 
[P^i,a_m(\vec{k})]=-k^i\,a_m(\vec{k})\ , \  i=1,2,3\ ,
\end{equation}
\begin{equation}
\label{hacom}
\text{and\qquad }
[P_0,a^\dagger_m(\vec{k})]=k_0\,a^\dagger_m(\vec{k})
-\frac{(\lambda-1)}{(\lambda +1)}\frac{k_m k^n}{k_0}a^\dagger_n(\vec{k})
\end{equation}
have to hold.
If we decompose $a^\dagger_m(\vec{k})$ according to (\ref{ahel}) then we
obtain
\begin{equation}
[P_0,a^\dagger_t(\vec{k})]=k_0\,a^\dagger_t(\vec{k})\ ,\  t=1,2\ ,
\end{equation}
for the transverse creation operators and also
\begin{equation}
[P_0,a^\dagger_{\bar{k}}(\vec{k})]=k_0\,a^\dagger_{\bar{k}}(\vec{k})
\end{equation}
for the creation operator in direction of $\bar{k}$. For the creation
operator in the direction of the four momentum $k$ one gets
\begin{equation}
[P_0,a^\dagger_k(\vec{k})]=k_0\, a^\dagger_k(\vec{k})
-2\,k_0\,\frac{\lambda-1}{\lambda +1}\,a^\dagger_{\bar{k}}(\vec{k})\ .
\end{equation}
In particular, for $\lambda \neq 1$, $a^\dagger_k(\vec{k})$ does not
generate energy eigenstates and the hermitean operator $P_0$ cannot be
diagonalized in Fock space because the commutation relations are
\begin{equation}
[P_0,a^\dagger]=M\,a^\dagger
\end{equation}
with a matrix $M$ which contains a nondiagonalizable Jordan block
\begin{equation}
M\sim k_0
\begin{pmatrix}
1&-2\frac{\lambda-1}{\lambda+1}\\
0&1
\end{pmatrix}\ .
\end{equation}

That hermitean operators are not guaranteed to be diagonalizable is of
course related to the indefinite norm in Fock space. For operators
$O_{\text{phys}}$ which correspond to measuring devices it is sufficient that
they can be diagonalized in the physical Hilbert space. This is
guaranteed if ${\cal H}_{\text{phys}}$ has positive norm. In Fock space it is
sufficient that operators $O_{\text{phys}}$  commute with the \textsc{brst} operator
$Q_s$ and that they satisfy generalized eigenvector equations
\begin{equation}
O_{\text{phys}}|\psi_{\text{phys}}\rangle = c |\psi_{\text{phys}}\rangle +
 Q_s |\chi\rangle \ ,\  c \in \Real\ ,
\end{equation}
from which the spectrum can be read off.

The Hamilton operator $H=P_0$ which results from the Lagrange density,
\begin{equation}
{\cal
L}=-\frac{1}{4e^2}F_{mn}F^{mn}-\frac{\lambda}{2e^2}(\partial_mA^m)^2\ ,
\end{equation}
\begin{align}
H=\frac{1}{2e^2}\! \int \!\dr^3x \,:\Bigl( &(\partial_0 A_i)^2
-(\partial_iA_0)^2 + \frac{1}{2}(\partial_jA_i-\partial_iA_j)(\partial_jA_i-\partial_iA_j) - \nonumber\\
& - \lambda(\partial_0 A_0)^2 + \lambda(\partial_iA_i)^2\Bigr ):\ ,\ i,j\in\{1,2,3\}\ ,
\end{align}
can be expressed in terms of the creation and annihilation operators,
\begin{equation}
H=\int\! \tilde{\dr}k\,k_0\,\Bigl (
\sum_{t=1}^{2}a^\dagger_ta_t  - \frac{2\lambda}{\lambda +1}
\bigl(a^\dagger_ka_{\bar{k}}+a^\dagger_{\bar{k}}a_k-2\frac{\lambda-1}{
\lambda + 1} a^\dagger_{\bar{k}}a_{\bar{k}}\bigr) \Bigr)\ .
\end{equation}
$H$ generates time translations (\ref{hacom}) because the creation and annihilation
operators fulfil  the commutation relations
\begin{equation}
\label{creannco}
[a_m(\vec{k}),a^\dagger_n(\vec{k}^\prime)]= 2k^0(2\pi)^3\delta^3(\vec{k}
- \vec{k}^\prime)\!\Bigl( -\eta_{mn}
+\frac{\lambda -1}{2\lambda k^0}
(\eta_{m0}k_n + \eta_{n0}k_m- \frac{k_m k_n}{k^0})\Bigr)
\end{equation}
which follow from the requirement that the propagator
\begin{equation}
\langle\Omega| {T}A^m(x)A_n(0)\Omega \rangle = -\ir\, e^2
\lim_{\varepsilon \rightarrow 0\scriptscriptstyle{+}}
\int\!
\frac{\dr^4p}{(2\pi)^4}\,\frac{ \e^{{\ir px}}}{(p^2+\ir \varepsilon)^2}
\Bigl (p^2\delta^m{}_n-\frac{\lambda-1}{\lambda}p^mp_n\Bigr )
\end{equation}
is the Green function corresponding to the equation of motion
(\ref{eqnln1}), which for positive (negative) times contains positive (negative) frequencies only.
If one decomposes the creation and annihilation operators according to
(\ref{ahel}) then the transverse operators satisfy
\begin{equation}
[a_i(\vec{k}),a^\dagger_j(\vec{k}^\prime)]=2k^0(2\pi)^3\delta^3(\vec{k}
- \vec{k}^\prime)\,\delta_{ij}\ ,\  i,j \in \{1,2\}\ .
\end{equation}
They commute with the other creation annihilation operators which have
the following off diagonal commutation relations
\begin{equation}
[a_{\bar{k}}(\vec{k}),a^\dagger_k(\vec{k}^\prime)]
=[a_{{k}}(\vec{k}),a^\dagger_{\bar{k}}(\vec{k}^\prime)]
=-\frac{\lambda +1}{2\lambda}2k^0(2\pi)^3\delta^3(\vec{k}-
\vec{k}^\prime)\ .
\end{equation}
The other commutators vanish.

Just as for $\lambda = 1$ the analysis of the \textsc{brst} transformations leads again to
the result that physical states are generated only by the transverse
creation operators.

\section{BRST Symmetry}\label{sec2}
\subsection{Graded Commutative Algebra}
To choose the physical states one could have proceeded like
Cinderella and could pick acceptable states by hand or have them picked by
doves. Prescribing the action of $Q_s$ on one particle states (\ref{sa},
\ref{scq}) is not really different from such an arbitrary approach.
{}From (\ref{sa}, \ref{scq}) we know nothing about physical multiparticle
states. Moreover we would like to know whether one can switch on interactions
which respect our definition of physical states. Interactions should
give transition amplitudes which are independent of the choice (\ref{equiv})
of the representative of physical states. The time evolution should
leave physical states physical.

All these requirements can be satisfied if the \textsc{brst}  operator $Q_s$
belongs
to a symmetry. We interpret the equation $Q_s^2=0$ as a graded commutator,
an anticommutator, of a fermionic generator of a Lie algebra
\begin{equation}
\label{qsq}
\{Q_s,Q_s\} = 0\ .
\end{equation}

To require that $Q_s$ be fermionic means that the \textsc{brst}  operator
transforms fermionic variables into bosonic variables and vice versa. In
particular we take the vectorfield~$A$ to be a bosonic field. Then the fields
$C$ and $\bar{C}$ have to be fermionic
though they are real scalar fields and carry no spin. They violate the
spin  statistics relation which requires physical fields with half-integer
spin to be fermionic and fields with integer spin to be bosonic. However, the corresponding particles do not
occur in physical states, they are ghosts. We call $C$ the ghost field and
$\bar{C}$ the antighost field. Because the ghost fields~$C$ and $\bar{C}$
anticommute they contribute, after introduction of interactions, to each loop
with the opposite sign as compared to bosonic contributions.
The ghosts compensate in loops for the unphysical bosonic degrees of freedom
contained in the vectorfield $A$.

We want to realize the algebra (\ref{qsq}) as local transformations
on fields and to determine actions which are invariant
under these transformations. {}From this invariant action one can construct the 
\textsc{brst}  operator as Noether charge corresponding to the symmetry of the action.

The transformations act on polynomials in bosonic and fermionic variables $\phi^i$. This means:
on the vector space of linear combinations of these variables, there acts a linear map, the Grassmann reflection
$\Pi$, $\Pi^2 = 1$. Each linear combination $\phi$ can be uniquely decomposed into its bosonic part,
$(\phi + \Pi \phi)/2$, which by definition is even (invariant) under Grassmann reflection, and into its fermionic part
$(\phi - \Pi \phi)/2$, which by definition changes sign under Grassmann reflection. 
For simplicity, we assume the variables $\phi^i$ chosen such, that they are either bosonic or fermionic
and introduce the grading $|\phi^i|$ modulo 2, such that $\Pi \phi^i = (-1)^{|\phi^i|}\phi^i$,
\begin{equation}
|\phi^i|=\left\{
\begin{array}{l}
0 \text{ if }\phi^i\text{ is  bosonic}\\
1 \text{ if }\phi^i\text{ is  fermionic}\ .
\end{array}
\right.
\end{equation}

By assumption the bosonic and fermionic variables have an associative product and are graded commutative,
\begin{equation}
\phi^i\phi^j=(-1)^{|\phi^i|\cdot|\phi^j|}\phi^j\phi^i =: (-)^{ij}\phi^j\phi^i\ ,
\end{equation}
i.e. bosons commute with bosons and fermions, fermions commute with bosons and
anticommute with fermions.

For readability we often use the shorthand notation
\begin{equation}
(-)^{ij}:=(-1)^{|\phi^i|\cdot|\phi^j|}\ .
\end{equation}

By linearity and the product rule $\Pi(AB) = \Pi(A)\Pi(B)$ the Grassmann reflection extends to polynomials. 
The grading of products is the sum of the gradings,
\begin{equation}
|\phi^i\phi^j|=|\phi^i|+|\phi^j| \modulo 2\ .
\end{equation}

Like the elementary variables, also each polynomial can be decomposed into its
bosonic and its fermionic parts. These parts have a definite grading and are 
graded commutative
\begin{equation}
A B = (-1)^{|A|\cdot|B|}B A\ .
\end{equation}

Transformations and symmetries are operations $O$ acting linearly, i.e. term
by term, on polynomials, \footnote{We deal with the graded commutative algebra of fields and their
derivatives and distinguish operations, acting on the algebra, from operators, acting in some Fock space.} 
\begin{equation}
O(\lambda_1 A + \lambda_2 B) = \lambda_1 O(A) + \lambda_2 O(B)\ .
\end{equation}
They are uniquely specified by their action on bosonic and on fermionic polynomials and can
be decomposed into bosonic operations, which map bosons to bosons and fermions to fermions,
\begin{equation}
|O_{\text{bosonic}}(A)|=|A|\ ,
\end{equation}
and fermionic operations, which maps bosons to fermions and fermions to bosons, 
\begin{equation}
|O_{\text{fermionic}}(A)|=|A|+ 1 \modulo 2\ .
\end{equation}
We consider only bosonic or fermionic operations. They have a natural grading,
\begin{equation}
|O|=|O(A)| - |A|\  \modulo 2\ .
\end{equation}
The grading of composite operations is the sum of the gradings
\begin{equation}
|O_1 \,O_2| = |O_1| + |O_2|\ \modulo 2\ .
\end{equation}
First order derivatives $v$ are linear operations with a graded
Leibniz rule\footnote{This Leibniz rule defines left
derivatives: the left factor $A$ is differentiated without a
graded sign.}
\begin{equation}
\label{leibniz}
v(AB)=(vA)B + (-)^{|v|\cdot|A|}A(vB)\ .
\end{equation}
They are completely determined by their action on elementary variables,
$\phi^i$: $v(\phi^i)=v^i$, i.e. $v=v^i\partial_i$. The partial derivatives
$\partial_i$ are naturally defined by 
\begin{equation}
\partial_i\phi^j=\delta_i^{\phantom{i}j}.
\end{equation}
They have the same grading as their corresponding variables,
\begin{equation}
\label{partial}
|\partial_i| = |\phi^i|\ , \ 
\partial_i\partial_j=(-)^{ij}\partial_j\partial_i\ .
\end{equation}
The grading of the components $v^i$ results naturally 
$|v^i|=|v|+|\phi^i|\modulo 2$.

An example of a fermionic derivative is given by the exterior derivative
\begin{equation}
\label{dext}
\dv =\dr x^m\partial_m\ , \ |\dv| = 1\ .
\end{equation}
It transforms coordinates $x^m$ into differentials $\dr x^m$ which have
opposite statistics
\begin{equation}
|\dr x^m|=|x^m| + 1 \modulo 2
\end{equation}
and which, considered as multiplicative operations, commute with $\partial_n$
\begin{equation}
\quad [\partial_n,\dr x^m] = 0\ .
\end{equation}
Therefore and because of (\ref{partial}) the exterior derivative is nilpotent
\begin{equation}
\dv ^2=0\ .
\end{equation}

The graded commutator of operations $O$ and $P$
\begin{equation}
\label{gradcom}
[O, P] = O\,P - (-1)^{|O|\,|P|} P\,O
\end{equation}
(i.e. the anticommutator, if both $O$ and $P$ are fermionic, or the commutator,
if $O$ is bosonic or $P$ is bosonic,) is linear in both arguments, graded antisymmetric
\begin{equation}
[O, P] =  - (-1)^{|O|\,|P|} [P,O]\ ,
\end{equation}
and satisfies the product rule
\begin{equation}
\label{gradprod}
[O,P\,Q] = [O,P]\,Q + (-1)^{|O|\,|P|}P\,[O,Q]\ .
\end{equation}
The graded commutator of first order derivatives is a first order derivative, i.e. satisfies
the Leibniz rule (\ref{leibniz}).

\subsection{Conjugation}

Lagrange densities have to be real polynomials to make the corresponding
$S$-matrix unitary. This is why we have to discuss complex conjugation. We
define conjugation such that hermitean conjugation of a time ordered operator
corresponding to some polynomial gives the
anti time ordered operator corresponding to the conjugate polynomial. We
therefore require for all variables $\phi^i$ and complex numbers $\lambda_i$
\begin{align}
(\phi^{i\,\ast})^\ast &= \phi^i \ ,\\
(\lambda_i \phi^i )^\ast &=
\lambda_i^\ast \phi^{i\,\ast} \ , \\
(\phi^i\phi^j)^\ast &= \phi^{j\,\ast}\, \phi^{i\,\ast} = (-)^{ij}\phi^{i\,\ast} \phi^{j\,\ast}\ .
\end{align}
As a consequence, conjugation preserves the grading,
\begin{equation}
|\phi^{i\,\ast}| = |\phi^i| \ ,
\end{equation}
and by additivity is defined on polynomials.

The conjugation of operations $O$ is defined by
\begin{equation}
\label{opcon}
O^\ast(A)=(-)^{|O||A|}(O(A^\ast))^\ast\ .
\end{equation}
This definition ensures that $O^\ast$ is linear and satisfies the Leibniz rule
if $O$ is a first order derivative. Both requirements have to hold in order to 
allow first order derivatives and their Lie-algebra to be real i.e. self conjugate. 

The exterior derivative $\dv$ is real, $\dv=\dv^\ast$, if the conjugate differentials are 
related to the differentials of the conjugate variables by
\begin{equation}
\quad (\dr x^m)^\ast = (-)^{|x^m|}\dr ((x^m)^\ast)\ .
\end{equation}

The partial derivative with respect to a real fermionic variable
is purely imaginary.  Also the operator $\delta$ 
is purely imaginary,
\begin{equation}
\delta=x^m\frac{\partial}{\partial (\dr x^m)} \ ,\  \delta^\ast = - \delta\ . 
\end{equation}

The anticommutator of $\dv$ and $\delta$ can be evaluated with the product rule of the
graded commutator (\ref{gradprod})  and with the elementary graded commutator 
\begin{equation}
[\frac{\partial}{\partial \phi^i},\phi^j] = \delta_i{}^j\ ,
\end{equation}
of the partial derivative and the operation, which multiplies with the variable $\phi^j$,
\begin{equation}
\label{nx}
\Delta=\{\dv ,\delta\}= x^m\frac{\partial}{\partial x^m} +
\dr x^m\frac{\partial}{\partial (\dr x^m)} = N_x + N_{\dr x}\ .
\end{equation}
The anticommutator counts the variables $x$ and $\dr x$ and is real as one can check with
\begin{equation}
(O_1O_2)^\ast=(-)^{|O_1||O_2|}O_1^\ast O_2^\ast
\end{equation}
which follows from (\ref{opcon}). 

Conjugation does \emph{not} reverse the order of two operations
$O_1$ and $O_2$.

We can now specify the main properties of the \textsc{brst}  transformation
$\s$: It is a real, fermionic, nilpotent first order derivative,
\begin{equation}
\s=\s^\ast \ ,\  |\s|=1\ ,\   \s^2=0\ ,\  \s(A\,B) = (\s A)\,B + (-1)^{|A|}A \s B\ .
\end{equation}
It acts on Lagrange densities and functionals of fields.
Space-time derivatives $\partial_m$ of fields are limits of differences of
fields taken at neighbouring arguments. It follows from the linearity
of $\s$ that it has to commute with space-time derivatives
\begin{equation}
\label{sder}
[\s,\partial_m]=0\ .
\end{equation}
Linearity  implies moreover that the \textsc{brst}  transformation of integrals
is given by the integral of the transformed integrand. Therefore the
differentials $\dr x^m$ are \textsc{brst}  invariant,\footnote{The first equation applies to the element $\dr x^m$
of the graded commutative algebra, the second to the operation, which multiplies elements of the algebra with $\dr x^m$.}
\begin{equation}
\s(\dr x^m) = 0 = \{\s,\dr x^m\} \ ,\  ( [\s,\dr x^m]=0 \text{ for fermionic }x^m )\ .
\end{equation}
Taken together the last two equations imply that $\s$ and $\dv$
(\ref{dext}) anticommute
\begin{equation}
\{\s,\dv \}=0\ .
\end{equation}

\subsection{Independence of the Gauge Fixing}

In the simplest multiplet $\s$ transforms a real anticommuting field
$\bar{C}=\bar{C}^\ast$, the antighost field, into $\sqrt{-1}$
times a real bosonic field
$B=B^\ast$, the auxiliary field. These denominations anticipate the roles
which the fields  will play in  Lagrange densities,
\begin{equation}
\label{scqt}
\s\bar{C}(x)=\ir B(x)\ ,\ \s B(x)=0\ .
\end{equation}
The \textsc{brst}  transformation which corresponds to an abelian gauge transformation
acts on a real bosonic vectorfield $A$ and a real, fermionic ghost
field~$C$ by
\begin{equation}
\label{sam}
\s A_m(x)=\partial_m C(x) \ ,\  \s C(x)=0\ .
\end{equation}

We can attribute to the fields 
\begin{equation}
\phi=(\bar{C},B,A,C)
\end{equation}
and to $\s$ and $\partial=(\partial_0,\partial_1,\dots,\partial_{D-1})$ a ghostnumber,
which adds on multiplication
\begin{gather}
\ghost(\bar{C})= -1 ,\  \ghost(B) =0  ,\  \ghost(A)=0 ,\  \ghost(C)=1 ,\ 
\ghost(\s)=1 ,\ \ghost(\partial)=0 ,\\
\ghost(M\,N) = \ghost(M) + \ghost(N)\ .
\end{gather}
Our analysis of the algebra (\ref{scqt}, \ref{sam}) in $D=4$ dimensions\footnote{%
In odd dimensions also Chern Simons
forms can occur.} will show: 

All Lagrangians of \textsc{brst} invariant  local actions 
\begin{equation}
W[\phi] =\int\! \dr^4x\, ({\cal L}\circ \hat{\phi})(x)
\end{equation}
with ghostnumber $0$ have the form
\begin{equation}
\label{lagr}
{\cal L} = {\cal L}_{\text{inv}}(F,\partial F,\dots)
+ \ir \s X(\phi,\partial \phi, \dots )\ .
\end{equation}
The part ${\cal L}_{\text{inv}}$ is real and depends only on the field
strength 
\begin{equation}
F_{mn}=-F_{nm}= \partial_m A_n -\partial_n A_m
\end{equation}
and its partial derivatives.
Therefore it is invariant under classical gauge transformations.
Typically it is given by (\ref{wmax})
\begin{equation}
\label{linv}
{\cal L}_{\text{inv}}(A,\partial A) = -\frac{1}{4e^2}F_{mn}F^{mn}\ .
\end{equation}

$X(\phi,\partial\phi,\dots)$ is a real, fermionic polynomial with ghostnumber
$-1$. Therefore, it has to contain a factor $\bar{C}$. In the simplest case it is
\begin{equation}
X=\frac{\lambda}{e^2}\,\bar{C}\,(-\frac{1}{2}B+\partial_m A^m)\ .
\end{equation}
$\lambda$ is the gauge fixing parameter. The piece $\ir \s X$ contributes the
gaugefixing for the vectorfield and contains the action of the ghostfields
$C$ and $\bar{C}$,
\begin{equation}
\label{lgf}
\ir \s X = \frac{\lambda}{2 e^2}(B - \partial_m A^m)^2
-\frac{\lambda}{2 e^2}(\partial_m A^m)^2 -\ir \frac{\lambda}{e^2}
\bar{C}\partial_m\partial^mC\ .
\end{equation}
This Langrange density makes $B$ an auxiliary field, its equation
of motion fixes it algebraically,  $B = \partial_m A^m$. $C$ and $\bar{C}$
are free fields (\ref{freec}, \ref{cq}).

The Lagrangian is invariant under scale transformations $T_a$, $a \in \mathbb R$,
\begin{equation}
T_a \bar{C} = \e^{-a} \bar{C} \ ,\ T_a {C} = \e^{a} {C}\ ,T_a A_m = A_m\ ,T_a B = B\ .
\end{equation}
The corresponding Noether charge is the ghostnumber.

To justify the name gauge fixing for the gauge breaking part
$-\frac{\lambda}{2 e^2}(\partial_m A^m)^2$ of the Lagrange density
we show that a change of the fermionic function $X$  cannot be measured in
amplitudes of physical states as long as such a change leads only to
a differentiable perturbation of amplitudes. This means that gauge fixing
and ghostparts of the Lagrange density are unobservable. Only the parameters
in the gauge invariant part ${\cal L}_{\text{inv}}$ are measurable.
\begin{theorem}{\quad}\\
Transition amplitudes of physical states are independent of the gauge fixing
within perturbatively connected gauge sectors.
\end{theorem}
Proof: If one changes $X$ by $\delta X$ then the Lagrange density and the
 action change by
\begin{equation}
\delta {\cal L} = \ir \s \delta X \ ,\  \delta W =
\ir \int\! \dr^4x\,\s \delta X\ .
\end{equation}
$S$-matrix elements of physical states $|\chi\rangle$ and $|\psi\rangle$
change to first order by
\begin{equation}
\delta \langle\chi_\text{in}|\psi_\text{out}\rangle =
\langle\chi_\text{in}|\ir\cdot \ir \int\! \dr^4x\,\s \delta X |\psi_\text{out}\rangle
\end{equation}
where $\s \delta X$ is an operator in Fock space.
The transformation $\s \delta X$ of the operator $\delta X$
is generated by $\ir$ times the
anticommutator of the fermionic operator $\delta X$ with the fermionic
\textsc{brst}  operator $Q_s$
\begin{equation}
\langle\chi_\text{in}| \s \int\! \dr^4x\, \delta X |\psi_\text{out}\rangle =
\langle\chi_\text{in}| \{\ir\,Q_ s,\int\! \dr^4x\,\delta X\} |\psi_\text{out}\rangle\ .
\end{equation}
This matrix element vanishes because $|\chi\rangle$ and $|\psi\rangle$ are
physical (\ref{phys}) and $Q_s$ is hermitean.

The proof does not exclude the possible existence of different sectors
of gauge fixing which cannot be joined smoothly by changing the parameters.

\subsection{Invariance and Anomalies}

Using this theorem we can concisely express the restriction which the
Lagrange density of a local, \textsc{brst}  invariant action in $D$ dimensions has
to satisfy.

It is advantageous to combine $\cal L$ with the
differential $\dr^Dx$ and consider the Lagrange density as a $D$-form
 \footnote{We indicate the ghostnumber by the
superscript and denote the form degree by the subscript.} $\omega^0_{D}= {\cal L}\,\dr^Dx$ with ghostnumber 0\,.
The \textsc{brst}  transformation of the Lagrange density $\omega^0_{D}$
has to give a (possibly vanishing) total derivative $\dv \omega_{D-1}^1$.

With this notation the condition for an invariant local action is
\begin{equation}
\label{invloc}
\s \omega^0_{D} + \dv \omega^1_{D-1} = 0\ .
\end{equation}

It is sufficient to determine this Lagrange density $\omega^0_{D}$ up to
a piece of the form $\s \eta^{-1}_{D}$, where $\eta^{-1}_{D}$ carries
ghostnumber -1. Such a piece contributes only to gaugefixing
and to the ghostsector and cannot be observed. It is trivially \textsc{brst}  invariant
because $\s$ is nilpotent. A total derivative part $\dr \eta^0_{D-1}$
 of the Lagrange
density contributes only boundary terms to the action and is also neglected.
This means that we look for the solutions of the equation
\begin{equation}
\label{relcoh}
\s \omega^0_{D} + \dv \omega^1_{D-1} = 0\ ,\ 
\omega^0_{D} \modulo\ ( \s\eta^{-1}_{D} + \dr \eta^0_{D-1} )\ .
\end{equation}

This is a cohomological equation, similar to (\ref{phys}) which
determines the physical states. The equivalence classes of
solutions $\omega^0_{D}$ of this equation span a linear space: the
relative cohomology of $\s$ modulo $\dv$ at ghostnumber~$0$ and form degree $D$\,.

If we use a Lagrange density which solves this equation, then the action
is invariant under the continuous symmetry
\begin{equation}
\phi \rightarrow \phi + \alpha \s \phi
\end{equation}
with an arbitrary fermionic
parameter $\alpha$. In classical field theory Noether's theorem
guarantees that there exists a current $j^m$ which is conserved as a
consequence of the equations of motion,
\begin{equation}
\partial_m j^m = 0\ .
\end{equation}
The Noether charge (from which we strip the parameter $\alpha$)
\begin{equation}
Q_s=\int\! \dr^3x\,j^0(t,x)
\end{equation}
is independent of the time $t$ and generates
the nilpotent \textsc{brst}  transformations
of functionals $A[\phi,\pi]$ of the phase
space variables $\phi^i(x)$ and
$
\pi_i(x)=\frac{\partial \cal L}{\partial \partial_0\phi^i}(x)
$
by the graded Poisson bracket
\begin{equation}
\nonumber
\{A,B\}_\text{P}=\! 
\int\!\! \dr^3x\,
\Bigl( (-1)^{|i|(|i|+|A|)}
\frac{\delta A}{\delta \phi^i(x)}\frac{\delta B}{\delta \pi_i(x)}
- (-1)^{|i||A|}
\frac{\delta A}{\delta \pi_i(x)}\frac{\delta B}{\delta \phi^i(x)}
\Bigr)
\end{equation}
\begin{equation}
\s A = \{ Q_s,A\}_\text{P}\ .
\end{equation}

If one investigates the quantized theory then in the simplest of all
conceivable worlds the classical Poisson brackets
would be replaced by (anti-) commutators of quantized operators.
In particular the \textsc{brst}  operator $Q_s$ would commute with the scattering
matrix $S$\,,
\begin{equation}
\label{s-matrix}
S=`` \,{T}\,\e^{\ir \int\! \dv^4x\,{\cal L}_\text{int}}\,\mbox{''}\quad , \quad [Q_s,S]=0\ ,
\end{equation}
and scattering processes would map physical states unitarily to physical
states
\begin{equation}
S {\cal H}_{\text{phys}} = {\cal H}_{\text{phys}}\ .
\end{equation}
Classically an invariant action is sufficient to ensure this property.
The perturbative evaluation of scattering amplitudes, however, suffers from
the problem, that the $S$-matrix (\ref{s-matrix}) has ill defined
contributions from products of ${\cal L}_\text{int}(x_1)\dots{\cal L}_\text{int}(x_n)$
if arguments $x_i$ and $x_j$ coincide. Though upon integration
$\int\! \dr^4 x_1\dots \dr^4 x_n$ this is a set of measure zero
these products of fields at coinciding space time arguments are the
reason for all ultraviolet divergencies which emerge upon the
naive application of the Feynman rules. More precisely the $S$-matrix is
a time ordered series in $\ir \int\! \dr^4x\,{\cal L}_\text{int}$ and a set of
prescriptions, indicated by the quotes in (\ref{s-matrix}), to define
in each order the products of ${\cal L}_\text{int}(x)$ at coinciding space-time
points. To analyze these divergencies it is sufficient to consider only
connected diagrams. In momentum space they
decompose into products of one particle irreducible
$n$-point functions $\tilde{G}_\text{1PI}(p_1,\dots,p_n)$ which define the
effective action. 
\begin{equation}
\begin{aligned}
\Gamma[\phi]& = \sum_{n=0}^\infty
\frac{1}{n!}\int\! \dr^4x_1\dots \dr^4x_n\,
\phi(x_1)\dots\phi(x_n)\,G_\text{1PI}(x_1,\dots,x_n)\\
& =  \int\! \dr^4x\, {\cal L}_0(\phi,\partial \phi, \dots ) +
\sum_{n\geq 1}\hslash^n\Gamma_n[\phi]
\end{aligned}
\end{equation}

To lowest order in $\hslash$ the effective action $\Gamma$ is the
classical action $\Gamma_0[\phi] = \int\! \dr^4x\, ({\mathcal L}_0\circ \hat \phi)(x)$\,. 
This is a local functional,
in particular ${\mathcal L}_0$ is a series in the fields and a polynomial in the
partial derivatives of the fields. The Feynman diagrams fix the expansion
of the nonlocal effective action $\Gamma = \sum \hslash^n \Gamma_n$
up to local functionals which can be chosen in each loop order, i.e. the Lagrange density 
can be chosen as a series in $\hslash$.
\begin{equation}
{\cal L}= {\cal L}_0 + \sum_{n\geq 1}\hslash^n{\cal L}_n
\end{equation}

The condition that the effective action be \textsc{brst} invariant 
\begin{equation}
\s\, \Gamma[\phi] =0
\end{equation}
has to be satisfied in each loop order. To lowest order it requires the Lagrange density
${\cal L}_0$ to be a solution of (\ref{relcoh}).

Assume the invariance condition to be satisfied up to $n$-loop order. The naive calculation of
$n+1$-loop diagrams contains divergencies which make it necessary to
introduce a regularization, e.g. the Pauli-Villars regularization,
and counterterms  (or use a prescription such as dimensional regularization
or the \textsc{bphz} prescription which is a shortcut for regularization and
counterterms).
No regularization respects locality, unitarity and symmetries simultaneously,
otherwise it would not be a regularization but an acceptable theory.
The Pauli-Villars regularization is local. It violates unitarity for energies
above the regulator masses and also because it violates \textsc{brst}  invariance.
If one cancels the
divergencies of diagrams with counterterms and considers the limit of infinite
regulator masses then unitarity is obtained if the \textsc{brst}  symmetry guarantees
the decoupling of the unphysical gauge modes. Locality was preserved for
all values of the regulator masses. What about \textsc{brst}  symmetry?

One cannot argue
that one has switched off the regularization and that therefore the symmetry
should be restored.  There is the phenomenon of hysteresis.
A spherically symmetric iron ball exposed to a symmetry breaking
magnetic field will usually not become spherically symmetric again if the magnetic field is
switched off. Analogously in the calculation of
$\Gamma_{n+1}$ we have to be prepared that the regularization and the
cancellation of divergencies
by counterterms does not lead to an invariant effective action but rather to
\begin{equation}
\s \Gamma = \hslash^{n+1}a + \sum_{k\geq n+2}\hslash^k a_k\ .
\end{equation}
If the  functional $a$ cannot be made to vanish by an appropriate
choice of ${\cal L}_{n+1}$ then the \textsc{brst}  symmetry is broken by the
anomaly $a$.

Because $\s$ is nilpotent the anomaly $a$ has to satisfy
\begin{equation}
\label{consis}
\s a=0\ .
\end{equation}

This is the celebrated consistency condition of Wess and Zumino
\cite{wesszumino}.
The consistency condition has acquired an outstanding importance because it
allows to calculate all possible anomalies $a$ as the general solution to
$\s a =0$ and to check in each given model whether the anomaly actually
occurs. At first sight one would not expect that the consistency equation
has comparatively few solutions. The \textsc{brst}  transformation $a=\s \Gamma$
of arbitrary  functionals $\Gamma$
satisfies $\s a=0$. The anomaly $a$, however, arises from the divergencies
of Feynman diagrams where all subdiagrams are finite and compatible with
\textsc{brst}  invariance. These divergencies can be isolated in parts of the
$n$-point functions which depend polynomially on the external momenta,
i.e. in local functionals. Therefore it turns out that the anomaly is a
local functional.
\begin{equation}
a = \int\! \dr^4x\, {\cal A}^1(x,\phi(x),\partial \phi(x),\dots)
\end{equation}
The anomaly density ${\cal A}^1$ is a jet function, i.e. a series in the
fields $\phi$ and a polynomial in the partial derivatives
of the fields comparable to a Lagrange density but with ghostnumber +1.
The integrand ${\cal A}^1$ represents an equivalence class. It is
determined only up to terms of the form
$\s {\cal L}$ because
we are free to choose contributions to the Lagrange density at each loop
order, in particular we try to choose ${\cal L}_{n+1}$ such that
$\s {\cal L}_{n+1}$ cancels ${\cal A}^1$ in order to make
$\Gamma_{n+1}$ \textsc{brst}  invariant. Moreover $\dr^4x\,{\cal A}^1$ is determined
only up to derivative terms of the form $\dr \eta^1$.

${\cal A}^1$ transforms into a derivative because the
anomaly $a$ satisfies the consistency condition. We combine the
anomaly density ${\cal A}^1$ with $\dr^Dx$ to a volume form
$\omega^1_{D}$ and denote the ghostnumbers as superscripts and the
form degree as subscript. Then the
consistency condition  and the description of the equivalence class read
\begin{equation}
\label{anomaly}
\s \omega^1_{D} + \dv\omega^2_{D-1} = 0 \ ,\  \omega^1_{D}
\modulo \bigl(\s \eta^0_{D} + \dr \eta^1_{D-1}\bigr)\ .
\end{equation}
This equation determines all possible anomalies.
Its solutions  depend only on the field content and the 
\textsc{brst}  transformations $\s$ and not on other particular properties of the model under consideration.

The determination of all possible
anomalies is again a cohomological problem just as the determination
of all \textsc{brst}  invariant local actions (\ref{relcoh}) but now with ghostnumbers
shifted by +1.
We will deal with both equations and consider the equation
\begin{equation}
\label{rel}
\s \omega^g_{D} + \dv\omega^{g+1}_{D-1} = 0 \ ,\  \omega^g_{D}
\,\modulo \bigl(\s\eta^{g-1}_{D} + \dr \eta^g_{D-1}\bigr)\ , \end{equation}
for arbitrary ghostnumber $g$.

\section{Cohomological Problems}\label{sec3}

\subsection{Basic Lemma}
In the preceding sections we have encountered repeatedly
the cohomological problem to solve the linear equation
\begin{equation}
\s\omega = 0\,,\ \omega \modulo \s\eta\ ,
\end{equation}
where $\s$ is a nilpotent operator $\s^2=0$, acting on the elements of an algebra $A$. 
The equivalence classes of solutions $\omega$ form a linear space, the cohomology $H(A, \s)$ of $\s$.
The equivalence classes of solutions $\omega^g_p$ of the problem
\begin{equation}
\s\omega^g_p + \dv \omega^{g+1}_{p-1}= 0 \ ,\  \omega^g_p \modulo \s\eta^{g-1}_{p} + \dr \eta^g_{p-1}\ , 
\end{equation}
where $\s^2=0=\dv^2=\{\s,\dv\}$ form
the relative cohomology $H^g_p(A, \s|\dv)$ of $\s$ modulo $\dv$ of ghostnumber
$g$ and form degree $p$.

Let us start to solve such equations and consider the problem to determine
the physical multiparticle states. Multiparticle states can be written as
a polynomial $P$ of the creation operators acting  on the vacuum
\begin{equation}
P(a^\dagger,c^\dagger,\bar{c}^\dagger)|\Omega \rangle \ ,
\end{equation}
if one neglects the notational complication that all these creation operators
depend on momenta $\vec{k}$ and have to be smeared with normalizable functions.
The \textsc{brst} operator $Q_s$ acts on these states in the same way as the
fermionic derivative
\begin{equation}
\s=\sqrt{2}|\vec{k}|
\Bigl(\ir a^\dagger_{\bar{k}}\frac{\partial}{\partial\bar{c}^\dagger} +
c^\dagger\frac{\partial}{\partial a^\dagger_k}\Bigr)
\end{equation}
acts on polynomials in commuting and anticommuting variables. For
one particle states, i.e. linear homogeneous polynomials $P$, we had
concluded that the physical states, the cohomology of $Q_s$ with particle number~1, 
are generated by the transverse creation operators
$a^\dagger_i$, i.e. by variables which are neither generated by $\s$ such
as  $a^\dagger_{\bar{k}}$
or $c^\dagger$ nor transformed as $\bar{c}^\dagger$ and
$a^\dagger_k$\,. 
 
To investigate the action of $\s$ on polynomials, we simplify our notation and
denote the variables with respect to which $\s$ differentiates collectively by $x$ and
their transformation by $\dr x$. Then the derivative $\s$
becomes the exterior derivative $\dv$ (\ref{dext}). 
It maps the variables $x$ to $\dr x$ with opposite statistics (grading),
\begin{equation}
\dv =\dr x^m\frac{\partial\phantom{x^m}}{\partial x^m}
\ ,\  |\dr x^m|= |x^m| +1\ \modulo 2\ .
\end{equation}
The cohomology of the exterior derivative $\dv$ acting on polynomials in $x$ and $\dr x$ 
is described by the basic lemma, 
\begin{theorem}{Basic Lemma}
\begin{equation}
\label{basic}
\dv f(x,\dr x)=0 \Leftrightarrow f(x,\dr x)= f_0 + \dv g(x,\dr x)\ .
\end{equation}
\end{theorem}
$f_0$ denotes the polynomial which is homogeneous of degree 0 in
$x$ and $\dr x$ and is therefore independent of these variables.

Applied to the  Fock space the basic lemma implies that physical
$n$-particle states are generated by polynomials $f_0$ of creation
operators which contain no operators
$a^\dagger_{\bar{k}},a^\dagger_k,c^\dagger,\bar{c}^\dagger $.
Physical multiparticle states are generated by physical (transverse) creation operators
$a^\dagger_i,\ i=1,2$. 

This result seems to be trivial, but it is strikingly different
from the consequences of a bosonic symmetry, e.g. a rotation of a vector with components $(x,y)$ 
leaves the polynomial  $x^2+y^2$ invariant though neither $x$ nor~$y$ are invariant.

The basic lemma determines all functions $\omega$ of the vector potential $A$,
the ghost $C$ and their derivatives, which are invariant under the abelian gauge transformation (\ref{sam})  
\begin{equation}
\label{slin}
\s A_m = \partial_m C\ ,\ \s C = 0\ , \ \s{}^2 = 0\ ,\ [\s, \partial_m] = 0\ .
\end{equation}
The symmetrized partial derivatives of the vectorfield (symmetrization is indicated by the braces) are not 
invariant and transform into the derivatives of $C$,
\begin{equation}
\s \partial_{(m_1}\dots \partial_{m_{k-1}}A_{m_{k})} = \partial_{m_1}\dots \partial_{m_{k}}C\ .
\end{equation}
By the basic lemma only trivial solutions of $\s \omega = 0$ can depend on these variables. The nontrivial
solutions depend on the remaining jet variables, the partial derivatives of antisymmetrized derivatives 
$F_{mn}= \partial_m A_n - \partial_n A_m$ and the undifferentiated ghost $C$. These variables are annihilated 
by $\s$ just as constants.
\begin{equation}
\label{sabel}
\s \omega = 0 \Leftrightarrow \omega = f(C, F , \partial F, \partial \dots \partial F)+ \s \eta\ . 
\end{equation}
The sum is direct, because $f$ and $\s \eta$ depend on different variables,
\begin{equation}
\label{slin0}
f(C, F , \partial F, \partial \dots \partial F)+ \s \eta = 0 \Leftrightarrow
f(C, F , \partial F, \partial \dots \partial F)= 0 \ \wedge \ \s \eta = 0\ .
\end{equation}

To prove the basic lemma (\ref{basic}) we introduce the operation
\begin{equation}
\delta = x^m\frac{\partial}{\partial (\dr x^m)}\ .
\end{equation}
The anticommutator $\Delta$ of $\dv$ and $\delta$ counts the
variables $x^m$ and $\dr x^m$ (\ref{nx}),
\begin{equation}
\label{delta}
\{\dv ,\delta\}=\Delta = x^m\frac{\partial}{\partial x^m} +
\dr x^m\frac{\partial}{\partial (\dr x^m)}= N_x + N_{\dr x}\ .
\end{equation}
Because $\dv$ is nilpotent it commutes with
$\{\dv,\delta\}$, no matter what $\delta$ is,
\begin{equation}
\label{danti}
\dv^2=0 \Rightarrow [\dv,\{\dv,\delta\}]=0\ .
\end{equation}
Of course we can easily check explicitly that $\dv$ does not change the overall
number of variables $x$ and $\dr x$ in a polynomial. We can decompose each polynomial
$f$ into pieces $f_n$ of definite homogeneity $n$ in the variables $x$ and
$\dr x$, i.e. $(N_x+N_{\dr x})f_n = n f_n$. Using (\ref{delta}) we can write
$f$ in the following form, 
\begin{align}
\nonumber
f&=f_0 + \sum_{n\ge 1} f_n =
 f_0 + \sum_{n\ge 1} (N_x+N_{\dr x})\frac{1}{n}f_n\\
\nonumber
&=f_0 + \dv \Bigl( \delta\, \sum_{n\ge 1} \frac{1}{n} f_n \Bigr)
+ \delta \Bigl( \dv \sum_{n\ge 1} \frac{1}{n} f_n \Bigr) \\
\label{hodge}
f &= f_0 +\dr \eta + \delta\, \chi \ .
\end{align}
This is the Hodge decomposition of an arbitrary polynomial in $x$ and $\dr x$ into
a zero mode~$f_0$, a $\dv$-exact\footnote{A polynomial $g$ is called $\dv$-exact (or, shorter, exact, 
if the nilpotent operator $\dv$ is evident)
if it is of the form $g=\dr \eta$ for some polynomial $\eta$.} part $\dr \eta$ and a $\delta$-exact part $\delta\,\chi$. If $f$ is $\dv$-closed, i.e. if it solves $\dv f=0$\,,
then the equations $\dv f_n = 0$ have to hold for each piece $\dv f_n$
separately because the pieces are eigenpolynomials of $N_x+N_{\dr x}$ with different 
eigenvalues and therefore linearly independent. But $\dv f_n =0 $ implies that the last term
in the Hodge decomposition, the $\delta$-exact term, vanishes. This
proves the lemma. If one writes $\frac{1}{n}$ as $\int_0^1 \dr t\, t^{n-1}$ 
one obtains Poincar\'e\,'s lemma for forms in a star shaped domain 
\begin{theorem}{Poincar\'e\,'s lemma}
\label{poinc}
\begin{equation}
\dv f(x,\dr x)=0 \Leftrightarrow f(x,\dr x)=f(0,0) + \dv
\delta\int_0^1\frac{\dr t}{t}(f(tx,t\dr x)-f(0,0))
\end{equation}
\end{theorem}
In this form the lemma is not restricted to polynomials but applies to
all differentiable differential forms $f$ which are defined along all rays $tx$ for
$0\le t \le 1$ and all $x$, i.e. in a star shaped domain.
Note that the integral is not singular at $t=0$ .

We chose to present the Poincar\'e lemma in the algebraic form --
though it applies only to polynomials and to analytic functions -- because
we will follow a related strategy to solve the cohomological problems to come:
given a nilpotent operation $\dv$ we inspect operations $\delta$ and the
anticommutators~$\Delta$. Only the zero modes
of $\Delta$ can contribute to the cohomology of $\dv$.

\subsection{Algebraic Poincar\'e Lemma}

The basic lemma for forms with component functions which are
functions of the base manifold does not apply to jet forms, i.e. differential forms $\omega$
with component functions which are functions of some jet space $\mathcal J_k$, $k<\infty$.
The jet forms, which we consider, are series in fields $\phi$, polynomials in derivatives of fields
$\partial \phi,\partial\partial \phi,\,\dots, \partial\dots \partial\phi$,
polynomials in $\dr x$ and series in the coordinates $x$,
\begin{equation}
\omega:(x,\dr x,\phi,\partial \phi, \partial \partial \phi,\dots )\rightarrow 
\omega(x,\dr x,\phi,\partial \phi, \partial \partial \phi,\dots )\ .
\end{equation}
Jet forms occur as integrands of local functionals. Because they depend
polynomially on derivatives of fields they contain only terms with a
finite number of derivatives, though there is no bound on the
number of derivatives which is common to all forms $\omega$.

We use curly brackets around a field to denote it and its derivatives
\begin{equation}
\{\phi\}=(\phi,\partial \phi, \partial \partial \phi,\dots )\ .
\end{equation}
For Lagrange densities $\omega = {\cal L}(x,\{\phi\})\,\dr^D x$ the basic lemma cannot hold: 
they satisfy $\dv \omega = 0$ because they are volume forms,
but they cannot be total derivatives, $\omega \neq \dr \eta$\,, if their
Euler derivative does not vanish.

Let us show that constants and Lagrange densities with nonvanishing Euler derivative constitute
the cohomology of the exterior derivative  $\dv$ in the space of jet forms.

The exterior derivative $\dv = \dr x^m\partial_m$ of jet forms differentiates the
coordinates~$x$. Acting on derivatives of a field, the partial derivatives $\partial_1,\partial_2\dots$ map 
them to the next higher
derivative with an additional label, just as a creation operator acts on a Fock state, 
\begin{equation}
\partial_m x^n = \delta_m{}^n\ ,\ \partial_m \dr x^m = 0
\ , \  \partial_k (\partial_l\dots \partial_m \phi) =
\partial_k \partial_l\dots \partial_m \phi\ .
\end{equation}
The jet variables satisfy no differential equation, i.e. 
$\partial_k \partial_l\dots \partial_m \phi$ are independent variables up to the fact that
partial derivatives commute
\begin{equation}
\partial_k \dots \partial_m \phi= \partial_m \dots \partial_k \phi\ .
\end{equation}
On these jet variables we define the operations $t^n $ which annihilate a derivative
and act like a derivative with respect to $\partial_n$, i.e. $t^n=\frac{\partial}{\partial(\partial_n)}$,
\begin{align}
& t^n(x^m)=0\ ,& t^n(\dr x^m)&=0\ , \\
\nonumber
&t^n(\phi)=0\ ,&  
t^n (\partial_{m_1}\dots\partial_{m_l}\phi)&=
\sum\limits_{i=1}^l \partial_{m_1}\dots \partial_{m_{i-1}}\,\delta_{m_i}^n\, \partial_{m_{i+1}}\dots \partial_{m_l}\phi\ .
\end{align}
The action of $t^n$ on polynomials in the jet variables is defined by linearity and
the Leibniz rule. Then $t^n$ are vector fields on the jet space $\mathcal J$ which act on jet functions.
 
Obviously the operations $t^n$ commute,  $[t^m,t^n] =0$\,. Less trivial is
\begin{equation}
\label{td}
[t^n,\partial_m]=\delta_m^n N_{ \{\phi\} }\ .
\end{equation}
$N_{ \{\phi\}}$ counts the (differentiated) fields. The
equation holds for linear polynomials, i.e. for the jet variables and
coordinates and differentials, and extends to arbitrary polynomials
because both sides of this equation satisfy the Leibniz rule.

To determine the cohomology of $\dv =\dr x^m\partial_m$ we consider
separately forms $\omega$ with a fixed form degree $p$\,,
\begin{equation}
N_{\dr x}=\dr x^m\frac{\partial}{\partial(\dr x^m)}\ ,\  N_{\dr x}\,\omega
= p \,\omega\ ,
\end{equation}
which are homogeneous of degree $N$ in  $\{\phi\}$. We assume $N > 0$,
the case $N=0$ is covered by Poincar\'e$\,$'s lemma
(theorem \ref{poinc}).

Consider the operation
\begin{equation}
\bv =t^m\frac{\partial}{\partial(\dr x^m)}
\end{equation}
and calculate its anticommutator with the exterior derivative $\dv$
as an exercise in graded commutators (\ref{gradcom}):
\begin{equation}
\begin{aligned}
\{\bv,\dv \}&=\{t^m\frac{\partial}{\partial (\dr x^m)},\dr x^n\}\,\partial_n
-\dr x^n[t^m\frac{\partial}{\partial (\dr x^m)},\partial_n]\\
&=t^m\delta_m{}^n\partial_n - \dr x^n\delta_n{}^{m}N
\frac{\partial}{\partial \dr x^m}\\
&= \partial_nt^n + \delta_n{}^n N - N\,N_{\dr x}\ .\\
\end{aligned}
\end{equation}
So we get
\begin{equation}
\label{db}
\{\dv,\bv\}=N(D-N_{\dr x}) + P_1\ .
\end{equation}
$D=\delta_n^n$ is the dimension of the base manifold, the operator $P_1$ is
\begin{equation}
P_1=\partial_kt^k\ .
\end{equation}

Consider more generally the operations $P_n$
\begin{equation}
P_n=\partial_{k_1}\dots\partial_{k_n}t^{k_1}\dots t^{k_n}
\end{equation}
which take away $n$ derivatives and redistribute them afterwards.
For each polynomial $\omega$ in the jet variables there exists a
$\bar{n}(\omega)$
such that
\begin{equation}
\label{maxn}
\forall n > \bar{n}(\omega):\ P_n \omega =0\ ,
\end{equation}
 because each monomial of
$\omega$ has a bounded number of derivatives.

Using the commutation relation (\ref{td}) one proves the recursion
relation
\begin{equation}
P_1P_k=P_{k+1}+kNP_k
\end{equation}
which can be used iteratively to express $P_k$ in terms of $P_1$ and $N$
\begin{equation}
\label{pk}
P_k=\prod\limits_{l=0}^{k-1}(P_1-lN)\ .
\end{equation}

Using the argument (\ref{danti}) that a nilpotent operation commutes
with all its anticommutators we conclude  from (\ref{db})
\begin{equation}
[\dv ,N(D-N_{\dr x}) + P_1] =0\ .
\end{equation}

Therefore $\dv \omega = 0$ implies $\dv (P_1\omega) = 0$ and from (\ref{pk})
we conclude $\dv (P_k\omega) =0$. We use the relation
(\ref{db})
to express these closed forms $P_k\omega$ as exact forms up to terms
$P_{k+1}\omega$.
\begin{equation}
\label{pkomega}
\begin{aligned}
\dv (\bv \omega)   &= P_1\omega + N(D-p)\,\omega\\
\dv (\bv P_k\omega) &= P_1P_k\omega + N(D-p)P_k\omega\\
&= P_{k+1}\omega +kNP_k\omega + N(D-p)P_k\omega\\
\dv (\bv P_k\omega) &= P_{k+1}\omega + N(D-p+k)P_k\omega \ ,\  k=0,1,\dots
\end{aligned}
\end{equation}
If $p<D$ then we can solve for $\omega$ in terms of exact forms
$\dv (\bv \omega)$ and $P_{1}\omega$ which can be expressed as exact form and
a term $P_2\omega$ and so on. This recursion terminates because
$P_n\omega =0 \;\forall n \ge \bar{n}(\omega)$ (\ref{maxn}).
Explicitly we have for $p<D$ and $N>0$\,:
\begin{equation}
\label{pkd}
\dv \omega = 0 \Rightarrow \ ,\ 
\omega = \dv \Bigl ( \bv 
\sum\limits_{k=0}^{\bar{n}(\omega)}\frac{(-)^k}{N^{k+1}}
\frac{(D-p-1)!}{(D-p+k)!}P_k\omega \Bigr ) = \dr \eta\ .
\end{equation}

To complete the investigation of the cohomology of $\dv$ we have to
consider volume forms $\omega = {\cal L}\,\dr^Dx$. We treat separately
pieces ${\cal L}_N$ which are homogeneous of degree $N>0$ in the jet
variables $\{\phi\}$. These pieces can be written as
\begin{equation}
\label{vari}
\begin{aligned}
N{\cal L}_N &= \phi^i \frac{\partial{\cal L}_N}{\partial \phi^i} +
\partial_m\phi^i\frac{\partial{\cal L}_N}{\partial (\partial_m\phi^i)}
+ \dots\\
&=  \phi^i \frac{\hat{\partial}{\cal L}_N}{\hat{\partial} \phi^i} +
\partial_m X_N^m \ ,\ 
X_N^m =  \phi^i\frac{\partial{\cal L}_N}{\partial (\partial_m\phi^i)} + \dots\ .
\end{aligned}
\end{equation}
Here we use the notation (\ref{eulerableit})
\begin{equation}
\label{eulder}
\frac{\hat{\partial}{\cal L}}{\hat{\partial} \phi^i}
= \frac{\partial{\cal L}}{\partial \phi^i}
-\partial_m \frac{\partial{\cal L}}{\partial (\partial_m\phi^i)}
+ \dots
\end{equation}
for the Euler derivative of the Lagrange density with respect to $\phi^i$.
The dots denote terms which come from higher derivatives.
The derivation of (\ref{vari})
is analogous to the derivation of the Euler Lagrange equations from the
action principle. Eq.(\ref{vari}) implies that the volume form
$\omega_N ={\cal L}_N\,\dr^Dx$ is an exact term and a piece proportional to the
Euler derivative
\begin{equation}
\label{recon}
{\cal L}_N\,\dr^Dx = \frac{1}{N}\phi^i
\frac{\hat{\partial}{\cal L}_N}{\hat{\partial} \phi^i}\, \dr^Dx +
\dv \Bigl ( \frac{1}{N}X_N^m\frac{\partial}{\partial (\dr x^m)}\dr^Dx\Bigr )\ .
\end{equation}
If we combine this equation with Poincar\'e$\,$'s lemma
(theorem \ref{poinc})
and with (\ref{pkd}), combine terms with different degrees of
homogeneity $N$ and different form degree $p$ we obtain the algebraic
Poincar\'e lemma for forms of the coordinates, differentials and jet
variables
\begin{theorem}{Algebraic Poincar\'e Lemma}
\label{algebraic}
\begin{equation}
\label{algebrai}
\begin{aligned}
\dv \omega(x,\dr x,\{\phi\}) & = 0 \quad \Leftrightarrow \\
\omega(x,\dr x,\{\phi\}) &= \text{\rm{const}} + \dr \eta(x,\dr x,\{\phi\}) + {\cal L}(x,\{\phi\})\,\dr^Dx\ .
\end{aligned}
\end{equation}
\end{theorem}
The Lagrange form ${\cal L}(x,\{\phi\})\,\dr^Dx$ is trivial, i.e. of the
form $\dr \eta$,
if and only if its Euler derivative vanishes
identically in the fields.

The algebraic Poincar\'e lemma does not hold if the base manifold is not
starshaped or if the fields $\phi$ take values in a
topologically nontrivial target space. In these cases the
operations $\delta = x\frac{\partial}{\partial (\dr x)}$ and
$\bv =t^n\frac{\partial}{\partial (\dr x^n)}$ cannot be
defined because
a relation like $x\cong x+2\pi$, which holds for the coordinates on a
circle, would lead to the contradiction
$0\cong 2\pi \frac{\partial}{\partial (\dr x)}$. Here we restrict our
investigations to topologically trivial base manifolds and topologically
trivial target spaces. It is the
topology of the invariance groups and the Lagrangian solutions in
the algebraic Poincar\'e lemma which give rise to a nontrivial cohomology of
the exterior derivative $\dv$ and the \textsc{brst} transformation $\s$.

The operations $t^m$ and $\bv$ used in the above proof of the algebraic Poincar\'e lemma (\ref{algebrai}) 
do not affect the dependence on the variables $x^m$. Therefore the algebraic Poincar\'e lemma holds
for $N>0$ also for the part of $\dv$ which differentiates the fields only but not the $x^m$. Hence, using the 
decomposition
\begin{equation}
\dv=\dv_x+\dvp\, ,\quad \dv_x=\dr x^m\frac{\partial}{\partial x^m} \, ,
\label{dvp}
\end{equation}
where $\dvp$ denotes the part of $\dv$ which differentiates only the fields, we obtain:

\begin{theorem}{Algebraic Poincar\'e Lemma for $\dvp$}
\label{algebraic2}
\begin{equation}
\begin{aligned}
\dvp \omega(x,\dr x,\{\phi\}) &= 0\quad \Leftrightarrow \\
\omega(x,\dr x,\{\phi\}) & = \chi(x,\dr x) + \dvp \eta(x,\dr x,\{\phi\}) + {\cal L}(x,\{\phi\})\,\dr^Dx .
\end{aligned}
\end{equation}
\end{theorem}

The algebraic Poincar\'e lemma is modified if the jet space
contains in addition variables which are space time constants. This
occurs for example if one treats rigid transformations as
\textsc{brst} transformations with constant ghosts $C$, i.e. $\partial_m C =0$.
If these ghosts occur as variables in forms $\omega$ then they are not
counted by the number operators $N$ which have been used in the proof of
the algebraic Poincar\'e lemma and can appear as variables in~$\eta$, 
in~${\cal L}$ and in the constant solution of $\dv \omega = 0$\,.

\subsection{Descent Equation}

We are now prepared to investigate the relative cohomology
and derive the so called descent equations. We recall that we deal with
two nilpotent derivatives, the exterior derivative $\dv$ and the
\textsc{brst} transformation $\s$, which anticommute with each other
\begin{equation}
\label{anticom}
\dv^2 =0 \ ,\  \s^2=0 \ ,\  \{\s,\dv\}=0\ .
\end{equation}
$\s$ leaves the form degree $N_{\dr x}$ invariant, $\dv$ raises it by 1
\begin{equation}
\label{Ndegree}
[N_{\dr x},\s] =0 \ ,\  [N_{\dr x},\dv]=\dv\ .
\end{equation}

To derive necessary conditions on the solution of (\ref{invloc})
\begin{equation}
\label{relcoh1}
\s \omega_D + \dv \omega_{D-1} = 0\ ,\ 
\omega_D \ \modulo (\s\eta_D + \dv\eta_{D-1})\ ,
\end{equation}
where the subscript denotes the form degree, 
we apply $\s$ and use (\ref{anticom})
\begin{equation}
0=\s(\s \omega_D + \dv \omega_{D-1})=\s\dv \omega_{D-1}=-\dv (\s \omega_{D-1})
\ .\end{equation}
By 
(\ref{algebrai}) $\s \omega_{D-1}$
is of the form $\text{const} + \dr \eta(\{\phi\}) + {\cal L}(\{\phi\})\,\dr^Dx$.
The piece ${\cal L}(\{\phi\})\,\dr^Dx$ has to vanish because $\omega_{D-1}$
has form degree $D-1$ and if $D>1$ then also the constant piece vanishes
because $\omega_{D-1}$ contains $D-1>0$ differentials and is not constant.
Therefore we conclude
\begin{equation}
\label{relcoh2}
\s \omega_{D-1} + \dv \omega_{D-2} = 0\ ,\ 
\omega_{D-1} \ \modulo\ ( \s\eta_{D-1} + \dr \eta_{D-2} )
\end{equation}
where we denoted $\eta$ by $\omega_{D-2}$ to indicate its form degree.
Adding to $\omega_{D-1}$ a piece of the form $\s \eta_{D-1}+\dr \eta_{D-2}$
changes $\omega_D$ only within its class of equivalent representatives.
Therefore $\omega_{D-1}$ is naturally a representative of an equivalence
class.
{}From (\ref{relcoh1}) we have derived (\ref{relcoh2}) which is nothing
but (\ref{relcoh1}) with  form degree  lowered by 1.
Iterating the arguments we lower the form degree step by step and obtain
the descent equations
\begin{equation}
\label{absteig}
\s \omega_{i} + \dv \omega_{i-1} = 0\ ,\  i=D,D-1,\dots,1\ ,\ 
\omega_{i} \ \modulo (\s \eta_{i} + \dr \eta_{i-1} )\
\end{equation}
until the form degree drops to zero. It cannot become
negative. For $i=0$ one has
\begin{equation}
\s \omega_0 = \text{ const}\ ,\ \omega_0 \ \modulo \s \eta_0
\end{equation}
because this is the solution to $\dv \s \omega_0= 0$ for $0$-forms.

If, however, the \textsc{brst} transformation is not spontaneously broken
i.e. if $\s$ does not transform fields into numbers,
$\s \phi  {}_{|(\phi=0)}=0$\,, then $\s \omega_0$ has to vanish.
This follows most easily if one evaluates  both sides of $\s \omega_0 =
\text{const}$ for vanishing fields. We assume for the following that the
\textsc{brst} transformations are not spontaneously broken,
\begin{equation}
\label{scoh}
\s \omega_0 = 0 \ ,\ 
\omega_0 \ \modulo \s \eta_0\ .
\end{equation}

We will exclude from  our considerations also
spontaneously broken rigid symmetries. There we cannot apply these
arguments because $\s \phi  {}_{|(\phi=0)}=C$ gives ghosts which are
space time constant and $\s \omega_0  = f(C) \neq  0$ can occur.

The descent equations (\ref{absteig}, \ref{scoh}) are just
another cohomological equation for a nilpotent operator
$\st$ and
a form $\tilde{\omega}$
\begin{equation}
\label{stilde}
\st = \s + \dv \ ,\  \st^2=0\ ,
\end{equation}
\begin{equation}
\label{otilde}
\tilde{\omega }= \sum\limits_{i=0}^{D}\omega_i\ ,\ \tilde{\eta}=\sum\limits_{i=0}^{D} \eta_i\ ,
\end{equation}
\begin{equation}
\label{stildecoh}
\st\ \tilde{\omega } = 0\ ,\ 
\tilde{\omega } \modulo \st \tilde{\eta}\ .
\end{equation}
That $\st$ is nilpotent follows from (\ref{anticom}). The
descent equations (\ref{absteig}, \ref{scoh})  imply
$\st \tilde{\omega } = 0$ and $\tilde{\omega}$ is  equivalent to
$\tilde{\omega} + \st \eta$. So (\ref{stildecoh}) is a consequence of the 
descent equations. On the other hand if (\ref{stilde}) holds then the 
equation (\ref{stildecoh})
implies the descent equations. This follows if one splits
$\st$, $\tilde{\omega}$ and  $\tilde{\eta}$ with respect to the
form degree (\ref{Ndegree}).

\begin{theorem}{\quad}\\
\label{absts}
Let $\st=\s+\dv$ be a sum of two nilpotent, anticommuting fermionic derivatives
where $\s$ preserves the form degree and $\dv $ raises it by one, then
each solution $ (\omega_0,\dots, \omega_D )$ of the descent equations
\begin{equation}
\s \omega_{i} + \dv \omega_{i-1} = 0\ ,\  i=0,1,\dots D\,,\ 
\omega_{i} \ \modulo (\s \eta_{i} + \dr \eta_{i-1} )\,,
\end{equation}
corresponds one to one to an element $\tilde{\omega}=\sum \omega_i$ of the cohomology
\begin{equation}
H(\st)= \{\tilde{\omega}: \ \st\ \tilde{\omega } = 0\ ,\ 
\tilde{\omega } \modulo \st \tilde{\eta}\}\ .
\end{equation}
The forms~$\omega_i$ are the parts of $\tilde{\omega}$ with form degree $i$.
\end{theorem}

The formulation of the descent equations as a cohomological problem of
the operator~$\st$ has several virtues. The solutions to
$\st\ \tilde{\omega } = 0$
can obviously be multiplied to obtain further solutions. They form an algebra, not just a vector space.
Moreover, for the \textsc{brst} operator in gravitational Yang Mills
theories we will find that the equation $\st \ \tilde{\omega } = 0$
can be cast into the form $\s \omega = 0$ by a change of variables,
where $\s$ is the original \textsc{brst} operator. This
equation has to be solved anyhow as part of the descent equations. Once
one has solved it one can recover the complete solution of the descent
equations, in particular one can read off $\omega_D$ as the $D$ form
part of $\tilde{\omega}$. These virtues
justify to consider with $\tilde{\omega}$ a sum of forms of different
form degrees which in traditional eyes would be considered to add
apples and oranges.

\subsection{K\"unneth's Theorem}

If the nilpotent derivative $\dv$ acts on a tensor product 
\begin{equation}
A=A_1\otimes A_2
\end{equation}
of vectorspaces which are separately invariant under $\dv$
\begin{equation}
\dv A_1 \subset A_1 \ ,\  \dv A_2 \subset A_2 \ ,
\end{equation}
then K\"unneth's theorem states  that the cohomology $H(A,\dv)$ of $\dv$
acting on $A$ is given by the product of the cohomology
$H(A_1,\dv)$ of $\dv$ acting on $A_1$
and $H(A_2,\dv)$ of $\dv$ acting on $A_2$.

\begin{theorem}{K\"unneth-formula}\\
\label{kunn}
Let $\dv=\dv_1+\dv_2$ be a sum of nilpotent differential
operators which leave their vectorspaces $A_1$ and $A_2$ invariant
\begin{equation}
\dv_1 A_1 \subset A_1\ ,\   \dv_2 A_2 \subset A_2
\end{equation}
and which are defined on the tensor product $A=A_1\otimes A_2$ by the Leibniz
rule 
\begin{equation}
\dv_1 (kl) = (\dv_1 k)l \ ,\  \dv_2(kl) = (-)^{|k|}k (\dv_2 l)\ ,\ 
\forall k\in A_1,\: l\in A_2\ .
\end{equation}
Then the  cohomology $H(A,\dv)$ of $\dv$ acting on $A$ is the tensor
product of the cohomologies of $\dv_1$ acting on $A_1$ and $\dv_2$ acting
on $A_2$
\begin{equation}
H (A_1\otimes A_2,\dv_1+\dv_2)=H (A_1,\dv_1)\otimes H (A_2,\dv_2)\ .
\end{equation}
\end{theorem}
The formula justifies to count numbers as nontrivial solution of $\dv \omega = 0$
rather than to exclude them for simplicity from the definition of $H(A,\dv)$.

To prove the theorem we consider an element $f\in H(d)$
\begin{equation}
f=\sum_ik_il_i
\end{equation}
given as a sum of products of elements $k_i\in A_1$ and $l_i\in A_2$.
Without loss of generality we assume that the elements $k_i$
are taken from a basis of $A_1$ and the elements $l_i$
are taken from a basis of $A_2$.
\begin{align}
\label{link}
\sum c_i k_i = 0 & \Leftrightarrow  c_i = 0  \ \forall\  i\\
\label{linl}
\sum c_i l_i = 0 & \Leftrightarrow  c_i = 0 \ \forall\  i
\end{align}
Otherwise one has a relation like
$l_1=\sum^\prime_i \alpha_i l_i$ or
$k_1=\sum^\prime_i \beta_i k_i$,
where $\sum^\prime$ does not contain $i=1$, and can rewrite $f$
with fewer terms
$f=\sum^\prime_i(k_i+\alpha_ik_1) \cdot l_i $ or
$f=\sum^\prime_i k_i\cdot
(l_i + \beta_il_1)$.
We can even choose $f\in H(\dv)$  in such a manner that the elements $k_i$
are taken from a basis of a complement to the space $\dv_1 A_1$. In other
words we can choose $f$ such that no linear combination of the elements
$k_i$ combines to a $\dv_1$-exact form.
\begin{equation}
\sum_i c_i k_i = \dv_1g \Leftrightarrow \dv_1g = 0 =c_i  \ \forall\  i
\end{equation}
Otherwise we have a relation like
$k_1=- \dv_1 \kappa + \sum^\prime_i \beta_i k_i $,
where $\sum^\prime$ does not contain $i=1$, and we can rewrite
$f\in H(\dv)$ up to an irrelevant piece as sum of products with
elements $k^\prime_i=\kappa,k_2,\dots $, where $\kappa$ is not in $\dv_1 A_1$, 
\begin{equation}
f=(-)^{|\kappa|}  \kappa \dv_2 l_1 + \sum_{i > 1} k_i\cdot (l_i + \beta_il_1)  
 - \dv (\kappa\  l_1)\ .
\end{equation}
We can  iterate this argument until no linear combination of the
elements $k^\prime_i$ combines to a $\dv_1$-exact form.

By assumption $f$ solves $\dv f\,=\,0$ which implies
\begin{equation}
\sum_i \Bigl ((\dv_1k_i)l_i + (-)^{k_i}k_i (\dv_2 l_i) \Bigr ) = 0 \ .
\end{equation}
In this sum $\sum_i (\dv_1 k_i)l_i$ and $\sum_i (-)^{k_i}k_i (\dv _2 l_i)$
have to vanish separately because the elements $k_i$ are linearly
independent from the elements $\dv_1 k_i\in \, \dv_1 A_1$.

$\sum_i (\dv_1 k_i)l_i =0 $, however, implies
\begin{equation} \dv_1 k_i = 0
\end{equation}
because the
elements $l_i$ are linearly independent and
$\sum_i (-)^{k_i}k_i (\dv_2 l_i)=0$ leads to
\begin{equation}
\dv_2l_i = 0
\end{equation}
analogously. So we have shown
\begin{equation}
\dv f=0 \Rightarrow f=\sum_i k_i l_i + \dv \chi \text{ where }
\dv_1 k_i = 0 = \dv_2 l_i \ \forall i \ .
\end{equation}
Changing $k_i$ and $l_i$ within their equivalence class
$k_i \modulo \dv_1\kappa_i$ and $l_i \modulo \dv_2\lambda_i$ does not change the
equivalence class $f \modulo \dv\chi$:
\begin{equation}
\sum_i(k_i + \dv_1\kappa_i)(l_i + \dv_2\lambda_i) = \sum_i k_i l_i +
\dv \sum_i \Bigl (\kappa_i (l_i + \dv_2 \lambda_i) + (-)^{k_i}k_i \lambda_i
\Bigr )
\end{equation}
Therefore $H(A, \dv )$ is contained in $H_1 (A_1, \dv_1)\otimes H_2 (A_2, \dv_2)$.
On the other hand, the inclusion $H_1 (A_1, \dv_1)\otimes H_2 (A_2, \dv_2) \subset H(A,\dv)$ is
trivial. This concludes the proof of K\"unneth's theorem.

\section{BRST Algebra of Gravitational Yang Mills Theories}\label{sec4}
\subsection{Covariant Operations}
Gauge theories such as gravitational Yang Mills theories 
rely on tensor analysis.
The set of tensor components is a subalgebra of
the polynomials in the graded commutative jet variables.
\begin{equation}
\Bigl ( \text{Tensors}\Bigr )
\subset
\Bigl ( \text{Polynomials\,}(\phi,\partial\phi, \partial \partial \phi,\dots
)\Bigr ) \end{equation}
The covariant operations $\Delta_M$ which occur in tensor analysis
\begin{equation}
\Delta_M: \Bigl (\text{Tensors}\Bigr ) \rightarrow \Bigl (\text{Tensors}\Bigr )
\end{equation}
map tensors to tensors and satisfy the graded Leibniz rule (\ref{leibniz}).
These covariant operations have a basis consisting of the real bosonic covariant
space time derivatives $D_a,\ a=0,\dots,D-1$, 
the complex covariant, fermionic spinor derivatives $D_\alpha,\  D_{\dot{\alpha}}=(D_\alpha)^\ast$, 
in supergravitational theories and real bosonic
spin and isospin transformations $\delta_i$, which correspond to a basis
of the Lie algebra of the gauge group and of the Lorentz group, possibly including dilatations 
and so-called $R$-transformations
\begin{equation}
( \Delta_M )= (D_a,\ D_\alpha,\  D^{\dot{\alpha}},\ \delta_i )\ .
\end{equation}
The grading of the covariant operations can be red off from the index, $|\Delta_M |=|M|$, 
only spinor derivatives are fermionic.

By assumption, the space of covariant operations is closed with respect to graded commutation:
the graded commutator
of covariant operations is a covariant operation which can be linearly combined from the basic
covariant operations,
\begin{equation}
\label{diffa}
[\Delta_M,\Delta_N] := \Delta_M\Delta_N - (-)^{MN}\Delta_N\Delta_M
={\cal F}_{MN}^{\phantom{MN}K}\Delta_K \ .
\end{equation}
with structure functions 
\begin{equation}
\F{MN}{K}= -(-)^{|M||N|}\F{NM}{K}
\end{equation}
which are components of graded antisymmetric tensorfields
and which are graded according to their index picture, $|\F{MN}{K}| = |M| + |N| + |K|$.

We  raise and lower spinor indices with $\varepsilon_{\alpha\beta}= -\varepsilon_{\beta\alpha}=\varepsilon_{\dot\alpha\dot\beta}$, 
$\varepsilon_{12}=1$, $Y_\alpha =\varepsilon_{\alpha\beta}Y^\beta $,
$Y_{\dot\alpha} =\varepsilon_{\dot \alpha\dot \beta}Y^{\dot\beta} $
and use the summation convention
\begin{equation}
X^M Y_M := X^a Y_a + X^\alpha Y_\alpha + X_{\dot{\alpha}}Y^{\dot{\alpha}}+ X^iY_i\ ,
\end{equation}
that in a spinor sum the first \emph{un}dotted index is \emph{up} and the first \emph{do}tted index is \emph{do}wn.
If then the components are graded according to their index picture 
\begin{equation}
|X^M| =  |X| + |M|\ ,\ |Y_M| =  |Y| + |M|
\end{equation}
and define components of the conjugate quantities $\bar X$ and $\bar Y$ by 
\begin{equation}
\bar X^{\bar M}= (-)^{|X|+|M|} (X^M)^\ast\ ,\ 
\bar Y_{\bar M}= (-)^{|Y|+|M|} (Y_M)^\ast
\end{equation}
then because of $(-)^{|M|} X^{\bar M}  Y_{\bar M}=X^M Y_M$ the conjugate of the sum  turns out to be
the sum over the conjugate products,
\begin{equation}
(X^M  Y_M)^\ast = (-)^{|X||Y|}\bar X^M \bar Y_M\ .
\end{equation}

The structure functions turn out to be graded real,
\begin{equation}
(\F{MN}{K})^\ast = (-)^{|K|(|M| + |N|) + |M||N|}\F{\bar M \bar N}{\bar K}\ ,
\end{equation}
i.e. conjugation maps the graded commutator algebra to itself, it is real.

Some of the structure functions have purely numerical values as for example
the structure constants in the commutators of infinitesimal Lorentz or isospin transformations
\begin{equation}
\label{diffiso}
[\delta_i,\delta_j] = f_{ij}{}^k\delta_k\ .
\end{equation}
Other constant structure functions are the elements of matrices $G_{i}$, 
which represent isospin or Lorentztransformations on the covariant space time derivatives
\begin{equation}
\label{repspin}
[\delta_i, D_a] = - G_{i\,a}{}^b D_b\ .
\end{equation}

Other components of the tensors $\F{MN}{K}$ are given by the Riemann
curvature, the Yang Mills field strength and in supergravity the
Rarita Schwinger field strength and auxiliary fields of the
supergravitational multiplet. We use the word field strength also to
denote the Riemann curvature and the Yang Mills field
strength collectively.

The commutator algebra (\ref{diffa}) implies the Jacobi identity.
If we denote the graded sum  over the cyclic permutations of an expression $X_{MNP}$ by
\begin{equation}
\label{osum}
\osum_{M\,N\,P}X_{MNP} := X_{MNP}+ (-)^{|M|(|N|+|P|)}X_{NPM} + (-)^{|P|(|M| + |N|)}X_{PMN}\ ,
\end{equation}
then the Jacobi identity can be written as 
\begin{equation}
\osum_{M\,N\,P} [\Delta_M,[\Delta_N,\Delta_P]] = 0\ .
\end{equation}
Inserting (\ref{diffa}) one obtains the first Bianchi identity for the structure functions
\begin{equation}
\label{bianchi}
\osum_{M\,N\,P} (\Delta_M \F{NP}{K}- \F{MN}{L}\F{LP}{K})=0\ .
\end{equation}

The covariant operations are not defined on arbitrary
polynomials of the jet variables. In particular one cannot realize the
commutator algebra (\ref{diffa}) on connections, on ghosts
or on auxiliary fields.

To keep the discussion simple we will not consider fermionic covariant derivatives in the following. 
Then the commutator algebra (\ref{diffa}) has more specifically the structure given by (\ref{diffiso}) 
and (\ref{repspin}) and 
\begin{equation}
[D_a,D_b] = -\T{ab}{c}D_c + \A{F}{ab}{i}\delta_i \ .
\end{equation}
We will simplify this algebra even more and choose the spin connection
by the requirement that the torsion $T_{ab}{}^c$ vanishes. 

\subsection{Transformation and Exterior Derivative}

The fields $\phi$ in gravitational Yang Mills theories are the
ghosts $C^N$, anti\-ghosts $\bar{C}^N$, auxiliary fields $B^N$, gauge
fields $A_m^{\phantom{m}N}\,, m=0,\dots,D-1$, also called connections, and
elementary tensor fields $T$. 
The gauge potentials, ghosts and auxiliary fields are real and correspond to
a basis of the covariant operations $\Delta_M$, i.e. there are
connections, ghosts and
auxiliary fields for translations, 
Lorentz transformations and isospin transformations. Matter
fields are tensors and denoted by~$T$.
\begin{equation}
\label{fields}
\phi = (C^N, \bar{C}^N,B^N,A_m^{\phantom{m}N}, T)
\end{equation}
We define the \textsc{brst} transformation of the antighosts and the auxiliary fields by
\begin{equation}
\s \bar{C}^N = \ir B^N \ ,\  \s B^N= 0\ .
\end{equation}
The \textsc{brst} transformation of tensors is  a sum of covariant
operations with ghosts as coefficients \cite{Brandt:1993xq}
\begin{equation}
\label{st}
\s T = - C^N \Delta_N  T\ .
\end{equation}
We require that partial derivatives $\partial_m$ of tensors can be expressed as a
linear combination of covariant operations with coefficients which by their definition are
the connections or gauge fields. For the exterior derivative this means
\begin{equation}
\begin{aligned}
\label{dten}
\dv T = \dr x^m\partial_m T &= -\dr x^m A_m^{\phantom{m}N}\Delta_N T = - A^N \Delta_N T\ ,\\
A^N  &= \dr x^m A_m^{\phantom{m}N}\ .
\end{aligned}
\end{equation}
$\s$ acts on tensors strikingly similar to $\dv$: $\s T$ contains ghosts $C^N$ where $\dv T$ contains
composite connection one forms $A^N$.

Let us check that (\ref{dten}) is nothing but the usual definition
of covariant derivatives. We spell out the sum over covariant
operations and denote the connection $-A_m^{\phantom{m}a}$ which correspond to
covariant space-time derivatives by  $e_m^{\phantom{m}a}$, the vielbein.
The index $i$ enumerates a basis of spin and isospin transformations,
\begin{equation}
\partial_m = -A_m^{\phantom{m}N} \Delta_M =
e_m^{\phantom{m}a}D_a -  A_m{}^{i}\delta_i\ .
\end{equation}
If the vielbein has an inverse $\A{E}{a}{m}$, which we take for
granted, 
\begin{equation}
e_m^{\phantom{m}a}E_a^{\phantom{a}n}=\delta_m^{\phantom{m}n}\ ,
\end{equation}
then we can solve for the covariant space time derivative and obtain as
usual
\begin{equation}
\label{covder}
D_a = E_a^{\phantom{a}m}(\partial_m + A_m{}^i\delta_i)\ .
\end{equation}

We require that $\s$ and $\dv$ anticommute and be nilpotent
(\ref{anticom}).
This fixes the \textsc{brst} transformation of the ghosts and  the connection and
identifies the curvature and field strength. In particular $\s^2 = 0 $
implies
\begin{equation}
0=\s^2 T = \s\, (-C^N\Delta_N T) = -(\s C^N)\,\Delta_NT + C^N \s(\Delta_N T)\ .
\end{equation}
$\Delta_N T$ is a tensor so
\begin{equation}
C^N \s\,(\Delta_N T) = -C^N C^M \Delta_M \Delta_N T =-
\frac{1}{2}C^N C^M [\Delta_M, \Delta_N ] T\ .
\end{equation}
The commutator is given by the algebra (\ref{diffa}) and we conclude
\begin{equation}
0= (\s C^N + \frac{1}{2}C^KC^L\F{LK}{N})\Delta_N T\ ,\  \forall T\ .
\end{equation}
This means that the operation
$(\s C^N + \frac{1}{2}C^KC^L\F{LK}{N})\Delta_N $ vanishes.
The covariant operations $\Delta_N$ are understood to be linearly
independent. Therefore $\s C^N$ is determined
\begin{equation}
\label{scn}
\s C^N = -\frac{1}{2}C^KC^L\F{LK}{N}\ .
\end{equation}
The \textsc{brst} transformation of the ghosts is given by a polynomial which is
quadratic in the ghosts with expansion coefficients given by the
structure functions $\F{LK}{N}$.
$\s$ transforms the algebra of polynomials generated by  ghosts (not
derivatives of ghosts) and tensors into itself (\ref{st}, \ref{scn}).

The requirement that $\s$ and $\dv$ anticommute fixes the transformation of the connection,
\begin{equation}
\begin{aligned}
0&=\{\s,\dv \}T=\s (-A^N\Delta_N T) + \dv (-C^N\Delta_N T)\\
 &=-(\s A^N)\Delta_NT - A^NC^M\Delta_M\Delta_NT
- (\dv C^N)\Delta_N T - C^NA^M\Delta_M\Delta_N T\\
&=-(\s A^N +\dv C^N + A^KC^L\F{LK}{N})\Delta_N T \ ,\  \forall T\ .
\end{aligned}
\end{equation}
So we conclude
\begin{equation}
\s A^N = -\dv C^N  - A^KC^L\F{LK}{N}
\end{equation}
for the connection one form $A^N$. For the gauge field
$A_m^{\phantom{m}N}$ we obtain
\footnote{Anticommuting $\dr x^m$ through $\s$ changes the signs.}
\begin{equation}
\s A_m^{\phantom{m}N} = \partial_mC^N  + A_m^{\phantom{m}K}C^L\F{LK}{N}\ .
\end{equation}
The \textsc{brst} transformation of the connection contains
the characteristic inhomogeneous piece $\partial_mC^N$.

$\dv^2=0$ identifies the field strength as curl of the connection,
\begin{equation}
\begin{aligned}
0&=\dv^2 T= \dr x^m \dr x^n\partial_m\partial_n T =
-\dr x^m \dr x^n\partial_m (\A{A}{n}{N}\Delta_N T)\\
&= -\dr x^m \dr x^n \Bigl( (\partial_m \A{A}{n}{N}) \Delta_N T +
\A{A}{n}{N} \partial_m (\Delta_N T) \Bigr) \\
&=- \dr x^m \dr x^n \Bigl ( (\partial_m \A{A}{n}{N}) \Delta_N T -
\A{A}{n}{N} \A{A}{m}{M}\Delta_M\Delta_N T \Bigr )\ .
\end{aligned}
\end{equation}
Because the differentials anticommute, the antisymmetric part of the bracket vanishes,
\begin{equation}
0= \partial_m \A{A}{n}{K} - \partial_n \A{A}{m}{K} -
\A{A}{m}{M}\A{A}{n}{N}\F{MN}{K}\ .
\end{equation}
We split the summation over $M\,N$, employ the definition
of the vielbein, denote by $i$ and $j$ collectively spin and isospin values
\begin{equation}
\begin{aligned}
0= \partial_m \A{A}{n}{K} - \partial_n \A{A}{m}{K} &-
\A{e}{m}{a}\A{e}{n}{b}\F{ab}{K}   +
\A{e}{m}{a}\A{A}{n}{i}\F{ai}{K} \\
&+
\A{A}{m}{i}\A{e}{n}{a}\F{ia}{K} -
\A{A}{m}{i}\A{A}{n}{j}\F{ij}{K}
\end{aligned}
\end{equation}
and solve for the structure functions $\A{\cal F}{ab}{K}$ with two space time indices.
Up to nonlinear terms, they are the antisymmetrized derivatives of the gauge fields,
\begin{equation}
\F{ab}{K}  =  \A{E}{a}{m}\A{E}{b}{n} \Bigl(
2\partial_{[m} \A{A}{n]}{K} 
+ 2\A{e}{[m}{c}\A{A}{n]}{i}
\F{ci}{K}  -
\A{A}{m}{i}\A{A}{n}{j}\F{ij}{K}  \Bigr)\ .
\end{equation}
They are the torsion, $\F{ab}{c}= -\A{T}{ab}{c}$,
if $K=c$ corresponds to space-time
translations, 
\begin{equation}
T_{ab}{}^c= E_a{}^m E_b{}^n\bigl( 
\partial_m e_n{}^c - \partial_n e_m{}^c + \omega_{m\,d}{}^c e_n{}^d  - \omega_{n\,d}{}^c e_m{}^d
\bigr)\ ,
\end{equation}
the Riemann curvature $\A{R}{ab}{cd}$, if $K=cd=-dc$
corresponds to Lorentz transformations,
\begin{equation}
R_{abc}{}^d= E_a{}^m E_b{}^n\bigl( 
\partial_m \omega_{n\,c}{}^d - \partial_n \omega_{m\,c}{}^d
- \omega_{m\,c}{}^e \omega_{n\,e}{}^d + \omega_{n\,c}{}^e \omega_{m\,e}{}^d
\bigr)\ ,
\end{equation}
and the Yang Mills field
strength $\A{F}{ab}{i}$, if $K=i$ ranges over isospin indices,
\begin{equation}
\A{F}{ab}{i}
=
E_a{}^m E_b{}^n\bigl( 
\partial_m A_{n}{}^i - \partial_n A_{m}{}^i - A_m{}^j A_n{}^k\,f_{jk}{}^i\bigr) 
\ .
\end{equation}
The formula applies, however, also to supergravity, which has a more complicated algebra (\ref{diffa}). 
It allows  in a surprisingly simple way to identify the Rarita Schwinger field strength
$\A{\Psi}{ab}{\alpha}$ when $K=\alpha$ corresponds to supersymmetry
transformations.

We choose the spin connection $\omega_{a\,bc}=E_a{}^k{\omega}_{k\,bc}$ such that the torsion vanishes, 
\begin{equation}
{\omega}_{a\,bc} =
\frac{1}{2}\bigl (
\eta_{ad}E_b{}^m E_c{}^n +\eta_{bd}E_a{}^m E_c{}^n-\eta_{cd}E_a{}^m E_b{}^n
\bigr )
\bigl (
\partial_m e_n{}^d -\partial_n e_m{}^d
\bigr )\ . 
\end{equation}
This choice simplifies the algebra. It does not restrict the generality of our considerations, because 
a different spin connection differs by a tensor only and leaves the algebra of all tensors unchanged.

We have used that $\s$ and $\dv$ are nilpontent and anticommute if applied to tensors. 
This has fixed the transformations of the ghosts and connections and identified the
structure functions ${\cal F}_{ab}{}^N$.  That $\s$ and $\dv$ are nilpontent and anticommute
also if applied to connections and ghosts follows from the Bianchi identity (\ref{bianchi}). 

The formulas
\begin{equation}
\s T = - C^N\Delta_N T \ ,\  \dv T = - A^N \Delta_N T
\end{equation}
for the nilpotent, anticommuting operations $\s$ and $\dv$ not only encrypt
the basic geometric structures. They allow also
to prove easily that the cohomologies of $\s$ and $\s+\dv$ acting on tensors
and ghosts, (\emph{not} on connections, derivatives of ghosts,
auxiliary fields and antighosts) differ only by a change of variables.
Inspection of  $(\s+\dv)$, acting on tensors~$T$, shows 
\begin{gather}
\label{stt}
\st T = (\s+\dv )\,T=-(C^N+A^N)\Delta_N T = -\tilde{C}^N\Delta_N T\ ,\\
\text{where\quad}
\tilde{C}^N = C^N + A^N = C^N + \dr x^m \A{A}{m}{N}\ ,
\end{gather}
that the $\st $-transformation of tensors is the
$\s$-transformation with the ghosts~$C$ replaced by~$\tilde{C}$.

The $\st $-transformation of $\tilde{C}$ follows from
$\st ^2=0$ and the transformation of tensors (\ref{stt}) by the
same arguments which
determined $\s C$ from $\s^2=0$ and from $(\ref{st})$. So we obtain
\begin{equation}
\label{stc}
\st \tilde{C}^N= -\frac{1}{2}\tilde{C}^K\tilde{C}^L\F{LK}{N}\ .
\end{equation}
This is just the tilded version of (\ref{scn}).

Define the map $\rho: C \mapsto \tilde{C}= C + A$ to translate the ghosts $C$ by the connection $1$-forms $A$
and to leave $A$ and tensors $T$ invariant. Jet functions~$P$, which depend on ghosts and tensors, and are 
constant as functions of $A$ are transformed by the corresponding pullback $\rho^*$ to
\begin{equation}
\label{rho}
\rho^*(P) = P \circ \rho\ .
\end{equation}
Then (\ref{stt},\ \ref{stc}) and (\ref{st},\ \ref{scn}) state that the \textsc{brst} cohomologies 
of  $\s$ and $\st$ acting on functions of ghosts and tensors are invertibly related
\begin{equation}
\st \circ \rho^* = \rho^* \circ \s\ \ .
\end{equation}
\begin{theorem}{\quad}\\
\label{sequiv}
A form $\omega(C,T)$ solves $\s \omega(C,T)=0$ if and only if
$\omega(\tilde{C},T)$ solves $\st \,\omega(\tilde{C},T)=0$\,.
\end{theorem}
If we combine this result with theorem \ref{absts} then the solutions
to the descent equations can be found from the cohomology of $\s$ if we
can restrict the jet functions, which contribute to the cohomology of $\st$, 
to functions of the ghosts and tensors.

\subsection{Factorization of the Algebra}\label{sec4.3}

If the base manifold and the target space of the fields have trivial topology,
then we can restrict the jet functions, which contribute to the cohomology of $\st$, 
to functions of the ghosts and tensors, because
the algebra of jet variables is a product of algebras on which
$\st $ acts separately and trivially on all factors, apart of the algebra of
ghosts and tensors. Using K\"unneth's formula (theorem \ref{kunn}) we can
determine nontrivial Lagrange densities and anomaly candidates
from solutions of $\st \omega(\tilde{C},T)=0$\,. To establish this result we prove the following theorem:
\begin{theorem}{\quad}\\
\label{Asplit}
The algebra $A$ of series in $x^m$ and the fields $\phi
$ {\normalfont(\ref{fields})} and of polynomials in $\dr x^m$ and the partial derivatives
of the fields is a product algebra
\begin{equation}
\label{aprod}
A= A_{\tilde{C},T}\otimes \prod_l A_{u_l,\,\st u_l}
\end{equation}
where the variables $u_l$ are enlisted by {\normalfont($k=1,2,\dots$)}
\begin{equation}
\label{bprod}
\bigl(
x^m,\A{e}{m}{a}, \omega_m{}^{ab}, \A{A}{m}{i},
\partial_{(m_k}\dots \partial_{m_1} \A{A}{{m_0)}}{N},
\bar{C}^N, \partial_{m_k}\dots \partial_{m_1} \bar{C}^N 
\bigr)
\end{equation}
$\st $ acts on each factor $A_{u_l,{\,\st u_l}}$ separately, 
$\st A_{u_l,{\,\st u_l}} \subset A_{u_l,{\,\st u_l}}$\,.
\end{theorem}

The braces around indices, $\partial_{(m_k}\dots \partial_{m_1} \A{A}{{m_0)}}{N}$, denote symmetrization.
The subscript of the algebras denote the generating elements, e.g.
$A_{\A{e}{m}{a},{\,\st \A{e}{m}{a}}}$ is the algebra of series in the
viel\-bein~$\A{e}{m}{a}$ and in $\st \A{e}{m}{a}$.
$\st $ leaves $A_{u_l,\,\st u_l}$ invariant by construction
because of $\st ^2=0$.

To prove the theorem we inspect the variables $u_l$ and $\st u_l$
to lowest order\footnote{We do not count powers of the vielbein  $\A{e}{m}{a}$ or its
inverse. They are not affected by the change of variables, which we investigate.
Derivatives of the vielbein, however, are counted.} in the differentials and fields.
To this order the variables $\st u_l$ are 
\begin{equation}
\bigl (
\dr x^m, \partial_m C^a,
\partial_m C^{ab},\partial_m C^i, \partial_{m_k}\dots \partial_{m_0}C^N ,
\ir B^N, \ir \partial_{m_k}\dots \partial_{m_1} B^N
\bigr )
\end{equation}
Also to lowest order the covariant derivatives of the field strengths are 
\begin{equation}
\bigl( T \bigr) \approx \bigl( \A{E}{a_k}{m_k}\dots \A{E}{a_0}{m_0}
\partial_{m_k}\dots \partial_{[m_1}\A{A}{m_0]}{N},\ k=1,2,\dots \bigr)\ . 
\end{equation}
The brackets denote antisymmetrization of the enclosed indices.
In linearized order we find all jet variables as linear combinations
of  the variables $\tilde{C},T, u_l$ and $\st u_l $: the
symmetrized derivatives of the connections belong to
$\bigl (u_l\bigr )$, the antisymmetrized derivatives of the connections
belong to the field strengths listed as $T$.
The derivatives of the vielbein are slightly tricky. The symmetrized
derivatives are the variables
$-\partial_{(m_k}\dots \partial_{m_1} \A{A}{{m_0)}}{N}$
for $N=a$. The antisymmetrized derivatives of the vielbein
are in one to one correspondence to the spin connection because we have chosen
it such that the torsion vanishes.

So the transformation of the jet variables $\psi=\{\phi\}=\phi,\partial\phi,\dots$ to the
variables $\psi^\prime = \bigl ( \tilde{C},T, u_l,\st u_l \bigr )$ has the structure
\begin{equation}
\label{invers}
\psi^{\prime\,i} = M^i{}_{j} \psi^j + O^i(\psi^2)\ ,
\end{equation}
where $M$ is invertible,
\begin{equation}
M^i{}_j=\frac{\partial \psi^{\prime\,i}}{\partial \psi^j}_{|_{\psi=0}}\,.
\end{equation}
Therefore, the map $\psi \rightarrow \psi^\prime$ is invertible
in a neigbourhood of $\psi=0$\,, analytic functions $f(\psi)$ are 
analytic functions $F(\psi^\prime)=f(\psi(\psi^\prime))$ of $\psi^\prime$ and the algebra, generated by $x, \dr x$ and the jet variables 
$\{\phi\}$, coincides with the algebra, generated by $\psi^\prime$\,,
\begin{equation}
A_{x,\dr x, \{\phi\}}= A_{\tilde{C},T}\otimes \prod_l A_{u_l,\st u_l}.
\end{equation}
Because $\st$ leaves each factor of the product algebra invariant,
K\"unneth's theorem (theorem \ref{kunn}) applies and 
the cohomology of~$\st $ acting on the algebra $A_{x,\dr x, \{\phi\}}$ 
of the jet variables is given by the product
of the cohomologies of~$\st $ acting on the
ghost tensor algebra $A_{\tilde{C},T}$ and on the
algebras $A_{u_l,\st u_l}$
\begin{equation}
\label{cohprod}
H(A,\st ) = H(A_{\tilde{C},T}\,,\st )\otimes \prod_l
H(A_{u_l,\st u_l},\st )\ .
\end{equation}
By the basic lemma (theorem \ref{basic}) the cohomology of $\dv$ acting
on an algebra $A_{x,\dr x}$ of differential forms $f(x,\dr x)$ which depend on
generating and independent variables $x$ and $\dr x$ is given by numbers
$f_0$. The algebra $A_{u_l,\st u_l}$ and the action of $\st$ on this algebra
differ only in the denomination. Therefore the cohomology  
$H(A_{u_l,\st u_l},\st )$ is given by numbers, at least as long as
the variables $u_l$ and $\st u_l$ are independent and not subject to constraints.

Whether the variables $u_l, \st u_l$ are subject to constraints is
a matter of choice of the theory which one considers. This choice
influences the cohomology. For example, one could require that two coordinates
$x^1$ and $x^2$  satisfy $(x^1)^2 +(x^2)^2= 1$ because one wants to consider a
theory on a circle. Then the differential
$\dv({\arctan}\frac{y}{x})=\dr \varphi$ is closed
($\dv \dr \varphi=0$) but not exact, because the angle $\varphi$ is not a
function on the circle, $\dr \varphi $ is just a misleading notation for a
one form which is not $\dv$ of a function $\varphi$. In this
example the periodic boundary condition
$\varphi \sim \varphi + 2\pi$ gives rise to a nontrivial cohomology
of $\dv$ acting on $\varphi$ and $\dr \varphi$. Nontrivial cohomologies also
arise if the fields take values in nontrivial spaces. For example if
in nonlinear sigma models one requires scalar fields $\phi^i$ to take
values on a sphere $\sum_{i=1}^{n+1}  {\phi^i}^2=1 $ then the volume form
$\dr^n \phi$ is nontrivial. More complicated is the case where scalar
fields are restricted to take values in a general coset $G/H$ of a group $G$ with a subgroup $H$\,. 
Also the relation 
\begin{equation}
\det \A{e}{m}{a}\neq 0
\end{equation}
restricts the vielbeine to take values in the group $GL(D)$ of
invertible real $D\times D$ matrices. This group has the nontrivial cohomology of $O(D)$.

In our investigation we neglect the cohomologies coming from a
nontrivial topology of the base manifold with coordinates $x^m$ or the
target space with coordinates~$\phi$ or $\A{e}{m}{a}$.
We have to determine the cohomology of $\st $ on the ghost tensor
variables anyhow and start with this problem. To obtain the complete
answer we can determine the cohomology of the base space and the target
space in a second step which we postpone.
%
%
%
So we choose to investigate topologically trivial
base manifolds and target spaces. We combine eq. (\ref{cohprod})
with theorem \ref{absts} and theorem \ref{sequiv} and conclude
\begin{theorem}{\quad}\\
If the target space and the base manifold have trivial topology then
the nontrivial solutions of the descent equations  in gravitational
theories are in one to one correspondence to the nontrivial solutions
$\omega(C,T)$ of the equation $\s \omega = 0$. Up to trivial terms 
the solution $\omega_D$ of the descent equation {\normalfont(\ref{relcoh1})}  
is given by the $D$-form part of the form $\omega(C+A,T).$
\end{theorem}

$\omega$ depends only on the ghosts, not on their derivatives. Therefore
the ghostnumber of $\omega$ is bounded by the number of ghosts
for translations and spin and isospin transformations
$D+\frac{D(D-1)}{2}+\text{dim}(G)$. If we take the $D$-form part of
$\omega(C+A,T)$ then $D$ differentials $\dr x^m$
rather than ghosts have to be picked. Therefore the ghostnumber of
nontrivial solutions of the relative cohomology is bounded by
$\frac{D(D-1)}{2}+\text{dim}(G)$. 

{}From the theorem one concludes that anomaly candidates which one
expresses in terms of the ghost variables $\hat{C}$\,, (the index $i$ enumerates
spin and isospin values collectively, translation ghosts are denoted by $c$, to distinguish them from
spin and isospin ghosts $C$)  
\begin{equation}
\label{chat}
\hat{C}^i=C^i-c^a\A{E}{a}{m}\A{A}{m}{i}\ ,\  \hat{c}^m=c^a\A{E}{a}{m}\ ,
\end{equation}
can be chosen such that they contain no ghosts $\hat{c}^m$ of coordinate 
transformations or, in other words, that coordinate transformations are not anomalous.

This holds, because the variables $\hat{C}^i$ are invariant under the shift 
$\rho: C \mapsto C + A$ (\ref{rho}), only the translation ghosts are shifted, 
\begin{equation}
\rho (\hat{C}^i) = C^i\ ,\ \rho (\hat{c}^m) = \hat c^m + \dr x^m\ .
\end{equation}
Therefore, if one expresses a form $\omega(C+A,T)$ by ghost variables
$\hat{c}$, $\hat{C}$ then $\omega$ depends on $\dr x^m$ only via the
combination $\hat{c}^m + \dr x^m$. The $D$ form part $\omega_D$
originates from a coefficient function multiplying
\begin{equation}
(\hat{c}^1 + \dr x^1)(\hat{c}^2 + \dr x^2)\dots (\hat{c}^D + \dr x^D)=
(\dr x^1\dr x^2\dots \dr x^D + \dots) \ .
\end{equation}
This coefficient function of $\dr^D x $ cannot contain a translation ghost
$\hat{c}^m$ because $\hat{c}^m$ enters only in the combination
$\hat{c}^m + \dr x^m$ and  $D+1$ factors of $\hat{c}^m + \dr x^m$ vanish.

In our formulation $\s$ maps the subalgebra of ghosts and tensors to itself,
\begin{equation}
-\s T =  C^N \Delta_N T = c^a \A{E}{a}{m}(\partial_m +
\A{A}{m}{i}\delta_i)\,T + C^i\, \delta_i T\ .
\end{equation}
In terms of the ghosts $\hat{C}$ this is a shift term
$\hat c^m\partial_m T$ and the \textsc{brst} transformation of a Yang Mills theory
\begin{equation}
- \s T = \hat{c}^m\partial_m T + \hat{C}^i \delta_i T\ .
\end{equation}
This formulation arises naturally if one enlarges the
\textsc{brst} transformation of Yang Mills theories to allow also general
coordinate transformations. However, $\partial_m T$ is not a tensor
and it is not manifest, that $\s$ leaves a subalgebra invariant. 


\section{BRST Cohomology on Ghosts and Tensors}\label{sec5}
\subsection{Invariance under Adjoint Transformations}
In the preceding section the problem to determine Lagrange densities
and anomaly candidates has been reduced
to the calculation of the cohomology of $\s$ acting
on tensors and ghosts, 
\begin{equation}
\s \omega(C,c,T) = 0\ ,\ \omega\  \modulo \s \eta(C,c,T)\ .
\end{equation}
Let us recall the transformation $\s$ explicitly
\footnote{By choice of the spin connection $\A{T}{ab}{c}$ vanishes. Translation ghosts
are denoted by~$c$ to distinguish them from spin and isospin ghosts $C$, which are enumerated by~$i$.}
\begin{align}
\s T&=-(c^aD_a + C^i\delta_i ) T\ ,\\
\s c^a&=-C^ic^b\A{G}{ib}{a}\ ,\\
\label{defs}
\s C^i&= \frac{1}{2}C^kC^l\A{f}{kl}{i}+\frac{1}{2}c^a c^b\A{F}{ab}{i}\ .
\end{align}
To determine the cohomology of $\s$, we proceed as in the derivation of the
basic lemma and investigate the anticommutator of $\s$ with
other fermionic operations. Here we consider the partial derivatives
with respect to the spin and isospin ghosts $C^i$. These anticommutators are
the generators  $\delta_i$ of spin and isospin transformations
\begin{equation}
\label{delant}
\delta_i=-\{\s,\frac{\partial}{\partial C^i}\}
\end{equation}
which on the ghosts $c$ are represented by $G_i$ and on the ghosts $C$ by the adjoint
representation
\begin{equation}
\delta_i c^a = \A{G}{ib}{a}c^b \ ,\ 
\delta_i C^j = \A{f}{ki}{j}C^k\ .
\end{equation}
Eq. (\ref{delant}) is easily verified on the elementary variables
$c,C$ and $T$. It extends to arbitrary polynomials because
both sides of the equation are linear operations with the same product
rule.

Arbitrary linear combinations $\delta = a^i \delta_i$ of the spin and isospin transformations 
commute with $\s$ because each anticommutator $\{\s,\ro \}$  of a nilpotent $\s$ commutes with
$\s$ no matter what operation $\ro$ is (\ref{danti}),
\begin{equation}
[\delta,\s] = 0\ .
\end{equation}
The representation of the isospin transformations on the algebra of
ghosts and tensors is completely reducible because the
isospin transformations belong to a semisimple group or to abelian
transformations which decompose the algebra into polynomials of definite
charge and definite dimension. Therefore the following theorem applies.
\begin{theorem}{\quad}\\
\label{sinv}
If the representation of $\delta_i$ on the ghost and tensor algebra is completely 
reducible then each
solution of $\s \omega~=~0$ is invariant under all $\delta = a^i \delta_i$ up to an
irrelevant piece,
\begin{equation}
\s \omega = 0 \Rightarrow  \omega = \omega_{\normalfont\text{inv}}+ \s \eta
\ ,\  \delta\, \omega_{\normalfont\text{inv}}=0\ .
\end{equation}
\end{theorem}
The theorem is proven by the following argument. The null space of $\s $\,,
\begin{equation}
Z=\{\omega :\ \s \omega = 0 \}\ ,
\end{equation}
is mapped by spin and isospin transformations to itself,
\begin{equation}
\s \delta_i \omega  = \delta_i \s \omega = 0 \ ,\ 
\delta_i Z \subset Z\ .
\end{equation}
$Z$ contains the subspace $Z_\delta$ of elements which can be written as  sum of isospin
transformations applied to some other elements $\kappa^i \in Z$,
\begin{equation}
Z_{\delta}= \{\omega \in Z:\ \omega = \delta_i ( \kappa^i )\ ,\ 
\s \kappa^i =0 \}\ .
\end{equation}
$Z_\delta$ is mapped by isospin transformations to
itself. A second invariant subspace is given by
$Z_{\text{inv}}$,
the subspace of $\delta$ invariant elements,
\begin{equation}
Z_{\text{inv}}=\{\omega \in Z:\  a^i\delta_i \omega = 0\ \}\ .
\end{equation}
If the representation of $\delta_i$ is completely reducible then
the space $Z$ decomposes as a sum 
\begin{equation}
Z = Z_{\text{inv}}\oplus Z_\delta\oplus Z_{\text{comp}}
\end{equation}
with a complement $Z_{\text{comp}}$ which is also mapped to itself. This
complement, however,
contains only $\omega=0$ because if there were a nonvanishing element
$\omega \in Z_{\text{comp}}$ it would
not be invariant because it is not  from $Z_{\text{inv}}$. $\omega$ would be
mapped to $\delta \omega \in Z_{\delta}$ and $Z_{\text{comp}}$ would not be
an invariant subspace,
\begin{equation}
Z = Z_{\text{inv}}\oplus Z_{\delta}\ .
\end{equation}
Each  $\omega $ which solves $\s \omega = 0 $ can therefore be
decomposed as
\begin{equation}
\omega = \omega_{\text{inv}}+  \delta_i  \kappa^i \ ,\  \s \kappa^i = 0\ , \ \delta_i \omega_{\text{inv}}= 0 \ .
\end{equation}
We replace $\delta_i$ by $-\{s,\frac{\partial}{\partial C^i}\}$
(\ref{delant}), use $\s \kappa^i = 0$  and verify the theorem,
\begin{equation}
\omega = \omega_{\text{inv}} + \s \eta \ ,\  
\eta= -\frac{\partial}{\partial C^i}\kappa^i\ .
\end{equation}

The theorem restricts nontrivial solutions to $\s \omega = 0$ to $\delta$-invariant~$\omega$. If we decompose it as a sum of products of 
polynomials $\Theta_l$ of the spin and isospin ghosts $C$ and of forms~$f^l$ of 
the translations ghosts $c$, which depend on tensors~$T$, it has the form
\begin{equation}
\omega(C,c,T) = \sum_l \Theta_l(C) f^l(c,T)\ , 
\end{equation}
where $\Theta_1, \Theta_2 \dots $ as well as  $f^1, f^2\dots$ can be taken to be linearly independent (otherwise one could
express $\omega$ as a sum with fewer terms).

The operation $\s$ decomposes into
\begin{equation}
\s = -C^i \delta_i + \s_c + \s_1 + \s_2 \ ,
\label{fb1}
\end{equation}
where $\s_c$ acts only on spin and isospin ghosts $C$ and preserves the number
of translation ghosts and 
where $\s_1$ and $\s_2$ increase this number by $1$ and $2$,
\begin{align}
\s_c T  &= 0\ , & \s_c c^a &= 0\ ,& \s_c C^i &= -\frac{1}{2}C^j C^k f_{jk}{}^i\ ,\\
\label{defs1}
\s_1 T &=-c^a D_a T\ , &\s_1 c^a &= 0\ ,& \s_1 C^i &= 0\ ,\\
\label{defs2}
\s_2 T &=0\ ,&\s_2 c^a &= 0\ ,& \s_2 C^i &=\frac{1}{2}c^a c^b\A{F}{ab}{i}\ .
\end{align}
Therefore, to lowest order in the translation ghosts $c$ the conditions $\s \omega = 0$ and 
$\delta\, \omega = 0$ lead to
\begin{equation}
0 = \s_c \omega =  \s_c \sum_l \Theta_l(C) f^l(c,T) = \sum_l (\s_c \Theta_l) f^l(c,T) \ ,
\label{fb2}
\end{equation}
which implies that each $\Theta_l$ is a solution to
\begin{equation}
\s_c \Theta = 0\ ,\Theta\, \modulo \s_c \eta\ ,
\label{fb3}
\end{equation}
because the functions $f^1, f^2\dots $ are linearly independent. If we change e.g. $\Theta_1$
by $\s_c \eta$, then $\omega$ is changed to lowest order in form degree
by $\s (\eta  f^1)-(-)^{|\eta|}\eta\s f^1 $, i.e. up to a trivial term by a form 
$\hat \omega = -(-)^{|\eta|}\eta\s f^1$. But the condition $\s \omega = 0$ and $\s (\omega + \hat{\omega})=0$,
differs only by denomination and one obtains the same set of solutions, whether one uses $\Theta_1$ or 
$\Theta_1+ \s_c \eta$. 

Therefore each $\Theta_i$ is an element of the Lie algebra cohomology 
of the spin and isospin ghosts $C$\,. In particular, it can be taken to be  $\delta$ invariant, 
because $-\{\s_c, \frac{\partial}{\partial C^i}\}$
generates the $\delta$ transformations of the ghosts $C$.  
Therefore all $f^l$ are also $\delta$ invariant.

The algebra of the spin and isospin ghosts is a product algebra of the ghosts of the simple and abelian 
factors of the spin and isospin Lie algebra. 
Each factor of the algebra is left invariant by $\s_c$. Therefore the space of invariant 
ghost polynomials can be determined  separately for each factor of the Lie algebra. 
By K\"unneth's formula (theorem \ref{kunn}) the 
cohomology of the product algebra is the product of the cohomologies of the factors. 

\subsection{Lie Algebra Cohomology}

The following results for simple Lie algebras can be found in the
mathematical literature \cite{greub3}:
the cohomology of $\s_c$ has dimension $2^r$ where $r$ is the rank of the Lie algebra. 
The cohomology is the algebra generated by $r$ primitive polynomials $\theta_\alpha(C), \ \alpha=1,\dots ,r$.
These primitive polynomials cannot be written as a sum of products of other invariant
polynomials. They have odd ghostnumber
$\ghost(\theta_\alpha(C))=2m(\alpha)-1$ and therefore are fermionic. They
can be obtained from traces of suitable matrices $M_i$
which represent a basis of the Lie algebra and are given with a suitable
normalization by
 \begin{equation}
\theta_\alpha (C)=
\frac{(-)^{m-1}m!(m-1)!}{(2m-1)!}
\tr(C^i M_i)^{2m -1}
\ ,\  m=m(\alpha) \ ,\  \alpha=1,\dots,r\ .
\end{equation}
The number $m(\alpha)$ is the degree of homogeneity of the
corresponding Casimir invariant
\begin{equation}
I_\alpha(X)=\tr (X^i M_i)^{m(\alpha)}\ .
\end{equation}
These Casimir invariants generate all invariant functions of a
set of commuting variables~$X^i$ which transform as an
irreducible multiplet under the adjoint representation.

The degrees $m(\alpha)$ for the classical simple Lie algebras are given by
\begin{equation}
\label{mlist}
\begin{array}{l l l l }
SU(n+1)&A_{n}&m(\alpha)=\alpha+1&\alpha=1,\dots,n\ ,n \ge 1\,,\\
SO(2n+1)&B_n&m(\alpha)=2\alpha&\alpha=1,\dots,n\ ,n \ge 2\,,\\
SP(2n)&C_n&m(\alpha)=2\alpha& \alpha=1,\dots,n\ ,n \ge 3\,,\\
SO(2n)&D_n&m(\alpha)=2\alpha&\alpha=1,\dots,n-1,\,m(n)=n,\,n \ge 4\,.
\end{array}
\end{equation}
With the exception of the last primitive element of $SO(2n)$
the matrices $M_i$ are the defining representation of the classical
Lie algebras.
The last primitive element $\theta_n$ and the last Casimir invariant
$I_n$ of $SO(2n)$ are constructed from the spin representation
$\Gamma_i$.
Up to normalization they are given by
\begin{equation}
\begin{gathered}
\theta_n \sim \varepsilon_{a_1b_1\dots a_nb_n}(C^{a_1}{}_{c_1}C^{c_1b_1})
\dots (C^{a_{n-1}}{}_{c_{n-1}}C^{c_{n-1}b_{n-1}} )\,C^{a_nb_n}\\
I_n \sim \varepsilon_{a_1b_1\dots a_nb_n}X^{a_1b_1}\dots X^{a_nb_n}\ .
\end{gathered}
\end{equation}
If $n$ is even then the element $\theta_n$ of $SO(2n)$ is
degenerate in ghostnumber with $\theta_{\frac{n}{2}}$.

The primitive elements for the exceptional simple Lie algebras
$G_2,$ $F_4,$ $E_6,$ $ E_7,$ $ E_8$ can also be found in the literature
\cite{oraif}.
Their explicit form is not important for our purpose. In each case the
Casimir invariant with lowest degree $m$  is quadratic ($m=2$).

For a one dimensional abelian Lie algebra the ghost $C$ is invariant
under the adjoint transformation. It generates the invariant
polynomials $\Theta(C) = a + b\, C$ which span a $2^r$ dimensional
space where $r=1$ is the rank of the abelian Lie algebra. The generator
$\theta$ of this algebra of invariant polynomials has odd ghostnumber
gh$(C)=2m-1$ with $m=1$.
\begin{equation}
\theta(C) =  C
\end{equation}
The Casimir invariant $I$ of the one dimensional, trivial
adjoint representation acting on a bosonic variable $X$ is homogeneous
of degree $m=1$ in $X$ and is simply given by $X$ itself,
\begin{equation}
I(X)= X\ .
\end{equation}

Polynomials of $r$ anticommuting variables $\theta_\alpha$ constitute 
a $2^r$ dimensional Grassmann algebra. The statement that the primitive elements $\theta=(\theta_1,\theta_2\dots)$
generate the Lie algebra cohomology asserts that 
\begin{equation}
\s_c \Theta(C)= 0 \Leftrightarrow
\Theta(C) =\Phi(\theta(C)) + \s_c \eta\ .
\end{equation}
Because the cohomology is $2^r$ dimensional  there are no algebraic relations
among the functions $\theta$ apart from the anticommutation relations
which result from their odd ghostnumber,
\begin{equation}
\Theta(C)=\Phi(\theta(C)) = 0
\Leftrightarrow \Phi(\theta) = 0\ .
\end{equation}

The Casimir invariants $I=(I_1,I_2\dots)$ generate the space of
$\delta$ invariant polynomials in commuting variables $X$ which
transform under the adjoint representation
\begin{equation}
\delta_i P(X) = 0 \Rightarrow
P(X)=f(I(X))\ .
\end{equation}

If there are no algebraic relations among the variables $X$ apart from their
commutation relations then there is no algebraic relation among the Casimir
invariants $I(X)$ up to the fact that the $I_\alpha$ commute
\cite{greub3}.
\begin{equation}
P(X)=f(I(X))=0 \Leftrightarrow f(I)=0
\end{equation}

To sum up: if we expand the solution $\omega$ of $\s \omega = 0$ into parts $\omega_n$ with 
definite degree~$n$ in translations ghosts $c$
\begin{equation}
\omega = \omega_{\underline n} + \sum_{n > \underline n} \omega_n
\end{equation}
then, up to a trivial solution, $\omega_{\underline n}$ is
a superfield~$\Phi$ in the anticommuting primitive invariant polynomials $\theta = (\theta_1, \dots \theta_r)$
\begin{equation}
\omega_{\underline  n}(C,c,T)=\Phi(\theta(C),c,T)
\label{fb4}
\end{equation}
with coefficients, which are spin and isospin invariant forms of the translation ghosts $c$ and tensors.

In next to lowest degree the equation $\s \omega = 0$ imposes the restriction
\begin{equation}
\label{co1}
\s_1 \omega_{\underline n}+ (- C^i \delta_i + \s_c) \omega_{{\underline n }+1 }=0\ .
\end{equation}
But $\s_1$ (\ref{defs1}) maps invariant functions of the translations ghosts and tensors to
invariant functions and treats spin and isospin ghost as constants,
\begin{equation}
s_1 T = c^aD_a T\ ,\  s_1 c^a = 0 \ ,\  s_1 C^i = 0\ .
\end{equation}
Therefore $s_1 \Phi$ is again a superfield in $\theta$ with coefficients $f(T)$, which are
$\delta$ invariant forms of the translation ghosts and which depend on tensors. Such a superfield
is not of the form $(-C^i\delta_i+\s_c) \eta$\,, unless it vanishes, therefore (\ref{co1}) implies
\begin{equation}
\s_1 \Phi(\theta,c, T) = 0\ ,
\label{fb6}
\end{equation}
and because $\s_1$ only acts on the coefficients $f$ of the $\theta$ expansion of $\Phi$, they have to
satisfy
\begin{equation}
\s_1 f(c,T) = 0\ , \ f \modulo \s_1 \eta\ .
\label{fb7}
\end{equation}
Indeed, we can neglect a contribution of the form
$(\s_1 \eta)\,g(\theta)$ because it can be written as $(\s - \s_2)(\eta\,g(\theta) )$ 
because $\eta$ and $g$ are $\delta_I$ and $\s_c$ invariant.
$\s (\eta\, \Phi(\theta)) $ changes $\omega=\omega_{\underline n}+ \dots$ only by an
irrelevant piece. $\s_2 (\eta\, g(\theta))$ can be absorbed in the parts $\dots$
with higher ghost degree. Therefore we can neglect contributions
$(s_1\eta)\, g(\theta) $  to $\omega_{\underline n}$.

To determine the cohomology of $\s_1$ on the algebra of undifferentiated translation ghosts
and tensors, 
we split the ghost form $f(c,T)$ and $\s_1 f$ according to a number $N$ of jet variables, where we do not 
count powers of the vierbein $e_m{}^a$ and its inverse $E_b{}^n$, but count the differentiated $e_m{}^a$,
count the connections for isospin transformations and their derivatives with a weight $2$ and count the ghosts
and remaining variables with normal weight,
\begin{equation}
N = N_{\{\partial e\}}+ 2 N_{\{A\}} + N_{\{C\}} + N_{\{\phi\}}\ .
\end{equation}
\begin{equation}
f = \sum_{n \ge \underline{n}}f_n\ ,\ N\,f_n = n \,f_n\ ,
\end{equation}
The operation $\s_1$ decomposes into pieces $\s_{1\,n}$, 
which increase the $N$ number by $n$
\begin{equation}
\s_1 = \sum_{n\ge 1} \s_{1, n}\ ,\ [N , \s_{1, n}] = n \,\s_{1, n}\ .
\end{equation}
Then the equation $\s_1 f = 0$ implies in lowest $N$ order 
\begin{equation}
\label{s1lin}
\s_{1, 1}f_{\underline n}= 0 \ , \ f_{\underline n}\modulo \s_{1,1} \eta\ .
\end{equation}
We can neglect contributions $\s_{1, 1} \eta$ to $f_{\underline{n}}$ because up to terms of higher $N$ number
they are of the form $\s_1 \eta$ and therefore trivial.

The ghosts are invariant under $\s_1$ (\ref{defs1}), on tensors $\s_1$ acts by 
\begin{equation}
\s_1 T = - c^a D_a T = - c^m (\partial_m + A_m^i \delta_i)\, T\ .
\end{equation}
Therefore $\s_{1 ,1}$ acts on ghosts and tensors as
\begin{equation}
\s_{1, 1} = - c^m \partial_m
\end{equation}
where the partial derivative $\partial_m$ only acts on tensors, not on ghosts and not on $e_m{}^a$\,,
because the differentiation of $e_m{}^a$ increases the $N$ number. 

The part $c^m A_m{}^i\delta_i$ increases the $N$ number by at least~$2$,
even if an isospin transformation decreases the number $N$ of fields and
transforms a field $\phi$ into $\delta_i \phi$ with a field independent part. 
If the field independent part $(\delta_i \phi)_{0}$ does not vanish, then
the field~$\phi$ is called a Goldstone field and the transformation
is said to be nonlinear or to be a spontaneously broken symmetry.

Lorentz transformations are not spontaneously broken, they transform fields into linear 
combinations of fields. Isospin transformation may be spontaneously broken, but $N$ counts their
connection with a weight $2$, so $c^m A_m{}^i\delta_i$ increases the $N$ number by at least~$2$
and $c^m A_m{}^i\delta_i$ does not contribute to $\s_{1, 1}$.

To lowest $N$ order, the partial derivative in $\s_{1, 1}$ does not differentiate ghosts and 
the vierbein, therefore the derivative $\partial_m c^n = \partial_m (c^a E_a^m)$ vanishes. This justifies to 
change the notation and denote  $c^m$ by $\dr x^m$. Then $\s_{1, 1}$ is the exterior derivative and,
changing the name $f_{\underline{n}}$ to $\omega$, it is a differential form, which 
to lowest order in the fields satisfies
\begin{equation}
\dv \omega(T,\dr x) = 0\ ,\ \omega\, \modulo \dr \eta(T,\dr x)
\label{cplproblem}
\end{equation}
by (\ref{s1lin}) and depends on the lowest order parts of tensors~$T$\,.

\subsection{Covariant Poincar\'e Lemma}\label{sec5.3}

Because we consider only the terms with lowest $N$ number, the covariant derivatives
of tensors, which appear in $\omega$, contribute only with their partial derivative.
By the same reason, the field strength and curvature enter only as antisymmetrized 
derivative of the connection,
\begin{equation}
D_{m_{_\text{ linearized}}} = \partial_m\ ,\ 
F_{mn{\text{ lin}}} = \partial_m A_n - \partial_n A_m\ .
\end{equation}
Collectively we call the linearized field strength 
and the remaining fields $T$ (not the ghosts which we 
are going to introduce in (\ref{shilf})), on which by assumption
the derivatives act freely, together with their (higher) derivatives linearized tensors. 

Differential forms 
with coefficients,  which are functions of linearized tensors  are termed linearized tensor forms.

The cohomology of $\dv$ acting on jet forms is given by the
algebraic Poincar\'e lemma (theorem \ref{algebraic}). This lemma,
however, does not apply, if the differential forms are restricted to
be linearized tensor forms,  because the derivatives do not act freely, i.e. without 
restriction apart from the property that they commute, but subject to the Bianchi identities 
(and their derivatives) 
that the antisymmetrized derivatives of the field strength vanish, 
\begin{equation}
\partial_k  F_{lm{\text{ lin}}}+ \partial_l  F_{mk{\text{ lin}}} + \partial_m  F_{kl{\text{ lin}}}= 0\ .
\end{equation}

The cohomology of $\dv$ acting on linearized tensor forms $\omega(T,\dr x)$ has been derived for Yang Mills theories \cite{Brandt:1989rd,Brandt:1989gy,DuboisViolette:1992ye}, Riemannian geometry \cite{gilkey} and 
gravitational Yang Mills theories \cite{Brandt:1989et}. We consider a slightly more general problem 
and analyse linearized tensor forms $\omega(T,x,\dr x)$ which may also depend on the 
coordinates $x^m$. On such forms we investigate the cohomology of the exterior 
derivative $\dvp$\,, which differentiates only the fields, and of 
$\dv=\dvp+\dv_x$ (\ref{dvp}) which also differentiates coordinates.
The results apply, in particular, if background fields occur which are given functions of the
coordinates. The solution of (\ref{cplproblem}) is spelled out at the end of this subsection.

\begin{theorem}{Linearized Covariant Poincar\'e Lemma for $\dv=\dvp+\dv_x$}\\
(i) Let $\omega$ be a linearized tensor form which may depend on the $x^m$, then
\begin{align}
\nonumber
\dv \omega(T,x,\dr x) = 0\,  &\Leftrightarrow & \omega(T,x,\dr x) &= {\cal L}(T,x)\,\dr^D x + 
P(\dr A)
+ \dr \eta(T,x,\dr x),\\
\omega(T,x,\dr x) = \dv \chi\,  &\Leftrightarrow &\omega(T,x,\dr x) &=  P_1(\dr A) + 
\dr \eta(T,x,\dr x)\,,\label{cpl1}
\end{align}
where the Lagrangian form $\mathcal L\, \dr^D x$ and $\eta$ are linearized tensor forms 
which may depend on the $x^m$
and where $P$ and $P_1$ are polynomials in the field 
strength two forms $\dr A^i = \frac{1}{2}\dr x^m \dr x^n (\partial_m A_n{}^i -\partial_n A_m{}^i)$ 
and where $P_1$ has no constant (field independent) part, $P_1(0)=0$\,.\\
(ii) A polynomial $P$ in the field 
strength two forms cannot be written as 
exterior derivative $\dv$ of a tensor form $\eta$ which may depend on the $x^m$,
\begin{equation}
\label{pplusdeta}
P(\dr A) + \dr \eta(T,x,\dr x) = 0\  \Leftrightarrow\ P(\dr A) = 0 = \dr \eta(T,x,\dr x)\ .
\end{equation}
\label{CPL1}
\end{theorem}
For the exterior derivative $\dvp$ which differentiates only the fields one gets:
\begin{theorem}{Linearized Covariant Poincar\'e Lemma for $\dvp$}\\
(i) Let $\omega$ be a linearized tensor form which may depend on the $x^m$, then
\begin{align}
\notag
\dvp \omega(T,x,\dr x) = 0\ &\Leftrightarrow & \omega(T,x,\dr x) =&\, {\cal L}(T,x)\,\dr^D x + 
P(\dr A,x,\dr x) \\ 
&&&+ \dvp \eta(T,x,\dr x)\ ,\label{cpl2}\\
\notag
\omega(T,x,\dr x) = \dvp \chi\ &\Leftrightarrow &\omega(T,x,\dr x) =&\,  P_1(\dr A,x,\dr x) + 
\dvp \eta(T,x,\dr x)\ , 
\end{align}
where the Lagrangian form $\mathcal L\, \dr^D x$ and $\eta$ are linearized tensor forms 
which may depend on the $x^m$
and where $P$ and $P_1$ are polynomials in the field 
strength two forms $\dr A^i = \frac{1}{2}\dr x^m \dr x^n (\partial_m A_n{}^i -\partial_n A_m{}^i)$ 
and the coordinate differentials $\dr x^m$ which may depend on the $x^m$ and where $P_1$ has no 
field independent part.\\
(ii) A polynomial $P$ in the field 
strength two forms and the coordinate differentials $\dr x^m$ which may depend on the $x^m$ cannot 
be written as exterior derivative $\dvp$ of a tensor form $\eta$ which may depend on the coordinates,
\begin{equation}
\begin{aligned}
\label{pplusdeta2}
& P(\dr A,x,\dr x) + \dvp \eta(T,x,\dr x) = 0  \Leftrightarrow \\
& P(\dr A,x,\dr x) = 0 = \dvp \,\eta(T,x,\dr x)
\end{aligned}
\end{equation}
\label{CPL2}
\end{theorem}
The part ${\cal L}(T,x)\,\dr^D x$ is not of the from $P(\dr A) + \dr \eta(T,x,\dr x)$ or 
$P(\dr A,x,\dr x) + \dvp \eta(T,x,\dr x)$, respectively, if and only if its Euler derivative
(\ref{eulder}) does not vanish.

We prove the theorems by induction with respect to the form degree, starting with the proof of 
(\ref{cpl1}). (\ref{cpl1}) holds
for $0$-forms: By the algebraic Poincar\'e lemma (theorem \ref{algebraic}), $0$-forms are closed if and 
only if they are 
constant. A constant, however, is a polynomial $P(\dr A)$.

Assume the theorems to hold for all forms with degree less than the form degree $p>0$ of the
form $\Omega(T,x,\dr x)$ which solves $\dv \Omega(T,x,\dr x) = 0$ or which in case $p=D$ has 
vanishing Euler derivative.
Then by the algebraic Poincar\'e lemma (theorem \ref{algebraic}) $\Omega$ is of the form
\begin{equation}
\Omega =\dv \omega_{p-1} \ ,
\label{leiter}
\end{equation}
where $\omega_{p-1}$ depends on the jetvariables 
$x$,$\dr x$,$A$,$\partial A$,$\partial\dots\partial A$,$T$,$\partial T$,$\partial\dots\partial T$\,.

As $\Omega$ depends only on tensors, it is invariant under the 
transformation $\s$ (\ref{slin}), which annihilates $x, \dr x$, the ghosts $C$ and tensors and
which anticommutes with $\dv$\,. 
\begin{equation}
\label{shilf}
\s \partial_{(m_1}\dots \partial_{m_{k-1}}A^i_{m_{k})} = \partial_{m_1}\dots \partial_{m_{k}}C^i\ ,\ 
k=1,2,\dots 
\end{equation}
(The parentheses denote symmetrization of the enclosed indices).

{}From $\s \Omega = \s \dv \omega_{p-1} = 0$ and $\{\s,\dv\}=0$ one concludes $\dv \s \omega_{p-1}=0$. So, by the 
algebraic Poincar\'e lemma, there is a jetform $\omega_{p-2}$ with ghostnumber 1  such that
\begin{equation}
\label{leiter0}
\s \omega_{p-1} + \dv \omega_{p-2} = 0\ .
\end{equation}
Applying $\s$ to this equation one concludes $\s \dv \omega_{p-2} = 0 $ and derives iteratively
the descent equations
\begin{equation}
\label{leiter2}
\s \omega_{p-k} + \dv \omega_{p-k-1} = 0\ .
\end{equation}

At some stage the iteration has to end
\begin{equation}
\label{unten}
\s \omega_{p-G} = 0\ ,
\end{equation}
because the form degree can not become negative. 

In terms of $\tilde{\omega}= \sum_k \omega_{p-k} $ and $\st = \s + \dv$ all equations
are summed up in
\begin{equation} 
\label{covtild}
\Omega = \st \tilde{\omega}\ .
\end{equation}
The operation $\st$ transforms the variables 
\begin{equation}
\tilde{C}^i=C^i + \dr x^m A^i_m
\end{equation}
into the field strength two form~$\dr A^i$ (making use of $\s \dr x^m A^i_m = - \dv C^i$)
\begin{equation}
\st \tilde{C}^i = \dr A^i=\frac{1}{2}\dr x^m\dr x^n(\partial_m A^i_n -\partial_n A^i_m)
\end{equation}
and defines variables $q^i_{m_1\dots m_k}$ as transformed symmetrized derivatives of $A^i$
\begin{equation}
\st \partial_{(m_1}\dots \partial_{m_{k-1}}A^i_{m_k )}= q^i_{m_1\dots m_k} ,\, 
q^i_{m_1\dots m_k} = \partial_{m_1}\dots\partial_{m_k}C^i + \dots ,\, k=1,2\dots
\end{equation}
Up to nonlinear terms, they are the derivatives of the ghosts.
Coordinates and tensors transform into
\begin{equation}
\st x^m = \dr x^m\ ,\ \st T = \dv_T T\ ,
\end{equation}
where the exterior derivative $\dv_T$ is defined to differentiate only the tensors and to vanish on
the ghosts, on $q^i_{m_1\dots m_k}$ and on the  symmetrized derivatives of $A^i$.
 
We perform a coordinate transformation in jet space and use the new variables $\tilde C^i$ and
$q^i_{m_1\dots m_k}$ in place of the $C^i$ and their partial derivatives as new coordinates. 
The transformation is invertible (\ref{invers}) because the linearized  transformation is invertible. 

On functions of these variables $\st$ acts as the derivative
\begin{equation}
\st = \dr x^m \frac{\partial}{\partial x^m}\, + \dr A^i \frac{\partial}{\partial \tilde{C}^i} +
q^i_{m_1\dots m_k}\frac{\partial}{\partial (\partial_{(m_{1}}\dots A^i_{m_k)})}+ \dv_T\ .
\end{equation}
It commutes with the number operator $N= N_{\partial_(A_)}+ N_q$, which counts the vectorfields~$A^i$,
its symmetrized derivatives and the $q$-variables. 
The tensor form $\Omega$ does not depend on these variables, therefore
$0 = N\Omega = \st N \tilde\omega $ and only the part of $\tilde\omega$ with $N\tilde\omega = 0$ can contribute
to $\Omega$. Therefore we can restrict $\tilde\omega$ to a form, which depends on coordinates $x$, undifferentiated 
ghosts $C^i$ and tensors,
\begin{equation}
\tilde\omega = \sum_{k=0}^m \frac 1{k!}\tilde C^{i_1}\dots \tilde C^{i_k}\chi^{(k)}_{i_1\dots i_k}(T,x,\dr x) .
\end{equation}
The crux of the matter is to prove that $\tilde\omega$ actually can be taken to be at most linear in 
the $\tilde C^i$. 
To prove this, we examine the case that the highest order in the $\tilde C^i$ which occurs in $\tilde\omega$ 
is non-zero, i.e. $m>0$. Since $\Omega$ does not
depend on the $\tilde C^i$, we obtain from
$\Omega = \st\tilde\omega$ at order $m$ in the $\tilde C^i$:
\begin{equation}
m>0:\quad \dv \chi^{(m)}_{i_1\dots i_m}(T,x,\dr x)=0 .
\end{equation}
By induction hypothesis (\ref{cpl1}) holds for all form degrees smaller than $p$.
$\chi^{(m)}_{i_1\dots i_m}(T,x,\dr x)$ has form degree $p-m-1$. Hence, 
we conclude
\begin{equation}
m>0:\quad \chi^{(m)}_{i_1\dots i_m}(T,x,\dr x)=P_{i_1\dots i_m}(\dr A) + \dr \eta_{i_1\dots i_m}(T,x,\dr x) .
\end{equation}
With no loss of generality we can ignore
the contribution $\dv\eta_{i_1\dots i_m}(T,x,\dr x)$ to $\chi^{(m)}_{i_1\dots i_m}$ because we can remove 
it by replacing $\tilde\omega$ with $\tilde\omega^\prime$ given by
\begin{equation}
m>0:\quad 
\tilde\omega^\prime=\tilde\omega-\st\Bigl(\frac {(-)^{m}}{m!}\tilde C^{i_1}\dots 
\tilde C^{i_m}\eta_{i_1\dots i_m}(T,x,\dr x)\Bigr).
\end{equation}
Assuming that this redefinition has been performed and dropping the prime on $\tilde\omega^\prime$, we obtain
\begin{equation}
\chi^{(m)}_{i_1\dots i_m}(T,x,\dr x)=P_{i_1\dots i_m}(\dr A).
\label{coeff2}
\end{equation}
At order $m-1$ in the $\tilde C^i$ we now obtain from $\Omega = \st\tilde\omega$:
\begin{align}
m=1:\quad & \Omega (T,x,\dr x)=\dr A^i\,P_i(\dr A)+\dv\chi^{(0)}(T,x,\dr x)
\label{coeff3}\\
\notag m>1:\quad & 0=\tilde C^{i_1}\dots \tilde C^{i_{m-1}}
\Bigl(\dr A^{i_{m}}P_{i_1\dots i_m}(\dr A)+\dv\chi^{(m-1)}_{i_1\dots i_{m-1}}(T,x,\dr x) \Bigr)
\end{align}
In the cases $m>1$, this imposes
\begin{equation}
\begin{aligned}
m>1:\quad  0=\dr A^{i_{m}}P_{i_1\dots i_m}(\dr A)+\dv\chi^{(m-1)}_{i_1\dots i_{m-1}}(T,x,\dr x).
\end{aligned}
\label{coeff4}
\end{equation}
$\dr A^{i_{m}}P_{i_1\dots i_m}(\dr A)$ has form degree $p-m+1$ and thus, in the cases 
$m>1$, $\dr A^{i_{m}}P_{i_1\dots i_m}(\dr A)$ has a form degree smaller than $p$. Hence, 
assuming that (\ref{pplusdeta}) holds for all form degrees smaller than $p$, we conclude 
from (\ref{coeff4}) by means of (\ref{pplusdeta}):
\begin{equation}
m>1:\quad  \dr A^{i_{m}}P_{i_1\dots i_m}(\dr A)=0=\dv\chi^{(m-1)}_{i_1\dots i_{m-1}}(T,x,\dr x).
\label{coeff5}
\end{equation}
Equation (\ref{coeff5}) implies that $\tilde C^{i_1}\dots \tilde C^{i_m}P_{i_1\dots i_m}(\dr A)$ 
does not contribute to $\Omega = \st\tilde\omega$ in the cases $m>1$ because it is $\st$-invariant and, 
therefore, the contribution 
$\tilde C^{i_1}\dots \tilde C^{i_m}\chi^{(m)}_{i_1\dots i_m}(T,x,\dr x)$ to $\tilde\omega$ 
can be assumed to vanish with no loss of generality whenever $m>1$. In other words: with no loss of 
generality we can assume $m\leq 1$. 

Now, in the case $m=1$, equation (\ref{coeff3}) yields $\Omega = \dv\eta(T,x,\dr x)+P(\dr A)$ with
 $\eta(T,x,\dr x)=\chi^{(0)}(T,x,\dr x)$ and $P(\dr A)=\dr A^iP_i(\dr A)$. If $m=0$ then
$\Omega =\st\tilde\omega$ directly gives $\Omega = \dv\eta(T,x,\dr x)$ with 
$\eta(T,x,\dr x)=\chi^{(0)}(T,x,\dr x)$. This completes the inductive proof of (\ref{cpl1}).

Analogously one can prove (\ref{cpl2}) inductively, with $\dvp$ in place of $\dv$ and using the 
algebraic Poincar\'e lemma for $\dvp$ (theorem \ref{algebraic2}).

The proof of (\ref{pplusdeta2}) is direct because 
$P(\dr A,x,\dr x)$ does not contain a derivative of a linearized tensor while every monomial 
contained in $\dvp \eta(T,x,\dr x)$ contains at least one such derivative.

The proof of (\ref{pplusdeta}) is technically somewhat more involved than the proof of 
(\ref{pplusdeta2}) because $\dv$ (in contrast to $\dvp$) contains the piece 
$\dv_x=\dr x^m\frac{\partial}{\partial x^m}$ which does not add derivatives of fields. 
To prove (\ref{pplusdeta}) we use induction with respect to the form degree of $P(\dr A)$. 
(\ref{pplusdeta}) is trivial in form degrees 0 and 1 because there is no $\dr \eta$ with 
form degree 0 and there is no $P(\dr A)$ with form degree 1.
In the cases that $P(\dr A)$ has a form degree $p$ with $p>1$ we decompose $\eta$ 
into pieces $\eta_k$ with definite degree $k$ in derivatives~$\partial$. 
As the linearized tensors have definite degrees in derivatives, the pieces $\eta_k$
are functions of $x$, $\dr x$ and linearized tensors $T$,
\begin{equation}
\eta(T,x,\dr x)=\sum_{k=0}^M\eta_k(T,x,\dr x),\quad N_\partial\, \eta_k(T,x,\dr x)=k\,\eta_k(T,x,\dr x).
\end{equation}
The decomposition terminates at some degree $M$ in derivatives since $\eta$ is a local form 
(by assumption). Each field monomial contained in $P(\dr A)$ has exactly $p/2=r$ derivatives. 

Assume now that $M\geq r$. In these cases $\dv\eta(T,x,\dr x)+P(\dr A)=0$ imposes $\dvp\eta_M=0$. 
The algebraic Poincar\'e lemma for $\dvp$ (theorem \ref{algebraic2}) implies $\eta_M=\dvp\chi_{M-1}$ 
for some $\chi_{M-1}$ with $M-1$ derivatives~$\partial$. (\ref{cpl2}) for form 
degree $p-1$ now implies $\eta_M=\dvp\chi'_{M-1}(T,x,\dr x)$ (a form $P'(\dr A,x,\dr x)$ cannot 
occur here because such a form would have form degree $p-1$ and thus would only contain terms 
of degree $\leq r-1$ in the $\dr A^i$ which do not contain more than $r-1$ derivatives $\partial$, 
in contradiction to $M\geq r$). Now we consider
$\eta'=\eta-\dr \chi'_{M-1}(T,x,\dr x)$ in place of $\eta$. $\eta'$ fulfills $\dr \eta'(T,x,\dr x)+P(\dr A)=0$
but contains only terms with less than $M$ derivatives. In this way we successively remove from $\eta$
all parts $\eta_k$ with $k\geq r$\,. 

Hence, we can assume $M<r$ with no loss of generality. In the case $M=r-1$ we obtain
$P(\dr A)=\dvp \eta_{r-1}(T,x,\dr x)$ which by  (\ref{pplusdeta2}) implies $P(\dr A)=0$\,. 
If $M<r-1$ we directly obtain $P(\dr A)=0$. So we conclude
$P(\dr A)=0=\dr \eta(T,x,\dr x)$. This ends the proof of theorems \ref{CPL1} and \ref{CPL2}.

Applied to tensor forms, which do not depend on the coordinates,  theorem \ref{CPL2} states the solution 
of (\ref{cplproblem}):
\begin{theorem}{Linearized Covariant Poincar\'e Lemma for $\dvp$ on tensor forms $\omega(T,\dr x)$}\\
(i) Let $\omega$ be a linearized tensor form which does not depend on the $x^m$, then
\begin{align}
\nonumber
\dvp \omega(T,\dr x) = 0\ & \Leftrightarrow & \omega(T,\dr x) &= {\cal L}(T)\,\dr^D x + 
P(\dr A,\dr x) + \dvp \eta(T,\dr x)\ ,\\
\omega(T,\dr x) = \dvp \chi\ & \Leftrightarrow &\omega(T,\dr x) &=  P_1(\dr A,\dr x) + 
\dvp \eta(T,\dr x)\ ,\label{cpl3}
\end{align}
where the Lagrangian form $\mathcal L \dr^D x$ and $\eta$ are linearized tensor forms 
which do not depend on the $x^m$
and where $P$ and $P_1$ are polynomials in the field 
strength two forms $\dr A^i = \frac{1}{2}\dr x^m \dr x^n (\partial_m A_n{}^i -\partial_n A_m{}^i)$ 
and the coordinate differentials $\dr x^m$ and where $P_1$ has no field independent part,
$P_1(0,\dr x)=0$\,.\\
(ii) A polynomial $P$ in the field 
strength two forms and the coordinate differentials $\dr x^m$ cannot be written as 
exterior derivative $\dvp$ of a tensor form $\eta$ which does not depend on the $x^m$,
\begin{equation}
\label{pplusdeta3}
P(\dr A,\dr x) + \dvp \eta(T,\dr x) = 0\  \Leftrightarrow\ P(\dr A,\dr x) = 0 = \dvp \eta(T,\dr x)\ .
\end{equation}
\label{CPL3}
\end{theorem}

\subsection{Chern Forms}

For differential forms, which depend on tensors, this includes field strengths 
\begin{equation}
F_{mn}{}^i = \partial_m A_n{}^i -\partial_n A_m{}^i - A_m{}^j A_n{}^k f_{jk}{}^i
\end{equation}
and the covariant derivatives of tensors, and which are Lorentz invariant and isospin invariant, 
invariant for short, one immediately concludes:
\begin{theorem}{Covariant Poincar\'e Lemma}\\
\label{covpointheo}
Let $\omega$ be an invariant differential form which depends on tensors, then
\begin{equation}
\begin{aligned}
\dv \omega &= 0 & &\Leftrightarrow & \omega &= {\cal L}\,\dr^D x + P + \dr \eta\ ,\\
\omega &= \dv \chi & &\Leftrightarrow &\omega &=  P_1  + \dr \eta\ .\\
\end{aligned}
\end{equation}
The Lagrangian $\mathcal L$ and the differential form $\eta$ are invariant and depend on tensors, 
$P$ and $P_1$ are invariant polynomials in the field strength two forms 
$F^i = \frac{1}{2}\dr x^m \dr x^n F_{mn}{}^i$ 
and $P_1$ has no constant, $F$ independent part.
\end{theorem}

The Lagrange density ${\cal L}\,\dr^D x$ cannot be written as
$P(F)+\dr \eta$ if its Euler derivative does not vanish.

We call the invariant polynomials $P$ Chern forms. They are
polynomials in commuting variables, the field strength two forms $F = (F^1, F^2,\dots)$
which transform under the adjoint representation of the Lie algebra. These
invariant polynomials are generated by the elementary Casimir invariants
$I_\alpha(F)$. 

The Chern forms comprise all topological densities which one can construct from
connections for the following reason. If a functional is to contain only
topological information its value must not change under continuous
deformation of the fields. Therefore it has to be gauge invariant and
invariant under general coordinate transformations. If it is a local
functional it is the integral over a density which satisfies the descent
equation and which can be obtained from a solution to $\s \omega=0$. If
this density belongs to a functional which contains only topological
information then the value of this functional must not change even under
arbitrary differentiable variations of the fields, i.e. its Euler
derivatives with respect to the fields must vanish. Therefore the
integrand must be a total derivative in the space of jet variables. But
it must not be a total derivative in the space of tensor variables
because then it would be constant and contain no information at all.
Therefore, by theorem \ref{covpointheo}, all topological densities
which one can construct from connections are given by Chern polynomials
in the field strength two form.

Theorem \ref{covpointheo} describes also the cohomology of $\s_1$
acting on invariant ghost forms because $\s_1$ acts on invariant ghost
forms exactly like the exterior derivative $\dv$ acts
on differential forms. We have to allow, however, for the additional
variables $\theta(C)$ in $\omega_{\underline n}$. They
generate a second, trivial algebra $A_2$ and can be taken into account
by K\"unneth's theorem (theorem \ref{kunn}). If we neglect the trivial
part $\s_1\eta_{\text{inv}}$ then the solution to (\ref{co1}) is given by
\begin{equation}
\label{sol1}
\omega_{\underline n}=
{\cal L}(\theta(C),T)c^1c^2\dots c^D +
P(\theta(C),I(F))
\end{equation}
The $\delta_I$ invariant Lagrange ghost density satisfies already the
complete equation $s\omega(C,T)=0$ because it is a $D$ ghost form. The
solution to $\st \tilde{\omega} = 0$ is given by
$\tilde{\omega}=\omega(C+A,T)$ and the
Lagrange density and the anomaly candidates are given by the part of
$\tilde{\omega}$ with $\dr^Dx$. The coordinate differentials come from
$c^a+\dr x^m\A{e}{m}{a}$
\footnote{We can use the ghosts variables $C$ or $\hat{C}$ (\ref{chat}).
The expressions remain unchanged because they are multiplied by $D$
translation ghosts.}.
If one picks the $D$ form part then one gets 
\begin{equation}
\dr x^{m_1}\dots \dr x^{m_D}\A{e}{m_1}{1} \dots \A{e}{m_D}{D}=
\det (\A{e}{m}{a})\,\dr^D x \ ,\    \det (\A{e}{m}{a})=:\sqrt{g}
\end{equation}
Therefore the solutions to the descent equations of Lagrange type are
\begin{equation}
\omega_D={\cal L}(\theta(C),T)\,\sqrt{g}\,\dr^D x\ .
\end{equation}
They are constructed in the well known manner from tensors $T$,
including field strengths and covariant derivatives of tensors, which
are combined to a Lorentz invariant and isospin invariant Lagrange
function. This composite scalar field is multiplied by the density
$\sqrt{g}$. Integrands of local gauge invariant actions are obtained
from this formula by restricting $\omega_D$ to vanishing ghostnumber.
Then the variables $\theta(C)$ do not occur. We indicate the
ghostnumber by a superscript and have
\begin{equation}
\omega_D^0={\cal L}(T)\sqrt{g}\,\dr^Dx\ .
\end{equation}
Integrands of anomaly candidates are obtained by choosing $D$ forms
with ghostnumber~$1$. Only abelian factors of the Lie algebra allow for
such anomaly candidates because the primitive invariants $\theta_\alpha$
for nonabelian factors have at  least ghostnumber $3$,
\begin{equation}
\omega_D^1=\sum_i C^i{\cal L}_i(T)\sqrt{g}\,\dr^Dx\ .
\end{equation}
The sum ranges over all abelian factors of the gauge group.
Anomalies of this form actually occur as trace anomalies or $\beta$
functions if the isospin algebra contains dilatations.

This completes the discussion of Lagrange densities and anomaly
candidates coming from the first term in (\ref{sol1}).

\section{Chiral Anomalies}\label{sec6}
\subsection{Chern Simons Forms}
It remains to investigate solutions which correspond to
\begin{equation}
\omega_{\underline n}=P(\theta(C),I(F))\ .
\end{equation}
Ghosts $C^i$ for spin and isospin transformations and ghost forms $F^i$
generate a subalgebra which is invariant under $\s$ and takes a
particularly simple form if expressed in terms of matrices $C=C^iM_i$
and $F=F^iM_i$, where $M_i$ represent a basis of the Lie algebra. 

For nearly all algebraic operations it is irrelevant that $F$ is a composite field. 
The transformation of $C$  (\ref{defs}) 
\begin{align}
\label{sc}
\s C &= C^2 + F
\intertext{can be read as definition of an elementary (purely imaginary) variable $F$\,. The transformation of 
$F$ follows from $\s^2 C = 0$ and turns out to be the adjoint transformation,}
\label{sf}
\s F &=-F\,C+C\,F \ .
\end{align}
Due to (\ref{sc}, \ref{sf}) $\s^2 F $ vanishes identically.

If one changes the notation and replaces $\s$ by $\dv =\dr x^m\partial_m$ and
$C$ by $A=\dr x^m A_{m}{}^{i}M_i$ then the same equations are the definition 
of the field strengths in Yang Mills theories and their Bianchi identities, 
\begin{equation}
F = \dv A - A^2\ ,\ \dv F + [A,F] = 0\ .
\end{equation}
The equations are valid whether or not the anticommuting variables
$C$ and the nilpotent operation $\s$ are composite.\footnote{This does not mean that there are no 
differences at all. For
example the product of $D+1$ matrix elements of the one form matrix $A$
vanish.}

The Chern polynomials $I_\alpha$ satisfy $\s I_\alpha=0$ because they are
invariant under adjoint transformations. All $I_\alpha$ are trivial i.e.
of the form $\s q_\alpha$. To show this explicitly we
define a one parameter deformation $F(t)$ of $F$\ ,
\begin{equation}
F(t)=tF-(t^2-t)C^2=t\,\s C-t^2\,C^2\ ,\  F(0)=0 \ ,\  F(1)=F\ ,
\end{equation}
which allows to switch on $F$.

All invariants $I_\alpha$ can be written as $\tr(F^{m(\alpha)})$ (if the
representation matrices $M_I$ are suitably chosen). We rewrite $\tr(F^m)$ in an artificially
more complicated form 
\begin{equation}
\tr (F^m) = \int_0^1\! \dr t\, \frac{\dv}{\dr t}\tr (F(t)^m)=m\int_0^1\! \dr t
\,\tr \bigl( (\s C+2t\,C^2)\,F(t)^{m-1}\bigr)\ .
\end{equation}
The integrand coincides with
\begin{equation}
\begin{aligned}
\s \tr\,(C F(t)^{m-1})&=\tr ((\s C)F(t)^{m-1}+t\,C\,[F(t)^{m-1},C])\\
&=\tr((\s C)\,F(t)^{m-1}- 2t\,C^2\,F(t)^{m-1})\ .
\end{aligned}
\end{equation}
The Chern form $I_\alpha$ is the $\s$ transformation of the Chern Simons
form $q_\alpha$, these forms generate a subalgebra,
\begin{equation}
\label{cherncoh}
\s q_\alpha=I_\alpha \ ,\  \s I_\alpha=0\ ,
\end{equation}
\begin{equation}
q_\alpha=m\int_0^1\!\dr t\,  \tr \Bigl( C\bigl (t\,F+(t^2-t)\,C^2\bigr )^{m-1}
\Bigr ) \ ,\  m=m(\alpha)\ .
\end{equation}     
Using the binomial formula and 
\begin{equation}
\int_0^1\! \dr t\,t^k\,(1-t)^l=\frac{k!\, l!}{(k+l+1)!}
\end{equation}
the $t$-integral can be evaluated. It gives the combinatorial coefficients of the Chern
Simons form.
\begin{equation}
\label{csimons}
q_\alpha(C,F)=\sum\limits_{l=0}^{m-1}\frac{(-)^lm!(m-1)!}
{(m+l)!(m-l-1)!} \tr_\text{sym}\bigl(C (C^2)^l  (F)^{m-l-1} \bigr)
\end{equation}
It involves the traces of completely symmetrized products of the $l$
factors $C^2$, the $m-l-1$ factors $F$ and the factor $C$. The part with
$l=m-1$ has form degree $0$ and ghostnumber $2m-1$ and agrees
with $\theta_\alpha$
\begin{equation}
\label{607}
q_\alpha(C,0)=\frac{(-)^{m-1}m!(m-1)!}{(2m-1)!}
\tr C^{2m-1}=\theta_\alpha(C)\ ,\  m = m(\alpha)\ . 
\end{equation}
For notational simplicity we write $\theta$ for $\theta_1,\theta_2\dots \theta_r$ and similarly 
$q$ for $q_1,q_2\dots q_r$ as well as $I$ for $I_1,I_2\dots I_r$. 

Each polynomial $P(\theta,I)$ defines naturally a polynomial $P(q,I)$ and a form
\begin{equation}
\omega(C,F)=P(q(C,F),I(F))
\end{equation}
which coincides with $P(\theta(C),I(F))$ in lowest form degree,
\begin{equation}
\omega_{\underline n}(C, F) = P(\theta(C),I(F))\ .
\end{equation}
On such polynomials $P(q,I)$ $\s$ acts simply as the operation 
\begin{align}
\s & = I_\alpha \frac{\partial}{\partial q_\alpha}\\
\label{sqi}
\s \omega(C,F) & = I_\alpha \frac{\partial}{\partial q_\alpha}
P(q,I)_{|_{q(C,F),I(F)}}
\end{align}
The basic lemma (\ref{basic}) implies that among the polynomials $P(q,I)$ the only
nontrivial solutions of $\s P=0$ are independent of $q$ and $I$, 
\begin{equation}
\label{610}
\s P(q,I)=0 \Leftrightarrow P(q,I)= P(0,0) + \s Q(q,I)\ .
\end{equation}
Though correct this result is misleading because we are not looking for
solutions of $\s P(q,I)=0$, which depend on variables $q$ and $I$, but solve
$\s \omega(q(C,F),I(F))=0$ in terms of functions of ghosts $C$
and field strength two forms $F$.
This equation has more solutions than (\ref{610}) because $D+1$ forms vanish. 

For $\s P(q(C,F),I(F))$ to vanish it is necessary and sufficient
that its lowest form degree is larger than $D$ and for $\omega$ not to vanish,
its lowest form degree has to be $D$ or smaller.
The lowest form degree of Chern Simons forms $q_\alpha(C,0)=\theta_\alpha(C)$ vanishes,
therefore the lowest form degree of polynomials $P(q,I)$ is counted by
\begin{equation} 
\label{nform}
N_\text{form}= \sum_\alpha 2 m(\alpha)\,I_\alpha \frac{\partial}{\partial I_\alpha}\ . 
\end{equation}
Because $I_\alpha= \s q_\alpha$ has form degree $2m(\alpha)$, 
$\s$ increases the lowest form degree. However, the increase has different values.

\subsection{Level Decomposition}

To deal with this situation, we introduce the notation $x_k$ for the variables
$q_\alpha$ and $I_\alpha$ with a fixed $m(\alpha)=k$
\begin{equation}
\{x_k\}=\{q_\alpha,I_\alpha : m(\alpha) = k \}
\end{equation}
and decompose the polynomial $P(q,I)$, which we consider as polynomial $P(x_1,x_2 \dots x_{\bar{k}})$, 
into pieces $P_k$ which
do not depend on the variables $x_1, \dots x_{k-1}$ and which vanish, if they do not depend 
on~$x_k$ \footnote{We assume without loss of generality $P(0,0\dots 0)$ to vanish. It is not affected 
by the shift
$C\mapsto C + A$ and does not contribute to a $D$-form which satisfies the descent equations.}, 
\begin{align}
P_1=P(x_1,x_2,\dots ,x_{\bar{k}}) & - P(0,x_2,\dots ,x_{\bar{k}})\ , \\
P_k = P(0,\dots,0,x_k,x_{k+1},\dots ,x_{\bar{k}}) & - 
    P(0,\dots,0,0,x_{k+1},\dots ,x_{\bar{k}})\ ,\\
P&=\sum_k P_k\ .
\end{align}
We decompose the polynomial $P_k$  \`a la Hodge (\ref{hodge}) with $\s$ and 
\begin{equation}
\ro_k=
\sum_{m(\alpha)=k}q_\alpha\frac{\partial}{\partial I_\alpha}\ ,\ 
\{\s,\ro_k\}= N_{x_k}\ ,
\end{equation}
as a sum of an $\s$ exact piece and an $\ro_k$ exact piece
\begin{equation} 
P_k= \s\rho + \ro_k \sigma
\end{equation}
No constant occurs in this Hodge decomposition of $P_k$, because $P_k$ vanishes 
by construction if as function of $x_k$ it is constant. Hodge decomposing $\sigma$ 
shows that it can be taken to be $\s$ exact without loss of generality.
Because the decomposition of $P_k$ is unique, $\ro_k \sigma$ is nontrivial and can 
be taken to represent the nontrivial  contribution to $\omega$ 
\begin{equation}
P_k=\ro_k \sigma\ , \ \sigma = \s \eta\ .
\end{equation}
It corresponds to a nontrivial solution $\omega(C,T) = P_k(q(C,F), I(F))$, if its lowest
form degree is not larger than $D$, and if the
lowest form degree of $\s P_k$ is larger than $D$. Because $\ro_k$ lowers the lowest form
degree by $2k$ and $\s$ increases the lowest form degree of $P_k$ by $2k$, this means, that
the lowest form degree $D^\prime$ of $\sigma$ has to satisfy
\begin{equation}
\label{dstrich}
D^\prime - 2 k \le D < D^\prime\ .
\end{equation}

If we want to obtain a solution $\omega$ with a definite ghostnumber
then we have to choose $\sigma$ as eigenfunction of the
ghost counting operator
\begin{equation}
N_C=\sum_\alpha \bigl (
2m(\alpha)I_\alpha\frac{\partial}{\partial I_\alpha} +
(2m(\alpha)-1)q_\alpha\frac{\partial}{\partial q_\alpha} \bigr )
\end{equation}
which counts the number of translation ghosts, spin
and isospin ghosts. The total ghostnumber of $\omega=\ro_k\sigma$ is $G$ if
the total ghostnumber of $\sigma$ is $G+1$, because $\ro_k$
lowers the total ghostnumber by~$1$.

We obtain the long sought solutions $\omega_D^g$ of the relative
cohomology (\ref{rel}) which for ghostnumber $g=0$ gives Lagrange densities of
invariant actions (\ref{relcoh}) and which for $g=1$ gives anomaly candidates
(\ref{anomaly}) if we substitute in $\omega$ the ghosts $C$
by ghosts plus connection one forms $C+A$ and if we pick the part with
$D$ differentials.
Therefore the total ghostnumber $G$ of $\sigma$ has to be chosen to be
$G=g+D+1$ to obtain a solution $\omega$ which contributes to
$\omega_D^g$. If the ghost variables $\hat{C}$ (\ref{chat}) are used to
express $\omega$ then $\omega_D^g$ is  simply obtained if all
translation ghosts $C^m$ are replaced by $\dr x^m$ and the part with the
volume element $\dr^D x$ is taken.
\begin{align}
\omega(C,F)=\bigl(\ro_k \sigma\bigr)_{|q(C,F),I(F)}& \,,\ 
N_C \,\sigma = (g + D + 1)\,\sigma\,,\ N_\text{form}\,\sigma = D^\prime \sigma
\nonumber
\\
\omega(C,F)&=f(\hat{C}^m,\hat{C}^i,F^i)\\
\nonumber
\omega_D^g&=f(\dr x^m,\hat{C}^i,{ \frac{1}{2}}\dr x^m \dr x^n F_{mn}{}^i)_{|\text{\tiny D form part}}
\end{align}
These  formulas end our general discussion of the \textsc{brst} cohomology of
gravitational Yang Mills theories. The general solution of the consistency
equations is a linear combination of the Lagrangian solutions and the
chiral solutions.

\subsection{Anomaly Candidates}

Let us conclude by spelling out the general formula for $g=0$ and $g=1$.
If $g=0$ then $\sigma$ can contain no factors $q_\alpha$ because the complete
ghostnumber $G\ge D^\prime$ is not smaller than the ghostnumber
$D^\prime$ of translation ghosts. $D^\prime$ has to be larger than $D$
(\ref{dstrich}) and not larger than $G=g+D+1=D+1$ which leaves
$D^\prime=D+1$ as only possibility. $D^\prime$ is even (\ref{nform}),
therefore chiral contributions to Lagrange densities occur only in odd
dimensions.

If, for example $D=3$, then $\sigma$ is an invariant 4 form.

For $k=1$ such a form is given by $\sigma=F_iF_j a^{ij}$ with $a^{ij}=a^{ji}
\in \Real$ if the isospin group contains abelian factors with the
corresponding abelian field strength $F_i$ and $i$ and $j$ enumerate
the abelian factors. The form $\omega=\ro_1\sigma=2q_iF_j a^{ij}$ yields the gauge
invariant abelian Chern Simons action in 3 dimensions which is remarkable
because it cannot be constructed from tensor variables alone and because
it does not contain the metric.

To construct $\omega_3^0$ one has to express
$q(C)=C$ by $C=\hat{C}+C^mA_m$. Then one has to replace all translation
ghosts by differentials $\dr x^m$ and to pick the volume form. One obtains
\begin{equation}
\omega_3^0{}_{\text{abelian}}=\dr x^mA_{m\,i}\dr x^k\dr x^lF_{kl\,j}
a^{ij}=\varepsilon^{klm}A_{m\,i}F_{kl\,j}a^{ij}\dr^3 x\ .
\end{equation}

For $k=2$ the form $\sigma=\tr F^2$ of each nonabelian factor contributes to
the nonabelian Chern Simons form. One has $I_1=\tr F^2=\s q_1$ and $\omega= r_2 I_1$ 
is given by the Chern Simons form $q_1$(\ref{csimons})
\begin{equation}
\omega=\tr(C\,F-\frac{1}{3}C^3)\ .
\end{equation}
The corresponding Lagrange density is
\begin{equation}
\omega_3^0{}_\text{nonabelian}=\tr(A\,F-\frac{1}{3}A^3)
=\frac{1}{2}(A_{m}{}^i F_{rs}{}^i - \frac{1}{3}
A_m{}^i A_r{}^j A_s{}^k f_{jk}{}^i )\,\varepsilon^{mrs}\,\dr^3 x
\end{equation}

The integrands of chiral anomalies $\omega_D^1$
have ghostnumber $g=1$. This fixes the total ghostnumber of $\sigma$ to be $G=D+2$ and because $G$ is not less
than $D^\prime>D$ we have to consider the cases $D^\prime=D+1$ and
$D^\prime=D+2$.

The first case can occur in odd dimensions only,
because $D^\prime$ is even, and only if the level $k$, the smallest value of $m(\alpha)$
of the variables occurring in $\sigma$, is 1 because the missing total ghostnumber
$D+2-D^\prime$, which is not carried by $I_\alpha(F)$, has to be
contributed by a Chern Simons polynomial $q_\alpha$ with
$2m(\alpha)-1=1$, i.e. with $m(\alpha)=1$. Moreover $\sigma=\s \eta$
has the form
\begin{equation}
\sigma=\sum_{ij\ \text{abelian}}a^{ij}(I_\alpha)\,q_i\,I_j \ ,\  a^{ij}=-a^{ji}\ ,
\end{equation}
where the sum runs over the abelian factors and the form degrees
contained in the antisymmetric $a^{ij}$ and in the abelian $I_j=F_j$
have to add up to $D+1$. In particular this anomaly can occur only if
the gauge group contains at least two abelian factors because $a^{ij}$
is antisymmetric. In $D=3$ dimensions $a^{ij}$ is linear in abelian
field strengths and one has
\begin{equation}
\sigma=\sum_{ijk\, \text{abelian}}a^{ijk}\,q_i\,I_j\,I_k \ ,\  a^{ijk}=a^{ikj} \ ,\ 
\osum_{ijk}a^{ijk}=0\ .
\end{equation}
This leads to
\begin{equation}
\omega=\ro_1\sigma=\sum_{ijk\, \text{abelian}}b^{ijk}\,q_i\,q_j\,I_k
=\sum_{ijk\,\text{abelian}}b^{ijk}C_iC_jF_k  \ ,\  b^{ijk}=-a^{ijk}+a^{jik}
\end{equation}
and the anomaly candidate is
\begin{equation}
\omega_3^1=2\sum_{ijk\,\text{abelian}}b^{ijk}\,\hat{C}_i\,A_j\,F_k=
\sum_{ijk\,\text{abelian}}b^{ijk}\,\hat{C}_i\,A_{m\,j}\,F_{rs\,k}\,\varepsilon^{mrs}\,\dr^3 x\ .
\end{equation}

If one considers $g=1$ and $D=4$ then $D^\prime=6$ because it is bounded
by $G=D+1+g=D+2$, larger than $D$ and even. This leaves
$D^\prime=G$ as only possibility, so the total ghostnumber is
carried by the translation ghosts contained in $\sigma=\sigma(I_\alpha)$ which is
a cubic polynomial in the field strength two forms $F$. Abelian two
forms can occur in the combination
\begin{equation}
\sigma=\sum_{ijk\,\text{abelian}}d^{ijk}\, F_i\, F_j\, F_k
\end{equation}
with completely symmetric coefficients $d^{ijk}$. These
polynomials are $\s$ exact. They lead to the abelian anomaly
\begin{equation}
\begin{aligned}
\omega_4^1{}_\text{abelian}&=\frac{3}{4}\sum_{ijk\,\text{abelian}}
d^{ijk}\, \hat{C}_i\, F_{mn\ j}\, F_{rs\ k}\, \varepsilon^{mnrs}\, \dr^4 x\\
&= 3\sum_{ijk\,\text{abelian}}d^{ijk}\, \hat{C}_i\, \dr A_j\, \dr A_k
\end{aligned}
\end{equation}
Abelian two forms $F_i$ can also occur in $\sigma$  multiplied with
$\tr (F_k)^2$ where $i$ enumerates abelian factors and $k$ nonabelian
ones. The mixed anomaly which corresponds to
\begin{equation}
\sigma=\sum_{ik}c^{ik}F_i \tr_k(F^2)
\end{equation}
is similar in form to the abelian anomaly
\begin{equation}
\omega_4^1{}_\text{mixed}=-\frac{1}{4}\sum_{ik}c^{ik}\,\hat{C}_i\,
(\sum_I F_{mn}{}^I F_{rs}{}^I)_{k}\, \varepsilon^{mnrs}\,\dr^4 x \ .
\end{equation}
The sum, however extends now over abelian factors enumerated by $i$ and
nonabelian factors enumerated by $k$. Moreover we assumed that the
basis, enumerated by $I$, of the simple Lie algebras is chosen such that
$\tr M_IM_J=-\delta_{IJ}$ holds for all $k$. Phrased in terms of $\dr A$ the
mixed anomaly differs from the abelian one because the nonabelian
field strength contains also $A^2$ terms \footnote{
The trace over an even power of one form matrices $A$ vanishes. 
}. 
\begin{equation}
\omega_4^1{}_\text{{mixed}}=\sum_{ik}c^{ik}\hat{C}_i
\tr_k \dv \Bigl ( A\dv A+\frac{2}{3}A^3 \Bigr )
\end{equation}

The last possibility to construct a polynomial $\sigma$ with form
degree $D^\prime=6$ is given by the Chern form $\tr(F)^3$ itself.
Such a Chern polynomial with $m=3$ exists for classical algebras only
for the algebras $SU(n)$ for $n\ge 3$ (\ref{mlist})
\footnote{ The Lie algebra $SO(6)$ is isomorphic to $SU(4)$.}. In
particular the Lorentz symmetry in $D=4$ dimensions is not anomalous.
The form $\omega$ which corresponds to the Chern form is the Chern
Simons form
\begin{equation}
\omega(C,F)=\tr \Bigl ( C\,F^2-\frac{1}{2}C^3\,F +
\frac{1}{10}C^5\Bigr )\ .
\end{equation}

The nonabelian anomaly follows after the substitution $C\rightarrow C+A$
and after taking the volume form
\begin{equation}
\begin{aligned}
\omega_4^1{}_\text{nonabelian}&=\tr\bigl(\hat{C}F^2-\frac{1}{2}(\hat{C}\,A^2F+
A \hat{C}\,A\,F+A^2\hat{C}\,F)+\frac{1}{2}\hat{C}\,A^4\bigr)\\
&=\tr\Bigl ( \hat{C}\dv \bigl ( A\dv A +\frac{1}{2}A^3\bigr )\Bigr )\ .
\end{aligned}
\end{equation}

\section{Inclusion of Antifields}\label{sec7}

\subsection{BRST-Antifield Formalism}

The \textsc{brst}-antifield formalism (or field-antifield formalism, or 
\textsc{bv} formalism) originated in the context of the renormalization of Yang-Mills theories 
using external sources for the \textsc{brst} transformations of the fields and ghost fields \cite{Zinn-Justin:1974mc}. Later it 
was realized that on the field antifield algebra one could extend the \textsc{brst} methods to gauge theories
with open algebras,  i.e. with commutator algebras which close only 
on-shell \cite{Kallosh:1978de,deWit:1978cd,Batalin:1981jr}. 
With fields and antifields one can treat the equations of motions, Noether identities and further reducibility 
identities as objects which occur in a cohomological problem \cite{Fisch:1990rp,Henneaux:1989jq}. 
Here, we restrict ourselves to discuss some selected features of this cohomology. For the general structure of the 
formalism we refer to the literature \cite{henteit,Gomis:1995he}. 

The formalism comprises ``fields'' and ``antifields''. The set of fields
contains the fields of the classical theory which we denote by $\varphi^i$ and
ghost fields denoted by $\hat C^\alpha$ which correspond to the gauge symmetries
($\alpha$ enumerates a generating set of gauge symmetries \cite{Henneaux:1989jq,henteit}). 
In addition the set of fields may contain
further fields, such as ``ghosts of ghosts'' when the gauge transformations are reducible,
or antighost fields $\bar C$ (which must not be confused with the antifields of the ghosts) used 
for gauge fixing but this is not relevant to the matters to be discussed later on. To distinguish 
fields from antifields we mark the latter by a superscript $\star$ (which must not be confused 
with the symbol $*$ used for complex conjugation). There is one antifield $\varphi^\star$ for 
each equation of motion of the classical theory and one antifield $C^\star$ for each nontrivial
 identity of these equations of motion. In a Lagrangian field theory with Lagrangian 
${\cal L}(x,\{\varphi\})$ there is one equation of motion for each field $\varphi^i$ given 
by $\frac{\hat{\partial}{\cal L}}{\hat{\partial} \varphi^i}=0$ which sets to zero the Euler 
derivative (\ref{eulder}) of ${\cal L}$ with respect to $\varphi^i$. Furthermore, the gauge 
symmetries of a Lagrangian correspond one-to-one to the nontrivial (Noether) identities 
relating the equations of motion (by Noethers second theorem). Therefore, in a Lagrangian 
field theory the fields $\varphi^i$ and $\hat C^\alpha$ correspond one-to-one to antifields
 $\varphi^\star_i$ and $C^\star_\alpha$.

We denote the infinitesimal gauge transformations of the fields $\varphi^i$ by 
$\delta_\varepsilon\varphi^i=R^i_\alpha\varepsilon^\alpha$ where $\varepsilon^\alpha$ 
are the parameters of gauge transformations and $R^i_\alpha$ are (in general field dependent)
 differential operators acting on $\varepsilon^\alpha$ according to
\begin{equation}
\delta_\varepsilon\varphi^i=R^i_\alpha\,\varepsilon^\alpha,\quad
R^i_\alpha\,\varepsilon^\alpha = \sum_k (\partial_{m_1}\dots\partial_{m_k}\,\varepsilon^\alpha )\, 
r^{i m_1\dots m_k}_\alpha(x,\{\varphi\}).
\end{equation}
By assumption these gauge transformations generate symmetries of a Lagrangian ${\cal L}$, 
i.e. the Lagrangian ${\cal L}$ transforms under $\delta_\varepsilon$ into a total derivative,
\begin{equation}
\delta_\varepsilon\,{\cal L}=\partial_m K^m.
\label{inv}
\end{equation}
The Euler derivative of (\ref{inv}) with respect to $\varepsilon^\alpha$ gives the 
Noether identity for the $\alpha$th gauge symmetry. This Noether identity reads
\begin{equation}
R^{i\dagger}_\alpha\ \frac{\hat{\partial}{\cal L}}{\hat\partial\varphi^i} = 0
\label{Noetherid}
\end{equation}
where $R^{i\dagger}_\alpha$ is an operation transposed to $R^i_\alpha$,
\begin{equation}
R^{i\dagger}_\alpha\,\chi = \sum_k (-)^k \partial_{m_1}\dots
\partial_{m_k}(r^{i m_1\dots m_k}_\alpha(x,\{\varphi\})\,\chi).
\end{equation}
By assumption the set of gauge transformations is closed under commutation up to 
trivial gauge transformations, i.e. the commutator of any two gauge transformations 
is again a gauge transformation at least on-shell,
\begin{equation}
[\delta_\varepsilon,\delta_{\varepsilon^\prime}]\approx \delta_f\, ,\ 
f^\alpha=f^\alpha(x,\{\varepsilon,\varepsilon^\prime,\varphi\})
\label{comalg}
\end{equation}
where $f^\alpha(x,\{\varepsilon,\varepsilon^\prime,\varphi\})$ are local 
structure functions of the parameters $\varepsilon^\alpha$, $\varepsilon^{\prime\alpha}$, 
the fields $\varphi^i$ and derivatives thereof, and $\approx$ denotes ``weak equality'' 
defined according to
\begin{equation}
F(x,\{\varphi\})\approx 0\quad \Leftrightarrow\quad F(x,\{\varphi\})=
\sum_k g^{i m_1\dots m_k}(x,\{\varphi\})\partial_{m_1}\dots\partial_{m_k}
\frac{\hat{\partial}{\cal L}}{\hat{\partial} \varphi^i}\, .
\end{equation}
Hence, two functions are weakly equal iff they differ only by terms which are at least linear in the 
Euler derivative $\frac{\hat{\partial}{\cal L}}{\hat{\partial} \varphi^i}$ or its derivatives. 
As the Euler derivative vanishes on-shell (i.e., for all solutions of the equations of motion), 
weak equality is ``equality on-shell''.

The \textsc{brst} transformations of the $\varphi,\hat C,\varphi^\star,C^\star$ take the form\footnote{We 
use conventions such that $\s: \chi \mapsto (S,\,\chi)$ where 
$S=\int \dr^Dx({\cal L}+(R^i_\alpha \hat C^\alpha)\varphi^\star_i+\dots)$ solves the 
master equation $(S,S)=0$ with the standard antibracket \cite{Batalin:1981jr,henteit,Gomis:1995he}.}
\begin{equation}
\begin{aligned}
\s\varphi^i&=-R^i_\alpha(x,\{\varphi\},\partial) \hat C^\alpha+\dots\ ,\\
\s \hat C^\alpha&=
\frac 12\,
f^\alpha(x,\{\varepsilon,\varepsilon^\prime,\varphi\})\Big|_{\varepsilon^\alpha
=\hat C^\alpha,\varepsilon^{\prime\alpha}=(-)^{|\hat C^\alpha|}\hat C^\alpha}+\dots\ ,\\
\s\varphi^\star_i&=(-)^{|\varphi^i|}\,\frac{\hat{\partial}{{\cal L}(x,\{\varphi\})}}{\hat{\partial} \varphi^i}+\dots\ ,\\ 
\s C^\star_\alpha
&=(-)^{|\hat C^\alpha|}R^{i\dagger}_\alpha(x,\{\varphi\},\partial)\varphi^\star_i+\dots
\label{sgeneral}
\end{aligned}
\end{equation}
with ellipsis indicating antifield dependent contributions. The grading $|\hat C^\alpha|$ 
of the ghost $\hat C^\alpha$ is opposite to the grading of the corresponding parameter
 $\varepsilon^\alpha$. Furthermore
the grading of an antifield is always opposite to the grading of the corresponding field:
\begin{equation}
|\hat C^\alpha|=|\varepsilon^\alpha|+1\,\modulo 2,\quad
|\varphi^\star_i|=|\varphi^i|+1\,\modulo 2,\quad
|C^\star_\alpha|=|\hat C^\alpha|+1\,\modulo 2.
\end{equation}
The ghostnumbers of a field and the corresponding antifield add up to minus one:
\begin{equation}
\ghost(\varphi^i)=0,\ \ghost(\varphi^\star_i)=-1,\ \ghost(\hat C^\alpha)=1,\ 
\ghost(C^\star_\alpha)=-2.
\end{equation}

A very useful concept for discussing various aspects of the \textsc{brst}-antifield formalism 
is the decomposition of the ghostnumber into a pure ghostnumber ($\pgh$) and an 
antifield number ($\af$) according to
\begin{equation}
\begin{aligned}
&\ghost=\pgh-\af,\\
&\af(\varphi^i)=0,\ \af(\hat C^\alpha)=0,\ \af(\varphi^\star_i)=1,\ \af(C^\star_\alpha)=2,\\
&\pgh(\varphi^i)=0,\ \pgh(\hat C^\alpha)=1,\ \pgh(\varphi^\star_i)=0,\ \pgh(C^\star_\alpha)=0.\\
\end{aligned}
\end{equation}
$\s$ decomposes into parts of various antifield numbers $\geq -1$,
\begin{equation}
\s=\delta+\gamma+\dots,\quad \af(\delta)=-1,\quad \af(\gamma)=0
\label{sdecomp}
\end{equation}
where the ellipsis indicates parts with antifield numbers $\geq 1$. The parts $\delta$ and 
$\gamma$ are the two crucial ingredients of $\s$. In particular they determine the structure 
of the \textsc{brst} cohomology. The part $\delta$ is often called the Koszul-Tate differential. 
It is the part of $\s$ which lowers the antifield number and therefore vanishes
on the fields 
\begin{equation}
\delta\varphi^i=0,\ \delta \hat C^\alpha=0\ ,\ 
\delta\varphi^\star_i=(-)^{|\varphi^i|}\frac{\hat{\partial}{\cal L}}{\hat{\partial} \varphi^i}\, ,\  
\delta C^\star_\alpha=(-)^{|\hat C^\alpha|}R^{i\dagger}_\alpha\varphi^\star_i\, .
\end{equation}
In particular the Koszul-Tate differential is the part of $\s$ which implements the 
equations of motion and the Noether identities in cohomology by the $\delta$-transformations 
of the $\varphi^\star$ and $C^\star$. It is nilpotent by itself. For instance, 
owing to (\ref{Noetherid}) one gets
\begin{equation}
\delta^2 C^\star_\alpha=(-)^{|\hat C^\alpha|+|\varepsilon^\alpha|+
|\varphi^i|}R^{i\dagger}_\alpha\delta\varphi^\star_i
=-R^{i\dagger}_\alpha\frac{\hat{\partial}{\cal L}}{\hat{\partial} \varphi^i}=0.
\end{equation}

In Yang Mills theories, Einstein gravity and gravitational Yang Mills theories with 
the standard gauge transformations the decomposition (\ref{sdecomp}) of $\s$ 
terminates with $\gamma$, i.e. in these cases one simply has $\s=\delta+\gamma$, 
because the commutator algebra of the gauge transformations closes even off-shell 
(i.e. (\ref{comalg}) holds with $=$ in place of $\approx$) and because the structure 
functions $f^\alpha$ do not depend on fields $\varphi$. Hence, on the fields 
$\varphi, \hat C$ one has in these cases $\s\varphi=\gamma\varphi$ and 
$\s \hat C=\gamma \hat C$. With respect to $\gamma$ the antifields 
$\varphi^\star, C^\star$ are tensors or, in the gravitational case, 
tensor densities with weight one. For instance, in pure Yang Mills 
theories in flat space-time, with semisimple isospin Lie algebra and 
Lagrangian ${\cal L}=-\frac 14 d_{ij}F_{mn}{}^iF^{mn\, j}$ 
(where $F_{mn}{}^i=\partial_m A_n{}^i-\partial_n A_m{}^i-A_m{}^j A_n{}^k f_{jk}{}^i$ and 
$d_{ij}$ is the Cartan-Killing metric of the isospin Lie algebra), the $\varphi$ are the 
components $A_m{}^i$ of the gauge fields and the $\hat C$ are the Yang Mills 
ghosts $\hat C^i$.\footnote{Depending on the context, $i$ numbers all fields when we refer 
to a general gauge theory, 
whereas in Yang Mills theories it enumerates a basis of the Lie algebra.} Denoting the corresponding
 antifields $A^{\star m}{}_i$ and
$C^\star{}_i$ one obtains
\begin{equation}
\begin{aligned}
\delta A_m{}^i&=0\, ,& \gamma A_m{}^i&=\partial_m\hat C^i+\hat C^jA_m{}^kf_{jk}{}^i\, ,\\
\delta \hat C^i&=0\, ,&\gamma\hat C^i&=\frac 12\,\hat C^j\hat C^k f_{jk}{}^i\, ,\\
\delta A^{\star m}{}_i&=d_{ij}D_n F^{nm j}\, ,&\gamma A^{\star m}{}_i
&=-\hat C^j f_{ji}{}^k A^{\star m}{}_k\, ,\\
\delta C^\star{}_i&=D_m A^{\star m}{}_i\, ,&\gamma C^\star{}_i&=-\hat C^j f_{ji}{}^kC^\star{}_k
\end{aligned}
\label{sYM}
\end{equation}
where $D_m=\partial_m+A_m{}^i\delta_i$ denotes the covariant derivative. 
Notice that one has $\gamma A^{\star m}{}_i=-\hat C^j\delta_j A^{\star m}{}_i$ 
and $\gamma C^\star{}_i=-\hat C^j\delta_j C^\star{}_i$, i.e. the antifields are 
indeed treated as tensors by $\gamma$.

\subsection{The Antifield Dependent BRST Cohomology}

We briefly indicate how one can compute the \textsc{brst} cohomology in gravitational Yang Mills 
theories in presence of antifields along the lines of sections \ref{sec3} to \ref{sec6} and discuss two 
different strategies to adapt the analysis in order to deal with the antifields. The first 
strategy eliminates the antifields by a suitable change
 of variables \cite{Brandt:1996mh,Brandt:2001tg}. The second strategy 
keeps the antifields throughout the analysis \cite{Barnich:1994ve,Barnich:1995mt,Barnich:2000zw,Barnich:1995ap}.
 Even though both strategies 
appear to be rather different, they are closely related 
and, of course, they lead to the same results.

\subsubsection{First Strategy}

Equations (\ref{sgeneral}) indicate that 
the antifields and all their derivatives might be removed from the cohomological 
analysis for $\st$ (and analogously for~$\s$) by the arguments given in section \ref{sec4.3} 
because each antifield variable ($\varphi^\star,C^\star,\partial \varphi^\star,\dots$) 
might be taken as a variable $u$ or be replaced by a variable $\st u$, respectively. 
Indeed, for a standard Lagrangian ${\cal L}$, $\st\varphi^\star_i$ contains a piece 
linear in the fields $\varphi$ and their derivatives given by the linearized
 Euler derivative $\frac{\hat{\partial}{\cal L}}{\hat{\partial} \varphi^i}$. 
Analogously, $\st C^\star_\alpha$ contains a piece that is linear in the  
$\varphi^\star$ and their derivatives given by the linearization of 
$(-)^{|\hat C^\alpha|}R^{i\dagger}_\alpha\varphi^\star_i$. 

As a consequence, one can eliminate the antifields and all their derivatives
 from the cohomological analysis for $\st$ provided one can construct a new set 
of variables replacing all the field and antifield variables and consisting of 
variables $u$ and $\st u$ and complementary variables $\tilde w=(\tilde C,\tilde T)$ 
such that $\st \tilde w=F(\tilde w)$ with a set of 
``generalized tensors'' $\tilde T$.\footnote{It turns out that in 
standard Yang Mills theories, Einstein gravity and gravitational 
Yang Mills theories one can use the same variables $\tilde C$ in 
the antifield dependent case as in the antifield independent case. 
Therefore we do not change the notation concerning these variables.} 
It can be shown quite generally that such a set of variables 
exists \cite{Brandt:2001tg}. In particular it exists for standard 
Yang Mills theories, Einstein gravity and gravitational Yang Mills theories. 
However, two important consequences of this strategy have to be pointed out. 

Firstly, the set of generalized tensors $\tilde T$ contains \emph{fewer} 
variables than the corresponding set of tensors $T$ in the antifield 
independent cohomology because along with the elimination of the 
antifields one also eliminates tensors $T$ that correspond to the 
Euler derivatives $\frac{\hat{\partial}{\cal L}}{\hat{\partial} \varphi^i}$ 
and their derivatives. For instance, (\ref{sYM}) shows that in pure 
Yang Mills theories the set of generalized tensors $\tilde T$ does not 
contain elements corresponding to the tensors $D_n F^{nm i}$ as these 
are eliminated along with the antifields $A^{\star m}{}_i$ and an 
analogous statement applies to all covariant derivatives of the $D_n F^{nm i}$. 
Hence, there are (combinations of) tensors $T$ which have no counterpart 
in the set of generalized tensors~$\tilde T$ because they are set to zero
 by the equations of motion and their derivatives (and, in fact, there 
are infinitely many such tensors). 

Secondly, each generalized tensor $\tilde T$ has an antifield independent 
part $\tilde T_0=\tilde T|_{\varphi^\star=0=C^\star}$ and the set of 
$\tilde T_0$ may be taken as a subset of the set of tensors $T$. 
However, some of the $\tilde T$ also contain antifield dependent 
contributions \cite{Brandt:2001tg}. As a consequence, even though 
the cohomology can be computed completely in terms of the variables 
$\tilde C,\tilde T$, some of the nontrivial representatives of the 
cohomology may depend on antifields through the dependence of 
variables $\tilde T$ on antifields. This is analogous to the way 
in which the undifferentiated gauge fields $A_m{}^i$ enter 
nontrivial representatives of the antifield independent cohomology. 
Namely, the $A_m{}^i$ are variables $u$ but are also used within 
the construction of the variables $\tilde C$ and $T$ through field 
strengths, covariant derivatives of tensors and $\tilde C=C+A$. 
Hence, even though the undifferentiated gauge fields $A_m{}^i$ 
are eliminated from the cohomological analysis as variables $u$, 
they nevertheless enter representatives of the cohomology through 
their occurrence within the variables $\tilde C$ and~$T$.

Let us now assume that the antifields have been eliminated.
In that case one is left with the 
computation of the cohomology in a space of functions $f(\tilde C,\tilde T)$. 
On the variables $\tilde C, \tilde T$ one has 
$\st \tilde T = -\tilde{C}^N\Delta_N \tilde T$ and 
$\st \tilde{C}^N= -\frac{1}{2}\tilde{C}^K\tilde{C}^L\tilde F_{LK}{}^N$, 
analogously to the antifield independent case. As a consequence, in order 
to analyse the cohomology of $\st$ in the space of functions $f(\tilde C,\tilde T)$ 
one can proceed exactly as in the antifield independent case until one arrives 
at the counterpart of equation (\ref{cplproblem}) which we write as
\begin{equation}
\dvt \omega(\tilde T,\dr x) = 0\ ,\ \omega\, \modulo \dvt \eta(\tilde T,\dr x)
\label{cplproblem2}
\end{equation}
where $\omega(\tilde T,\dr x)$ and $\eta(\tilde T,\dr x)$ are spin and isospin invariant forms 
which depend on the generalized tensors $\tilde T$. 
Here we  use the notation $\dvt$ to stress 
that this operation is the exterior derivative on forms
 $\omega(\tilde T,\dr x)$ of the generalized tensors~$\tilde T$. 
$\dvt$ acts on forms $\omega(\tilde T,\dr x)$ 
substantially different from $\dv$ acting on forms $\omega(T,\dr x)$ 
because the ideal of tensors $T$, which contain an Euler derivative or derivatives of Euler derivatives, 
have been eliminated together with the antifields from the tensor algebra.

To make this point clear, let us compare $\dv \tilde T_0$ to the 
antifield independent part $(\dvt \tilde T)_0$ of $\dvt \tilde T$. $\dv \tilde T_0$ 
in general contains tensors $T$ which are eliminated along with the antifields 
whereas $(\dvt \tilde T)_0$ does not contain any of these tensors $T$. 
Now recall that the tensors $T$ which are eliminated along with the antifields 
are just those that are set to zero by the equations of motion (or by the 
linearized equations of motion when one uses linearized tensors) and derivatives thereof. 
Hence, in general $\dv \tilde T_0$ is only weakly equal to $(\dvt \tilde T)_0$! 
Therefore, the problem posed by equation (\ref{cplproblem2}) is equivalent to 
the {\em weak cohomology} of $\dv$ (i.e., the cohomology of $\dv$ on-shell) 
on invariant tensor forms $\omega(\tilde T_0,\dr x)$, with the cocycle condition
\begin{equation}
\dv \omega(\tilde T_0,\dr x) \approx 0
\label{cplproblem3}
\end{equation}
and coboundaries fulfilling $\omega(\tilde T_0,\dr x)\approx \dr \eta(\tilde T_0,\dr x)$. 
The solution of this cohomological problem is the analog of the covariant Poincar\'e lemma 
in the antifield independent cohomology and may therefore be termed 
``weak covariant Poincar\'e lemma'' \cite{Barnich:1995ap}
and will be briefly discussed below. By means of the weak covariant Poincar\'e lemma 
one can finish the computation of the \textsc{brst} cohomology in presence of the antifields 
along the lines applied in the antifield independent case. We shall briefly sketch 
the results below.

\subsubsection{Second Strategy}

The second strategy treats
the antifields as additional tensors $T^\star$ (``antitensors''). 
This is possible because, as we have pointed out above, the antifields $\varphi^\star, C^\star$ 
transform as tensors or (in the gravitational case) as tensor densities 
under the part $\gamma$ of $\s$. Hence, the undifferentiated antifields
 $\varphi^\star, C^\star$ or (in the gravitational case) $\varphi^\star/e, C^\star/e$ 
(with $e=\det \A{e}{m}{a}$) can be viewed as tensors. Standard covariant derivatives 
of these antifield variables transform again as tensors under $\gamma$ and are used 
as antitensors $T^\star$ that substitute for the derivatives of the 
$\varphi^\star, C^\star$. The difference of the $T^\star$ and the $T$ is 
that the Koszul-Tate part $\delta$ of $\s$ acts nontrivially on the $T^\star$. 
However $\delta$ maps each antitensor $T^\star$ to (a combination of) tensors $T$ or antitensors $T^\star$.
Therefore the space of functions $f(C,T,T^\star)$ is invariant under $s$, 
just as the space of functions $f(C,T)$. 

This allows one to extent the methods used in the antifield independent case straightforwardly 
 to the antifield dependent case, with $T,T^\star$ in place of $T$ and with $\s=\delta+\gamma$,
 until one arrives at equation (\ref{fb1}). In the latter equation one now gets $\delta+\s_c$ 
in place of $\s_c$. In place of equation (\ref{fb2}) one therefore gets $(\delta+\s_c)\omega=0$ 
for the part of lowest order in the translation ghosts. As $\delta$ acts nontrivially only on 
the antitensors $T^\star$ and does not affect the dependence on the $C$ at all, and as $\s_c$ 
acts nontrivially only on the $C$ and does not affect the dependence on tensors or antitensors 
at all, one gets (using K\"unneth's Theorem) $\omega=\sum_l \Theta_l(C) f^l(c,T,T^\star)$ with 
$\s_c\Theta_l(C)=0$ and $\delta f^l(c,T,T^\star)=0$. $\delta f^l(c,T,T^\star)=0$ implies that 
one may take $f^l=f^l(c,T_0)$ where the $T_0$ form a subset of the tensors $T$ which corresponds
 to the set of generalized tensors $\tilde T$ of the first strategy. The reason is the following: 
just as one can eliminate the antifields from 
the cohomology of $\st$ along with a subset of weakly vanishing tensors $T$, one can also 
eliminate the antifields from the cohomology of $\delta$ along with the same subset of weakly
 vanishing tensors $T$. The remaining sets of generalized tensors $\tilde T$ and tensors $T_0$ 
correspond to each other, and one may actually take the set of $T_0$ identical to the 
set of $\tilde T_0$. The conditions $\s_c\Theta_l(C)=0$ are treated exactly as in the 
antifield independent case. As a consequence one gets $\omega_{\underline  n}=\Phi(\theta(C),c,T_0)$
in place of equation (\ref{fb4}).

The next change is in equation (\ref{co1}) where again $\delta+\s_c$ replaces $\s_c$. 
Consequently one gets a $\delta$-exact term $\delta(\dots)$ in the subsequent equations
 (\ref{fb6}), (\ref{fb7}) and (\ref{s1lin}). In particular, in place of equation 
(\ref{s1lin}) one gets 
$\s_{1, 1}f_{\underline n}(c,T_0)+\delta f_{\underline n+1}(c,T,T^\star)= 0$ 
(with $f_{\underline n}\modulo \s_{1, 1}\eta_{\underline n-1}+\delta \eta_{\underline n}$). 
Eventually equation (\ref{cplproblem}) is replaced by 
\begin{equation}
\begin{gathered}
\dv \omega_p(T_0,\dr x) +\delta \omega_{p+1}(T,T^\star,\dr x)= 0\,,\\
\omega_p(T_0,\dr x)\ \modulo \dr \eta_{p-1}(T,\dr x) +\delta \eta_{p}(T,T^\star,\dr x)\,.
\label{cplproblem4}
\end{gathered}
\end{equation}
Now, the problem posed by (\ref{cplproblem4}) is exactly the same as the problem associated 
with equation (\ref{cplproblem3}). Namely, in (\ref{cplproblem4}) $\omega_{p+1}(T,T^\star,\dr x)$
 has antifield number one because $\omega_p(T_0,\dr x)$ has antifield number zero 
(since $\omega_p(T_0,\dr x)$ does not depend on antifields at all). Hence, 
$\omega_{p+1}(T,T^\star,\dr x)$ is linear in those $T^\star$ that correspond to the 
$\varphi^\star_i$ and their derivatives. As the $\delta$-transformations of these 
antifields vanish weakly, one has $\delta \omega_{p+1}\approx 0$ and thus
 $\dv \omega_p(T_0,\dr x)\approx 0$ which reproduces precisely (\ref{cplproblem3}) 
because, as we mentioned above, the set of tensors $T_0$ corresponds to the set of 
tensors $\tilde T_0$. Analogous statements apply to the coboundary conditions.

\subsection{Characteristic Cohomology and Weak Covariant Poincar\'e Lemma}\label{charcoh}

The cohomological problem posed by (\ref{cplproblem3}) and (\ref{cplproblem4}) correlates 
the \textsc{brst} cohomology to the weak cohomology of $\dr$ on forms $\omega(\{\varphi\},x,\dr x)$ 
(without restricting these forms to tensor forms or invariant tensor forms). This cohomology
 has been termed characteristic cohomology (of the equations of motion) \cite{Bryant} and is
 interesting on its own because it generalizes the concept of conserved currents. To explain this,
 we write a $p$-form $\omega(\{\varphi\},x,\dr x)$ as
\begin{equation}
\omega_p=\frac{1}{p!(D-p)!}\,\dr x^{m_1}\dots\dr x^{m_p}\,\epsilon_{m_1\dots m_D}\,
j^{m_{p+1}\dots m_D}(\{\varphi\},x)
\label{char1}
\end{equation}
where the $\epsilon$-symbol is completely antisymmetric and $\epsilon_{0\dots (D-1)}=1$. 
The condition $\dv \omega_p\approx 0$ of the characteristic cohomology in form degree $p<D$ 
is equivalent to
\begin{equation}
\partial_{m_1}j^{m_{1}\dots m_{D-p}}(\{\varphi\},x)\approx 0\ .
\label{char2}
\end{equation}
For $p=D-1$ this gives $\partial_m j^m\approx 0$ which determines conserved currents. 
The representatives of the characteristic cohomology with $p<D$ are thus conserved 
differential $p$-forms of the fields which for $p=D-1$ provide the conserved currents.
Coboundaries of the characteristic cohomology are weakly $\dr$-exact forms 
$\omega_p\approx \dv\eta_{p-1}$ which are equivalent to 
$j^{m_{1}\dots m_{D-p}}(\{\varphi\},x)\approx \partial_{m_0}k^{m_{0}\dots m_{D-p}}(\{\varphi\},x)$ 
where $k^{m_{0}\dots m_{D-p}}=k^{[m_{0}\dots m_{D-p}]}$ is completely antisymmetric.

A remarkable feature of the characteristic cohomology is that the reducibility order $r$ of a 
theory gives a bound on the formdegrees $p$ below which the characteristic cohomology is 
trivial, provided the theory is a ``normal theory'' \cite{Barnich:1994db}. In this context 
one assigns reducibility order $r=-1$ to a theory which has no nontrivial gauge symmetries,
 $r=0$ to a gauge theory with irreducible gauge transformations (such as standard Yang Mills
 theories or Einstein gravity), $r=1$ to a gauge theory with gauge transformations which are
 reducible of first order etc.
It was proved 
for ``normal theories'' and $D>r+2$ that the characteristic 
cohomology is trivial in all form degrees smaller than $D-r-2$ \cite{Barnich:1994db}:
\begin{equation}
\begin{aligned}
0<p<D-r-2:&\quad
&\dv \omega_p&\approx 0\quad \Leftrightarrow\quad \omega_p\approx \dv\eta_{p-1}\, ,\\
p=0:&\quad
&\dv \omega_0&\approx 0\quad \Leftrightarrow\quad \omega_0\approx \text{constant}\ .\\
\end{aligned}
\label{char3}
\end{equation}
For standard Yang Mills theories, Einstein gravity and gravitational Yang Mills theories 
this result implies that the characteristic cohomology is trivial in formdegrees $p< D-2$ if $D>2$.
Furthermore for these theories the characteristic 
cohomology in formdegree $p=D-2$ is represented by $(D-2)$-forms corresponding one-to-one 
to ``free Abelian gauge symmetries'' which act nontrivially only on the corresponding Abelian
gauge fields and leave invariant all other fields \cite{Barnich:1994db}.
In formdegree $p=D-1$ the characteristic cohomology is represented by ``Noether forms'' $J$ 
involving the nontrivial Noether currents $j^m$,
\begin{equation}
J=\frac{1}{(D-1)!}\,\dr x^{m_1}\dots\dr x^{m_{D-1}}\,\epsilon_{m_1\dots m_D}\,j^{m_D}\,,\quad
\partial_m\, j^m\approx 0\ .
\label{char5}
\end{equation}

We shall now sketch a derivation of the result (\ref{char3}) for irreducible gauge theories ($r=0$).
 In these theories there are only antifields $\varphi^\star$ with antifield number 1 and antifields 
$C^\star$ with antifield number 2 (but no antifields with antifield numbers $k>2$). The derivation 
relates the characteristic cohomology to the cohomology of $\delta$ modulo $\dv$ on forms 
$\omega(\{\varphi,\varphi^\star,C^\star\},x,\dr x)$ of fields and antifields via descent equations 
for $\delta$ and $\dv$ and analyses these descent equations. In order to simplify the notation we 
shall assume that all fields $\varphi$ and antifields $C^\star$ are bosonic.

As discussed above, the cocycle condition $\dv\omega_p^0(\{\varphi\},x,\dr x)\approx 0$ can be written as
 $\dv\omega_p^0+\delta\, \omega_{p+1}^1=0$ where the subscript of a form $\omega_p^k$ denotes the formdegree
 and the superscript denotes the antifield number. Applying $\dv$ to $\dv\omega_p^0+\delta\, \omega_{p+1}^1=0$
 one obtains $\delta\, (\dv\omega_{p+1}^1)=0$. As $\dv\omega_{p+1}^1$ has antifield number 1 and the cohomology
 of $\delta$ is trivial in positive antifield numbers this implies $\dv\omega_{p+1}^1+\delta\, \omega_{p+2}^2=0$
 for some form $\omega_{p+2}^2$ with antifield number 2. Repeating this reasoning one obtains descent equations
 for $\delta$ and $\dv$ related to $\dv\omega_p^0\approx 0$,
\begin{equation}
\begin{aligned}
0=&\dv \omega_p^0 +\delta\, \omega_{p+1}^1\ ,\\
0=&\dv \omega_{p+1}^1 +\delta\, \omega_{p+2}^2\ ,\\
\vdots& \\
0=&\dv \omega_{D-1}^{D-p-1} +\delta\, \omega_{D}^{D-p}\ .
\label{wcpl1}
\end{aligned}
\end{equation}
We now analyse the last equation in (\ref{wcpl1}), i.e. the equation at formdegree $D$, using
 $\omega_{D}^{D-p}=\dr^D x\, a^{D-p}$ where $a^{D-p}$ has antifield number $D-p$. This gives
\begin{equation}
\delta\,a^{D-p}=\partial_m a^{m,\, D-p-1}
\label{wcpl2}
\end{equation}
with $a^{m,\, D-p-1}$ arising from $\omega_{D-1}^{D-p-1}$.
We analyse (\ref{wcpl2}) by considering the linearized
Koszul-Tate differential 
$\delta_0$ acting as
\begin{equation}
\delta_0\,\varphi	^i=0\, ,\ 
\delta_0\,\varphi^\star_i=D_{ij}\,\varphi^j\, ,\ 
\delta_0\, C^\star_\alpha=U^{i\dagger}_\alpha\,\varphi^\star_i
\label{wcpl3}
\end{equation}
where $D_{ij}\,\varphi^j$ are the linearized Euler derivatives 
$\frac{\hat{\partial}{\cal L}}{\hat{\partial} \varphi^i}$ of the Lagrangian and 
$U^{i\dagger}_\alpha\,\varphi^\star_i$  
is the linearized $\delta$-transformation of $C^\star_\alpha$ (with 
$D_{ij}=\sum_{k}d_{ij}^{m_1\dots m_k}\partial_{m_1}\dots \partial_{m_k}$ etc.).

At lowest order in the fields and antifields, (\ref{wcpl2}) imposes
\begin{equation}
\delta_0\,\underline{a}=\partial_m b^{m}
\label{wcpl3a}
\end{equation}
where $\underline{a}$ is the part of 
$a^{D-p}$ with lowest degree of homogeneity in the fields and antifields, and $b^{m}$ 
is the corresponding part of $a^{m,\, D-p-1}$.

Taking Euler derivatives of (\ref{wcpl3a}) with respect to the $C^\star$, $\varphi^\star$ 
and $\varphi$ we obtain (with $D_{ji}^\dagger=\sum_{r}(-)^kd_{ji}^{m_1\dots m_k}\partial_{m_1}\dots 
\partial_{m_k}$ etc.)
\begin{equation}
\delta_0\, \frac{\hat{\partial}\underline{a}}{\hat{\partial} C^\star_\alpha}=0,\quad
\delta_0\, \frac{\hat{\partial}\underline{a}}{\hat{\partial} \varphi^\star_i}=U^{i}_\alpha\,
\frac{\hat{\partial}\underline{a}}{\hat{\partial} C^\star_\alpha}\, ,\quad
\delta_0\, \frac{\hat{\partial}\underline{a}}{\hat{\partial} \varphi^i}=-D_{ji}^\dagger\,
\frac{\hat{\partial}\underline{a}}{\hat{\partial} \varphi^\star_j}\, .\label{wcpl6}
\end{equation}
Assume now that $p<D-2$. In this case $a^{D-p}$ and $\underline{a}$ have antifield number 
$D-p>2$. Hence, all Euler derivatives in equations (\ref{wcpl6}) have 
positive antifield numbers. Since the cohomology of $\delta_0$ vanishes for positive 
antifield numbers, the first equation (\ref{wcpl6}) implies
\begin{eqnarray}
\frac{\hat{\partial}\underline{a}}{\hat{\partial} C^\star_\alpha}=\delta_0 f^\alpha
\label{wcpl7}
\end{eqnarray}
for some $f^\alpha$ with antifield number $D-p-1$. Using (\ref{wcpl7}) in the second equation (\ref{wcpl6}) one gets 
\begin{eqnarray}
\delta_0\, \Big(\frac{\hat{\partial}\underline{a}}{\hat{\partial} \varphi^\star_i}-U^{i}_\alpha f^\alpha\Big)=0
\label{wcpl8}
\end{eqnarray}
where the expression in parentheses has antifield number $D-p-1>1$. We conclude that 
this expression is the $\delta_0$-transformation of some $f^i$ with antifield number $D-p$ which yields
\begin{eqnarray}
\frac{\hat{\partial}\underline{a}}{\hat{\partial} \varphi^\star_i}=U^{i}_\alpha f^\alpha+\delta_0 f^i.
\label{wcpl9}
\end{eqnarray}
Using (\ref{wcpl9}) in the third equation (\ref{wcpl6}) we obtain, owing to the operator identity
 $U^{i\dagger}_\alpha D_{ij}=0$ which follows from (\ref{Noetherid}):
\begin{eqnarray}
\delta_0\, \Big(\frac{\hat{\partial}\underline{a}}{\hat{\partial} \varphi^i}+D_{ji}^\dagger\,f^j\Big)=0
\label{wcpl10}
\end{eqnarray}
where the expression in parentheses has positive antifield number $D-p>2$. We conclude that this 
expression is the $\delta_0$-transformation of some $f_i$ with antifield number $D-p+1$ which yields
\begin{eqnarray}
\frac{\hat{\partial}\underline{a}}{\hat{\partial} \varphi^i}=-D_{ji}^\dagger\,f^j+\delta_0 f_i\ .
\label{wcpl11}
\end{eqnarray}
Analogously to equation (\ref{recon}) one can reconstruct $\dr^D x\,\underline{a}$ from the Euler 
derivatives of $\underline{a}$ with respect to the $C^\star$, $\varphi^\star$ and $\varphi$ up to 
a $\dv$-exact form, 
\begin{eqnarray}
\dr^D x\,\underline{a}=\frac 1N\, \dr^D x\,\Big( C^\star_\alpha\, 
\frac{\hat{\partial}\underline{a}}{\hat{\partial} C^\star_\alpha}+
\varphi^\star_i\, \frac{\hat{\partial}\underline{a}}{\hat{\partial} \varphi^\star_i}+
\varphi^i\, \frac{\hat{\partial}\underline{a}}{\hat{\partial} \varphi^i}\Big)+\dv (\dots) 
\label{wcpl12}
\end{eqnarray}
where $N$ is the degree of homogeneity of $\underline{a}$ in the fields and antifields. Using now 
equations (\ref{wcpl7}), (\ref{wcpl9}) and (\ref{wcpl11}) in (\ref{wcpl12}) one obatins
\begin{eqnarray}
\dr^D x\,\underline{a}&=&\delta_0\,k_D^{D-p+1}+\dv k_{D-1}^{D-p}\ ,\nonumber\\
 k_D^{D-p+1}&=&\frac {(-)^D}{N}\,\dr^D x\,(C^\star_\alpha f^\alpha -\varphi^\star_i f^i+\varphi^i f_i).
\label{wcpl13}
\end{eqnarray}
One now considers $\omega_D^{\prime\, D-p}=\omega_D^{D-p}-\delta\,k_D^{D-p+1}-\dv k_{D-1}^{D-p}$.
 If $\omega_D^{\prime\, D-p}$ vanishes one gets $\omega_D^{D-p}=\delta\,k_D^{D-p+1}+\dv k_{D-1}^{D-p}$.
 Otherwise $\omega_D^{\prime\, D-p}$ is treated as $\omega_D^{D-p}$ before and the procedure is iterated.
 For ``normal theories'' the linearized theory contains the maximum number of derivatives and the
 iteration can be shown to terminate resulting in 
\begin{eqnarray}
\omega_D^{D-p}=\delta\,\eta_D^{D-p+1}+\dv \eta_{D-1}^{D-p}\ .
\label{wcpl14}
\end{eqnarray}
Using (\ref{wcpl14}) in the last equation (\ref{wcpl1}) the latter gives 
\begin{eqnarray}
0=\dv ( \omega_{D-1}^{D-p-1} -\delta\, \eta_{D-1}^{D-p}).
\label{wcpl15}
\end{eqnarray}
Using the algebraic Poincar\'e lemma (theorem \ref{algebraic}) one concludes that the form in parentheses is equal to 
$\dv\eta_{D-2}^{D-p-1}$ for some $(D-2)$-form $\eta_{D-2}^{D-p-1}$ with antifield number $D-p-1$. 
This gives
\begin{eqnarray}
\omega_{D-1}^{D-p-1} =\delta\, \eta_{D-1}^{D-p}+\dv\eta_{D-2}^{D-p-1}\ .
\label{wcpl16}
\end{eqnarray}
In the same way one derives that all the forms $\omega_{p+k}^{k}$ with $k>0$ in (\ref{wcpl1}) are
 $\delta$-exact modulo $\dv$ in the case $p<D-2$:
\begin{eqnarray}
p<D-2,\ k>0:\ \omega_{p+k}^{k} =\delta\, \eta_{p+k}^{k+1}+\dv\eta_{p+k-1}^{k}\ .
\label{wcpl17}
\end{eqnarray}
Using this result for $k=1$ in the first equation (\ref{wcpl1}) one eventually gets
\begin{eqnarray}
p<D-2:\ 0=\dv (\omega_p^0 -\delta\, \eta_{p}^1)\ .
\label{wcpl18}
\end{eqnarray}
The algebraic Poincar\'e lemma now implies that the form in parentheses is constant if $p=0$ and 
$\dv$-exact if $p>0$,
\begin{equation}
p<D-2:\ \omega_p^0 =\left\{
\begin{array}{lr}
\delta\, \eta_{p}^1+\dv\eta_{p-1}^0&\ \text{if}\ p>0,\\
\delta\, \eta_{0}^1+\text{constant}&\ \text{if}\ p=0.
\end{array}
\right.
\label{wcpl19}
\end{equation}
Owing to $\delta\, \eta_{p}^1\approx 0$
this yields (\ref{char3}) for $r=0$ .

In the case $p=D-2$ the form $\omega_{D}^{2}$ can be taken as
\begin{eqnarray}
\omega_{D}^{2}=\dr^D x\,(C^\star_\alpha g^\alpha(\{\varphi\},x)+h(\{\varphi,\varphi^\star\},x))
\label{wcpl20}
\end{eqnarray}
where $h(\{\varphi,\varphi^\star\},x)$ is quadratic in the antifields $\varphi^\star$ and their derivatives.
The last equation in (\ref{wcpl1}) now gives
\begin{eqnarray}
0=\dv \omega_{D-1}^{1}+(-)^D\dr^D x\,\bigl (( -R^{i\dagger}_\alpha\varphi^\star_i) g^\alpha(\{\varphi\},x)+
\delta\,h(\{\varphi,\varphi^\star\},x)\bigr).
\label{wcpl21}
\end{eqnarray}
The Euler derivative of this equation with respect to 
$\varphi^\star_i$ yields $R^{i}_\alpha g^\alpha(\{\varphi\},x)\approx 0$
which imposes that the functions $g^\alpha(\{\varphi\},x)$ are (possibly field dependent) 
parameters of weakly vanishing gauge transformations.  Without going into further detail we note that 
in Yang Mills theories, Einstein gravity and gravitational Yang Mills theories and dimensions $D>2$ 
this implies $g^\alpha=0$ for all $\alpha$ except for $g^\alpha=\text{constant}$ when $\alpha$ labels 
a ``free Abelian gauge symmetry'', and that the characteristic cohomology for $p=D-2$ is in these 
theories represented by forms $\omega_{D-2}^0$ related via the descent equations (\ref{wcpl1}) to 
volume forms given simply by $\dr^D x\,C^\star_{i'}$ where $i'$ enumerates the ``free Abelian gauge 
symmetries''.

For $p=D-1$ the descent equations (\ref{wcpl1}) reduce to the single equation 
$0=\dv \omega_{D-1}^{0}+\delta\,\omega_{D}^{1}$ which directly provides (\ref{char5}) with 
$\omega_{D-1}^{0}\equiv J$. 

Using (\ref{char3}) one can derive the invariant characteristic cohomology, i.e. 
the characteristic cohomology on spin and isospin invariant tensor forms
$\omega_p(T,\dr x)=\dr x^{a_1}\dots \dr x^{a_p}f_{a_1\dots a_p}(T)$, in formdegrees $p<D-2$ 
similarly as the linearized covariant Poincar\'e lemmas in section \ref{sec5.3}. One obtains that in ordinary and 
gravitational Yang Mills theories and in Einstein gravity the invariant characteristic cohomology
is in formdegrees $p<D-2$ represented solely by Chern forms $P(F)$. 
This result extends for $D>2$ to formdegree $p=D-2$ if there are no ``free Abelian gauge symmetries'' 
and provides the weak covariant 
Poincar\'e lemma for these theories \cite{Barnich:1995ap}:
\begin{equation}
\dv \omega(T,\dr x)\approx 0\ \Leftrightarrow\ \omega(T,\dr x)\approx P(F)
+J_{\text{inv}}+\dr^Dx\, e\, f_{\text{inv}}(T)+\dv\eta(T,\dr x)
\label{char6}
\end{equation}
where $P(F)$ can contain a constant,
$J_{\text{inv}}$ is of the form (\ref{char5}) with $j^m=e\, E^m_a\, j^a(T)$ and $\eta(T,\dr x)$ is invariant.

\subsection{Antifield Dependent Representatives of the BRST Cohomology}

Each Noether form $J_{\text{inv}}$ in (\ref{char6}) gives rise to a cocycle of the 
cohomology of $\st$. We denote these cocycles by $\tilde J_{\text{inv}}$. Explicitly 
one may use
\begin{equation}
\begin{aligned}
\tilde J_{\text{inv}}&=\frac{1}{(D-1)!}\,\tilde c^{m_1}\dots\tilde c^{m_{D-1}}
\epsilon_{m_1\dots m_D}j^{m_D}\\
&+\frac{1}{D!}\,\tilde c^{m_1}\dots\tilde c^{m_{D}}\epsilon_{m_1\dots m_D}\, e\, G(T,T^\star)
\end{aligned}
\end{equation}
where $j^m$ are the components of the Noether current occurring in $J_{\text{inv}}$,
 $G(T,T^\star)$ is a spin and isospin invariant function of the tensors $T$ and antitensors
 $T^\star$ which has antifield number 1 and fulfills $\partial_m j^m=\delta (e\, G)$, and 
$\tilde c^m=\hat c^m+\dr x^m$ is the sum of the translation ghost $\hat c^m$ and the 
coordinate differential $\dr x^m$.

Owing to $\st \tilde J_{\text{inv}}=0$ one can finish the investigation of the antifield 
dependent \textsc{brst} cohomology along the lines of section \ref{sec6} by considering the $\tilde J_{\text{inv}}$ 
as Chern forms which have no Chern-Simons forms. Each $\tilde J_{\text{inv}}$ provides a 
representative $\omega^{-1}_D$ with ghostnumber $-1$ of the cohomology of $\s$ modulo $\dv$ 
given by
\begin{equation}
\omega_D^{-1}=\dr^Dx\,\, e\, G(T,T^\star)\ .
\end{equation}
Further antifield dependent representatives $\omega^g_D$ of the cohomology of $\s$ modulo $\dv$ 
arise from products of $\tilde J_{\text{inv}}$ and Chern Simons forms $q_\alpha$ (with the 
latter written in terms of the $\tilde C$). Among others this provides representatives $\omega^g_D$ 
with ghostnumbers $g=0$ and $g=1$ given by
\begin{gather}
\omega_D^{0}=\dr^Dx\,\bigl(e\,\hat C^i G(T,T^\star)-A_m{}^i \, j^m\bigr)\ ,\\
\omega_D^{1}=\dr^Dx\, \bigl(e\, \hat C^i\hat C^j G(T,T^\star)+(A_m{}^i \hat C^j-A_m{}^j \hat C^i)\,j^m\bigr)
\end{gather}
where $\hat C^i,\hat C^j$ are Abelian ghosts and $A_m{}^i,A_m{}^j$ are the corresponding Abelian 
gauge fields. 

Finally we comment on the case that ``free Abelian gauge symmetries'' are present. As discussed 
in subsection \ref{charcoh}, each of these symmetries gives rise to a nontrivial cohomology class 
of the characteristic cohomology in form degree $D-2$. Accordingly it induces corresponding 
modifications of the \textsc{brst} cohomological results. These modifications depend on the Lagrangian 
of the respective theory under consideration. We shall not discuss them in general here but 
restrict our comments to the cases that the gauge fields $A_m{}^{i'}$ of ``free Abelian 
gauge symmetries'' occur in the Lagrangian solely via terms 
$-\frac 14\,e\,F_{mn}{}^{i'}F^{mn\, j'}\delta_{i'j'}$ where $i'$ numbers the 
``free Abelian gauge symmetries'' and $F_{mn}{}^{i'}=\partial_m A_n{}^{i'}-\partial_n A_m{}^{i'}$ 
denote the corresponding field strengths. In this case the representatives of the 
characteristic cohomology in form degree $D-2$ are the Poincar\'e duals ${^*F}_{i'}$ of the 
field strength $2$-forms $\dr A^{i'}$ of the $A_m^{i'}$,
\begin{equation}
{^*F}_{i'}=\frac{1}{2!(D-2)!}\,\dr x^{m_1}\dots\dr x^{m_{D-2}}\epsilon_{m_1\dots m_D}
\,e\,F^{m_{D-1}m_D\,j'}\delta_{i'j'}\ .
\label{char4}
\end{equation}
Accordingly the weak covariant Poincar\'e lemma for $D>2$ (\ref{char6}) gets additional
 contributions $\lambda^{i'}{^*F}_{i'}$ which are linear combinations of the dual $(D-2)$-forms
 ${^*F}_{i'}$ with numerical coefficients $\lambda^{i'}$. These $(D-2)$-forms give rise 
to cocycles of $\st$ given by
\begin{equation}
\begin{aligned}
{^*\tilde F}{}_{i'}=&\frac{1}{2!(D-2)!}\,\tilde c^{m_1}\dots\tilde c^{m_{D-2}}
\epsilon_{m_1\dots m_D}\,e\,F^{m_{D-1}m_D j'}\delta_{i'j'}\\
&+\frac{1}{(D-1)!}\,\tilde c^{m_1}\dots\tilde c^{m_{D-1}}\epsilon_{m_1\dots m_D}A^{\star m_D}{}_{i'}\\
&+\frac{1}{D!}\,\tilde c^{m_1}\dots\tilde c^{m_{D}}\epsilon_{m_1\dots m_D}C^{\star}{}_{i'}\ .
\end{aligned}
\end{equation}
Each of these $\st$-cocycles ${^*\tilde F}{}_{i'}$ contains a representative 
$\omega^{-2}_D$ with ghostnumber $-2$ of the cohomology of $\s$ modulo $\dv$ given by
\begin{equation}
\omega_D^{-2}=\dr^Dx\,C^{\star}{}_{i'}\ .
\end{equation}
Antifield dependent representatives $\omega^g_D$ with ghostnumbers $g>-2$ of the 
cohomology of $\s$ modulo $\dv$ arise from products of ${^*\tilde F}{}_{i'}$ and 
Chern Simons forms $q_\alpha$ written in terms of the $\tilde C$.


\end{document}